\documentclass[prb,aps,twocolumn,showpacs,amsmath,amssymb,floatfix,eqsecnum,superscriptaddress]{revtex4-1}

\usepackage{graphicx}
\usepackage{times}
\usepackage{slashed} %Feynman slash notation
\usepackage{color}

\begin{document}

\def \beq {\begin{equation}}
\def \eeq {\end{equation}}
\def \beqa {\begin{eqnarray}}
\def \eeqa {\end{eqnarray}}
\newcommand \dg {\dagger}
\newcommand \up {\uparrow}
\newcommand \down {\downarrow}
\newcommand \al {\alpha}
\newcommand \be {\beta}
\newcommand \sg {\sigma}
\newcommand \ran {\rangle}
\newcommand \lan {\langle}
\newcommand \un {\underline}
\newcommand \ep {\epsilon}
\newcommand \lam {\lambda}
\newcommand \pd {\partial}
\newcommand \pdb {\bar{\partial}}
\newcommand \mb {\mathbf}
\newcommand \mbs {\boldsymbol}
\newcommand \yhat {\mathbf{\hat{y}}}
\newcommand \tb {\bar{t}}
\newcommand \nnb {\nonumber}
\newcommand \D {\Delta}
\newcommand \T {\mathcal{T}}
\newcommand \I {\mathcal{I}}
\newcommand \TI {\mathcal{TI}}
\newcommand \K {\hat{K}}
\newcommand \II {\mathbb{I}}
\newcommand \gamt {\tilde{\gamma}}
\newcommand \gam {\gamma}
\newcommand \dt {\tilde{d}}
\newcommand \Om {\Omega}
\newcommand \J {\mathcal{J}}
\newcommand \vphi {\varphi}
\newcommand \vth {\vartheta}
\newcommand \XX {\mathbb{X}}
\newcommand \ZT {Z^{\mathcal{T}}_2}
\newcommand \ZI {Z^{\mathcal{I}}_2}
\newcommand \ZTI {Z^{\mathcal{TI}}_2}
\newcommand \V {\mathcal{V}}
\newcommand \tmu {\bar{\mu}}
\newcommand \tnu {\bar{\nu}}

\title{A Bosonic Analog of a Topological Dirac Semi-Metal: Effective Theory, Neighboring Phases, and Wire Construction}

\author{Matthew F. Lapa}
\affiliation{Department of Physics and Institute for Condensed Matter Theory, University of Illinois at Urbana-Champaign, 61801-3080}
\author{Gil Young Cho} 
\affiliation{Department of Physics and Institute for Condensed Matter Theory, University of Illinois at Urbana-Champaign, 61801-3080}
\affiliation{Department of Physics, Korea Advanced Institute of Science and Technology, Daejeon 305-701, Korea}
\author{Taylor L. Hughes}
\affiliation{Department of Physics and Institute for Condensed Matter Theory, University of Illinois at Urbana-Champaign, 61801-3080}

\date{\today}

\begin{abstract}

We construct a bosonic analog of a two-dimensional topological Dirac Semi-Metal (DSM). The 
low-energy description of the most basic 2D DSM model consists of two Dirac cones at positions $\pm\mathbf{k}_0$ in 
momentum space. The local stability of the Dirac cones is guaranteed by a composite symmetry 
$Z_2^{\mathcal{TI}}$, where $\mathcal{T}$ is time-reversal and $\mathcal{I}$ is inversion. This model also exhibits interesting 
time-reversal and inversion symmetry breaking electromagnetic responses. In this work we construct a bosonic version by replacing each Dirac cone with a copy of the $O(4)$ Nonlinear Sigma Model (NLSM) with topological theta term and 
theta angle $\theta=\pm \pi$. One copy of this NLSM also describes the gapless surface termination of the 3D Bosonic 
Topological Insulator (BTI). We compute the time-reversal and inversion symmetry breaking electromagnetic responses for our 
model and show that they are twice the value one gets in the DSM case matching what one might expect from, for example, a bosonic Chern insulator. We also investigate the stability of the BSM model and find that the composite $\ZTI$ symmetry again plays an important role. Along the way we 
clarify many aspects of the surface theory of the BTI including the electromagnetic response, the charges and
statistics of vortex excitations, and the stability to symmetry-allowed perturbations. We briefly comment on the 
relation between the various descriptions of the $O(4)$ NLSM with $\theta=\pi$ used in this paper (a dual
vortex description and a description in terms of four massless fermions) and the recently proposed dual description of the 
BTI surface in terms of $2+1$ dimensional Quantum Electrodynamics with two flavors of fermion ($N=2$ QED$_3$). In
a set of four Appendixes we review some of the tools used in the paper, and also derive some of the more technical results.

\end{abstract}


\maketitle

\section{Introduction}
Massless 2+1-d Dirac fermions are one of the most well-studied systems in condensed matter physics. Such fermions often appear in relativistic field theories~\cite{Redlich}, but more importantly are known to be the low-energy description of the electronic structure some 2D materials, e.g., graphene~\cite{graphene}, and as the effective theory of the surface states of time-reversal invariant 3+1-d topological insulators~\cite{kanehasan}. In fact, in the latter two contexts alone, there have been thousands of articles in the past decade that discuss the properties of this fermion system.

The impetus for the intense focus on 2+1-d Dirac fermions was the experimental discovery of graphene~\cite{graphene}. Years earlier~\cite{wallace1947,semenoff1984} it had been theoretically predicted that the electronic band structure of graphene near the Fermi-level would be linear dispersing, gapless cones, i.e., massless Dirac fermions. Indeed, the unique signature of the Dirac fermions was quickly confirmed in quantum Hall measurements on graphene~\cite{graphene}. Graphene itself has four Dirac cones, two more than the minimum of two required to satisfy the Fermion doubling theorem in 2+1-d systems with time-reversal invariance. This theorem implies that a 2+1-d material with time-reversal symmetry cannot harbor an odd number of gapless Dirac cones. Hence, the system will have a semi-metallic nature with an even number of point-like Fermi surfaces, and is often referred to as a topological Dirac semi-metal (DSM). Remarkably, this 2+1-d (semi-)metal is relatively stable upon the requirement of some additional constraints: (i) inter-cone scattering across the Brillouin zone is suppressed (translation symmetry is sufficient for non-interacting fermions), (ii) intra-cone gapping terms are forbidden (minimally we need the composite symmetry of time-reversal combined with inversion), and  (iii) the system does not form a superconductor (we need to preserve $U(1)_c$). With these conditions the 2+1-d DSM forms a robust topological semi-metal phase. Interestingly, if we relax condition (ii) then the system will form a gapped insulator, but will typically have an unusual electromagnetic response (e.g., a quantum anomalous Hall effect\cite{haldane1988} or a charge polarization\cite{Ramamurthy2014}).

We can find examples of systems with an odd number of massless Dirac cones as well.  If we do not require time-reversal symmetry then there exist 2+1-d lattice models which have an odd number of Dirac cones, e.g., a Chern insulator model tuned to the topological critical point represents such a system~\cite{haldane1988}. 
On the other hand, there is  another way to avoid the Fermion doubling theorem while maintaining time-reversal (${\mathcal{T}}$). However, this requires something more drastic, i.e., we can produce an odd number of 2+1-d Dirac cones, and maintain $\mathcal{T},$ by considering the surface of a 3+1-d $\mathcal{T}$-invariant (electron) topological insulator (TI). The non-trivial $Z_2$ 3+1-d topological phase is known to have an odd number of massless Dirac cones on its surface with a characteristic spin-momentum locking feature of the states on the Fermi surface. Additionally, there must be at least one massless Dirac cone located at a time-reversal invariant momentum in the Brillouin zone. This is unlike the generic 2+1-d DSM for which the Dirac cones can exist at arbitrary points in the Brillouin zone~\footnote{In graphene the Dirac nodes lie at special points in the Brillouin zone, but these locations are required by the addition of spatial symmetries, not the stabilizing symmetry of time-reversal combined with inversion. At any rate, these special points $K$, and $K'$ are not time-reversal invariant momenta.}  It is well-known that theories with an odd number of 2+1-d massless Dirac cones typically exhibit the parity anomaly~\cite{Redlich}, and there are usually subtle features that must be carefully examined when considering the properties of such systems.

More recently there have been rapid developments in understanding symmetry-protected topological (SPT) phases with interactions~\cite{fidkowski2010,fidkowski2011,chen2011,turner2011,levin2012,tang2012,SenthilLevin,VL2012,VS2013,MKF2013,chan2013,xu2013wave,fidkowski2013non,SenthilWang2013,wang2014,wang2014int,kapustin2014,senthil2014symm}. One development in which we are particularly interested is the prediction that there could be bosonic analogs of the electron topological insulators. Some examples are the Bosonic Integer Quantum Hall Effect (BIQHE)~\cite{SenthilLevin,VL2012,grover2013,regnault2013,he2015,moller2009,moller2015,ye2013,liu2014} and the 3D $\mathcal{T}$-invariant Bosonic Topological Insulator (BTI) ~\cite{VS2013,MKF2013,SenthilWang2013,ye2015vortex}. The former is characterized by its quantized Hall conductance, which must come in integer multiples of $2e^2/h$, while the latter is characterized by a quantized magneto-electric polarizability with a $\Theta$-angle of $2\pi$ instead of the usual value of $\pi$ for the non-trivial phase of the electron topological insulator~\cite{QHZ2008}. These bosonic phases are not topologically ordered, but they are SPTs that require interactions to exist; at zeroth order the interactions serve to prevent the system of bosons from forming a trivial Bose condensate.

Consider, for a moment, the 3+1-d BTI. In analogy to the electron TI we expect the surface states to exhibit unusual properties. Indeed, for one example, the surface theory can exhibit an effectively 2+1-d ${\mathcal{T}}$-breaking phase with a Hall conductance of $\pm e^2/h$ which is forbidden for a purely 2+1-d BIQHE phase. To understand the properties of this exotic surface state several equivalent representations of the surface theory have been given in the literature: (i) a network model of quasi-1D strips that are arrayed to form a surface and coupled, (ii) a dual description of the surface bosonic theory in terms of dual vortices, and (iii) an effective field-theory description in terms of the $O(4)$ Nonlinear Sigma Model (NLSM) with a topological theta term with coefficient $\theta=\pi$. All three of these representations of the surface
were discussed in Ref.~\onlinecite{VS2013}. The description in terms of the $O(4)$ NLSM with $\theta=\pi$ was also discussed in Ref.~\onlinecite{xu2013AF}. Very recently, inspired by new developments in the description of the electron TI surface
\cite{son2015composite,metlitski2015particle,wang2015dual}, a new dual description of the BTI surface in terms of $2+1$-d Quantum Electrodynamics with two fermion flavors 
($N=2$ QED$_3$) was proposed~\cite{xu2015}. This new dual description was then derived in a coupled wires construction
in Ref.~\onlinecite{mross2015}. When any one of these theories is tuned to criticality it represents a surface state in a 
symmetry-preserving gapless phase. 

In this article our goal is to develop a thorough understanding of the surface of the 3+1d BTI, and then to subsequently combine multiple copies of the theory to form a symmetry preserving bosonic semi-metal state that can exist \emph{intrinsically} in 2+1-d without breaking some requisite symmetries. This type of semi-metal represents the bosonic analog of a 2+1-d DSM. We will present an effective theory for the bosonic semi-metal and explore in detail the requirements for its stability, the resulting electromagnetic responses, and possibilities for neighboring gapped phases with and without intrinsic topological order. We then provide an explicit coupled wires construction of this semi-
metal model. 

Our article is organized as follows: in Sec.~\ref{sec:overview} we give an overview of our main results, and in 
Sec.~\ref{sec:DSM-review} we review the properties of the 2+1-d fermion Dirac semi-metal. Next, in Sec.~\ref{sec:BTI-surface}, we review some properties of the surface theory of the 3+1-d $\mathcal{T}$-invariant BTI and provide new results and a synthesis of previous work. In Sec.~\ref{sec:BSM-model} we discuss our effective theory for the 2+1-d bosonic semi-metal built from multiple copies of the bosonic TI surface states, including the quasi-topological electromagnetic response, and the stability/instabilities of this critical state. 
In Sec.~\ref{sec:pol} we derive a criterion for identifying a gapless semi-metal phase from the value of its
polarization response.
Finally, in Sec.~\ref{sec:1D-bosonic-wire} we provide the details of the appropriate wire bundles and couplings to generate the bosonic semi-metal using a coupled-wire array. 
Following the conclusions we have a set of detailed Appendixes that review some of the technical tools used in the paper, and 
also contain explicit derivations of some of our more technical results.

\section{Motivation and Overview of Results}\label{sec:overview}

In this section we provide additional background motivation, describe the logic behind our construction of a bosonic analog 
of a topological DSM, and give an overview of our results. Henceforth, we call such a system a Bosonic Semi-Metal (BSM). 
 Readers interested in the technical details of the paper can refer
to the specific sections for more information. As mentioned above, the main goal of this paper is to construct a model of gapless bosons in 
2+1-d which shares many of the properties of the minimal two-node DSM of free fermions studied, for example,  in 
Ref.~\onlinecite{Ramamurthy2014}.
The main properties we will be interested in are: (1) the electromagnetic response of the system to perturbations which 
break time-reversal or inversion symmetry, and (2) the perturbative stability of the gapless, low-energy 
effective theory. As for any topological semi-metal, translation
symmetry is an important ingredient as it prevents any scattering processes between the different Dirac or bosonic ``cones"  (which are generically located at different points in momentum space). Indeed, in our BSM effective theory, translation symmetry
will forbid perturbations which could drive the system into a gapped state with only a trivial electromagnetic response.

Before we begin let us make a note about units. In this paper we consider systems constructed from fermions or bosons which all carry a single unit of
electric charge $e$. For most of the paper we work in units where $e=1$, but will restore the charge $e$ in 
all final response formulas. We also take $\hbar=1$, which means that the conductance quantum 
$\frac{e^2}{h} = \frac{1}{2\pi}$ in our units. We always express Hall conductances in units of $\frac{e^2}{h}$.

We start out in Sec.~\ref{sec:DSM-review}
by reviewing the continuum description of the two-node DSM.
We focus our review on the time-reversal and inversion symmetry breaking electromagnetic responses of the
DSM, and also the local (in momentum space) stability of the Dirac cones in the DSM. If the DSM is perturbed by gap-inducing terms that break $\mathcal{T}$ or $\mathcal{I}$ then the respective electromagnetic responses of the DSM take 
the forms
\begin{subequations}
\label{eq:DSM-responses}
\beqa
	\mathcal{L}_{\T} &=& \frac{e^2}{4\pi}\ep^{\mu\nu\lam}A_{\mu}\pd_{\nu}A_{\lam} \\
	\mathcal{L}_{\I} &=& \frac{e}{2\pi}\ep^{\mu\nu\lam}B_{\mu}\pd_{\nu}A_{\lam}\ ,
\eeqa
\end{subequations}
where $A_{\mu}$ is the potential for the external electromagnetic field, and $B_{\mu}$ is another 3-vector field
whose meaning is as follows. The two Dirac cones of the DSM are located at different points in the Brillouin zone with a wave-vector difference of $2B_i$ (for simplicity we choose their locations to be at $\mb{k}_{\pm} = \pm (B_x,B_y)$), and the two cones are separated in energy by an amount $2B_t$. The effective Lagrangian $\mathcal{L}_{\T}$ represents
a 2+1-d quantum Hall response with a Hall conductance of $1$, while $\mathcal{L}_{\I}$ represents a charge
polarization and orbital magnetization response whose precise meaning was discussed in Ref.~\onlinecite{Ramamurthy2014}, and which we 
review in Sec.~\ref{sec:1D-bosonic-wire}. We note that we have suppressed a sign in these terms that tracks the nature of the inversion or time-reversal breaking. 

Typically point-node semi-metals are unstable in 2+1-d unless extra symmetries are imposed. The stability of the DSM is due in part to the translation symmetry of the system. This symmetry prevents scattering processes
between Dirac cones at different locations in the Brillouin zone. The local stability (in momentum space) of each Dirac cone in the DSM 
(at the level of free fermions) is then guaranteed 
by $U(1)_c$ charge conservation symmetry and a composite symmetry $\ZTI$, consisting of a time-reversal
transformation combined with an inversion transformation. These two symmetries forbid translation invariant terms which could
gap out a single Dirac cone independently of any of the others. 
 
Having reviewed the DSM, we can make the following observation about a minimal, two-cone DSM which directly informs
our construction of a BSM model. Since a single Dirac cone
is the surface theory for the 3+1-d Electron Topological Insulator (ETI)\cite{kanehasan} 
the degrees of freedom in the two-node DSM can be
viewed as being constructed from two copies of the ETI surface theory, but with the two copies separated in momentum
space. We are therefore motivated to construct a model for a BSM from two copies of 
the surface theory for the 3D Bosonic Topological Insulator (BTI), but with those two copies also separated in momentum
space. According to Ref.~\onlinecite{VS2013}, one representation of the surface theory of the BTI is the $O(4)$ NLSM with
theta term and $\theta=\pi$, and it is this theory which we discuss next.

In Sec.~\ref{sec:BTI-surface} we give a lengthy review of the properties of the $O(4)$ NLSM with $\theta=\pi$ as it appears 
on the surface of the BTI. 
There are two reasons for giving an extended discussion of the BTI surface theory: (1) understanding
just one copy of this theory is a prerequisite for understanding our BSM effective theory, which consists of two copies of the surface theory,
and (2) we provide alternate derivations (and also proofs in Appendixes~\ref{app:vortices} and \ref{app:path-integral})  
for some of the properties
of this model. These discussions, and some additional new results, lend further support to many of the claims about this model that have already appeared in the literature. In particular we provide an extended discussion on the stability of the gapless nature of the $O(4)$ surface theory that we will require for our discussion of the BSM theory. 

To begin, we recall that the $O(4)$ NLSM can be equivalently formulated in terms of an $SU(2)$ matrix field
\beq
	U= \begin{pmatrix}
	b_{1} & -b_{2}^{*} \\
	b_{2} & b_{1}^{*}
\end{pmatrix}\ , 
\eeq
where the components $b_1$ and $b_2$ are interpreted as representing physical bosons on the surface of the BTI. As
such, they transform under the physical $U(1)_c$ charge conservation symmetry as $b_I \to e^{i\chi}b_I$ for $I=1,2$ 
(in units where the boson charge $e=1$).
This theory also has a time-reversal symmetry $\ZT$ under which $b_1$ and $b_2$ are separately invariant. The action for
this model includes the conventional NLSM ``kinetic energy" term, and the topological theta term,
\beq
	S_{\theta}[U]=\frac{1}{24 \pi^2} \int d^3 x\ \ep^{\mu\nu\lambda} \text{tr}[(U^{\dg}\pd_{\mu} U)( U^{\dg}\pd_{\nu} U)( U^{\dg}\pd_{\lambda} U)]\ ,
\eeq
where $\text{tr}[\cdots]$ is the usual trace operation. In the action, the theta term is multiplied by a parameter
$\theta$, which is an angular variable defined modulo $2\pi$. For the surface theory of the BTI we have $\theta=\pi$ 
\cite{VS2013}.
In Sec.~\ref{sec:BTI-surface} we review the calculation of the time-reversal breaking electromagnetic response of this theory
via its dual vortex description (developed in Refs.~\onlinecite{SenthilFisher,VS2013})
and also discuss an alternate method for calculating this response that confirms this result. We also
comment on the relation between the descriptions of the BTI surface used in this paper and the recently proposed
dual description of the BTI surface in terms of $N=2$ QED$_3$ \cite{xu2015,mross2015}.
We then go on to give a careful discussion of the effects that perturbations allowed by the $U(1)_c$ and $\ZT$
symmetries have on the surface theory. These perturbations were only discussed briefly in Ref.~\onlinecite{VS2013}. Finally, 
we review the construction of the symmetry-preserving $\mathbb{Z}_2$ 
topologically ordered surface phase of the BTI which was
first derived in Ref.~\onlinecite{VS2013}. 

After all of this setup we are ready to introduce an effective theory of the BSM. In Sec.~\ref{sec:BSM-model} we introduce a system with two copies of the $O(4)$ NLSM with theta term. One copy has $\theta=\pi$ and the other copy has $\theta=-\pi$, and  just as in the case of the fermionic DSM, the two copies of the
$O(4)$ NLSM are located at positions $\mb{k}_{\pm}= \pm (B_x,B_y)$ in momentum space.
In our description of the effective theory we discern how charge conservation, translation, time-reversal,
and inversion symmetries act on the fields in the model, and then compute the time-reversal and inversion breaking 
electromagnetic responses analogous to those found in the fermion DSM. We find that these responses also exist in the BSM case, and have exactly twice the value of the 
responses in Eq.~\eqref{eq:DSM-responses} for the free fermion DSM. This doubling of the response for bosonic vs. fermionic
systems is similar to what happens for the case of the ETI and BTI in 3D, 
and also the integer quantum Hall effects for fermions and bosons in 2D \cite{VS2013,SenthilLevin}. 

We then go on to give a partial discussion of the stability of our theory. We argue that the translation symmetry prevents us 
from coupling one copy of the theory to the other copy in order
to drive the system into a trivial insulating state, and that the combined $\ZTI$ symmetry ensures
the stability of each individual $O(4)$ NLSM. Finally, we discuss some 2D topologically ordered phases which can be 
accessed from our BSM model by condensing suitable bound states of the vortices in the theory. 
In particular, we find a phase with $\mathbb{Z}_2\times \mathbb{Z}_2$ topological order which breaks
either the time-reversal or the inversion symmetry of the original BSM model. This phase is essentially two copies of the 
$\mathbb{Z}_2$ topologically ordered 
phase found in Ref.~\onlinecite{VS2013}, but in which the time-reversal and the inversion symmetry of the BSM exchange the 
two copies. We also discuss phases with $\mathbb{Z}_2$ topological order which break either
the inversion or the time-reversal symmetry of the BSM model.

In Sec.~\ref{sec:pol} we give a different perspective on the stability of \emph{any} semi-metal phase by
relating the gaplessness of the semi-metal to its polarization response. In particular,
we consider three broad classes of 2D gapped phases with translation symmetry which can have a polarization response, and 
we show that for these classes of gapped phases the polarization in the $x$ or $y$ direction (in the presence of inversion 
symmetry) is 
always of the form $r\frac{e}{2 a_0}$, where $r\in\mathbb{Q}$ is a \emph{rational} number and $a_0$ is the lattice spacing. 
In addition, for each class of gapped phase we are able to relate the number $r$ to simple measurable properties of that
phase. A crucial point is that the three classes of gapped phases that we consider are representative of all gapped 2D phases 
with translation symmetry which could be expected to exhibit a polarization response. Our result then implies that 
a generic (i.e., non-rational) value of the polarization in a system with translation symmetry, and in particular a 
\emph{continuously tunable} polarization, indicates a gapless semi-metal phase. This shows that the gaplessness of the
semi-metal is directly related to its physically measurable polarization response. Since this response is expected to be
reasonably robust, this also provides additional evidence for the stability of the semi-metal phase itself.

Finally, in Sec.~\ref{sec:1D-bosonic-wire} we give an explicit construction of the BSM model using an array of 
coupled 1D wires. This construction is motivated by the fact that  2+1-d fermion DSMs can be constructed out of arrays of coupled wires~\cite{Ramamurthy2014}. A building block for the simplest two-node DSM of fermions is a wire with a single 1+1-d massless Dirac fermion. When coupled, arrays of these wires may exhibit three related phases: (i) if an array of these wires is dominated by an intra-wire topological tunneling term then the system becomes a 2+1-d weak topological insulator that exhibits a charge polarization parallel to the wires, (ii) if the array is dominated by an inter-wire topological tunneling term then the array forms a Chern insulator phase with an integer Hall conductivity of $\pm e^2/h$, or (iii) if there is significant competition between an intra-wire and inter-wire tunneling there can be a parent critical \emph{phase}, i.e., a DSM, which is unstable to the formation of phase (ii) if  time-reversal is broken, and unstable to phase (i) if the Dirac nodes meet at the boundary of the Brillouin zone and annihilate. 

A key observation of this construction is that a wire with a single 1+1-d Dirac fermion can be thought of as a narrow strip of a 
$\sigma_{xy}=e^2/h$ integer quantum Hall system. Hence, by analogy, we can immediately propose a 1+1-d bosonic wire 
model to serve as the building block for a 
coupled-wire construction of our 2+1-d BSM state: a narrow strip of the BIQHE, which will contain the degrees of freedom from 
both edges. An edge of the BIQHE can be described by an $SU(2)_1$ Wess-Zumino-Witten (WZW)
conformal field theory (CFT) \cite{SenthilLevin,VS2013}, 
therefore our 1+1-d bosonic wires will consist of two (time-reversed) copies of an $SU(2)_1$ WZW theory. The fields in 
each wire consist of bosons which carry charge $1$ under the $U(1)_c$ symmetry.

It has been known for some time that one copy of the $O(4)$ NLSM with $\theta=\pi$ can be obtained from an array of coupled
wires in which each wire contains a \emph{single} $SU(2)_1$ WZW theory \cite{SenthilFisher,TanakaHu,VS2013}. 
After giving a brief review of this result, we then show how our BSM model can be derived starting with 1+1-d bosonic
wires containing two $SU(2)_1$ WZW theories. We construct inter-wire tunneling terms which not only give the 
desired $O(4)$ NLSM's with theta angles $\pi$ and $-\pi$, but also shift the two copies of the $O(4)$ NLSM's to the locations
$\mb{k}_{\pm}= \pm (B_x,B_y)$ in momentum space (our specific construction gives the case with $B_x=0$). We then
show how to assign transformations under time-reversal and inversion symmetry to the fields in the coupled wires model  
so that the transformations of the fields in the BSM model are recovered in the continuum limit. We conclude 
Sec.~\ref{sec:1D-bosonic-wire} with a discussion of the different physical interpretations of the coupled wire constructions of
the DSM and BSM models, and we also indicate how inversion and time-reversal breaking perturbations of the BSM model
can be explored within its quasi-1D coupled wire description.

In Appendix~\ref{app:canonical} 
we review the canonical quantization of the $O(4)$ NLSM, and also work out the commutators for
this theory when expressed in terms of the constrained bosonic variables $b_1$ and $b_2$. This information is used in 
Sec.~\ref{sec:BTI-surface} to investigate the effects of symmetry-allowed perturbations on the BTI surface theory, and 
also in Sec.~\ref{sec:BSM-model} to discuss the stability of the BSM model to symmetry-allowed perturbations.
In Appendix~\ref{app:vortices} we study a family of exact, finite energy vortex solutions to the NLSM equations of 
motion, and we compute the quantum numbers carried by global excitations in the background of a single vortex. In 
particular, we are able to prove the result, first argued for in Ref.~\onlinecite{VS2013},
that the main effect of the theta term in the $O(4)$ NLSM is to attach a charge 
$\frac{\theta}{2\pi}$ of the boson $b_1$ to vortices in the phase of $b_2$, and vice-versa. 
In Appendix~\ref{app:path-integral} we discuss the role of the theta term of the $O(4)$ NLSM in the Minkowski spacetime
path integral of the theory.
Finally, in Appendix~\ref{app:fermion-puzzle} we resolve an apparent paradox associated with our alternative calculation, via 
auxiliary fermions, of the time-reversal breaking electromagnetic response of the BTI surface theory.

%Finally, in 
%Appendix~\ref{app:VO-rep}
%we review the vertex operator representation of the $SU(2)_1$ WZW model. This
%vertex operator representation is our main tool for studying our 1+1-d bosonic wire model in Sec.~\ref{sec:1D-bosonic-wire}.

\section{Review of the free-fermion Dirac Semi-Metal}
\label{sec:DSM-review}

Before going into the details of our construction of a bosonic analog of a Dirac semi-metal (DSM), we first give a review
of the free fermion DSM for the simple case of two Dirac points in the Brillouin zone. Our review closely follows
the discussion in Ref.~\onlinecite{Ramamurthy2014}, in which the 
electromagnetic responses of various topological semi-metals were derived. Specifically, we discuss a square lattice model of a DSM, its symmetry requirements, its
low-energy description, the time-reversal and inversion symmetry breaking electromagnetic
responses of the system, and finally the \emph{local} stability of the Dirac nodes.
A particularly important point is that the local stability
of the DSM, which means the stability against perturbations that can gap out individual Dirac cones, is guaranteed by 
enforcing a composite symmetry $\ZTI$, whose action consists of a time-reversal transformation composed with an inversion 
transformation. This composite symmetry will also play an important role in our bosonic semi-metal model.

We also note here that the surface theory of the 3D Electron Topological Insulator (ETI) is a single Dirac fermion. We may 
therefore view the simple two cone DSM as a theory constructed from similar degrees of freedom as two copies of the surface theory of the 3D ETI (but with
the two copies of the theory having opposite helicity). This observation is the motivation for our construction of a bosonic
semi-metal from two copies of the surface theory of the 3D Bosonic Topological Insulator (BTI), which is an $O(4)$ NLSM
with theta term and theta angle $\theta=\pm \pi$. We return to this point in later sections.  

\subsection{Lattice model of a DSM}

We now describe a lattice model, discussed in more detail in Ref.~\onlinecite{Ramamurthy2014}, 
which realizes a DSM phase for a certain range of parameters. 
The model consists of two species/orbitals of spinless fermions at half-filling
on the square lattice. We therefore have a two-component complex fermion operator $\vec{c}_{\mb{n}}$ at each site
$\mb{n}= (n_x,n_y)$ of the square lattice. We take the lattice spacing $a_0=1$. 
A number of symmetries play an important role in this system. 
They are discrete translation symmetry, $U(1)_c$ charge conservation 
symmetry, $\ZT$ time-reversal symmetry, and $\ZI$ inversion symmetry.
The fermions carry charge $1$, so they transform under 
$U(1)_c$ as
\beq
	U(1)_c: \vec{c}_{\mb{n}} \to e^{i\chi }\vec{c}_{\mb{n}}\ ,
\eeq
where $\chi$ is a constant phase. The action of the time-reversal and inversion symmetries on the 
fermions is given in terms of the anti-unitary operator $\mathcal{T}$ and the unitary operator 
$\mathcal{I}$, respectively. We will specify the action of these operators on the complex fermions after introducing the DSM 
model.

%Assuming periodic boundary conditions, we may Fourier transform to momentum space by defining
%\beq
%	\vec{c}_{\mb{k}}= \frac{1}{\sqrt{N}}\sum_{\mb{n}}\vec{c}_{\mb{n}}e^{-i \mb{k}\cdot \mb{n} a_0}\ ,
%\eeq
%where $\mb{k}$ is a wavevector in the first Brillouin zone, $a_0$ is the lattice spacing, and $N$ is the number of unit cells in the 
%system. We take $a_0=1$ for the rest of this section.
%The Hamiltonian for the system 
%can be written in terms of the operators $\vec{c}_{\mb{k}}$ as
%\beq
%	H= \sum_{\mb{k}}\vec{c}^{\ \dg}_{\mb{k}}\mathcal{H}(\mb{k})\vec{c}_{\mb{k}}\ ,
%\eeq
%where $\mathcal{H}(\mb{k})$ is the Bloch Hamiltonian.

The Bloch Hamiltonian of this model takes the form
\beq
	\mathcal{H}_{2D}(\mb{k})= \sin(k_x )\sigma^x + (1-m-\cos(k_x)-t_y\cos(k_y ))\sigma^z\ , \label{eq:DSM-lattice}
\eeq
where $\sigma^a$, $a= x,y,z$ are the Pauli matrices acting on the orbital space.
The $\sin(k_x )$ term represents a complex hopping for fermions in the $x$ direction, while the 
terms multiplying $\sigma^z$ represent a mass term as well as real hopping terms in the $x$ and $y$ directions.
This system has time-reversal and inversion symmetry where $\mathcal{T}$ and $\mathcal{I}$  act on the fermions
as
\beq
	\mathcal{T}\vec{c}_{\mb{n}}\mathcal{T}^{-1}= \sigma^z\vec{c}_{\mb{n}}\ ,
\eeq
and 
\beq
	\mathcal{I}\vec{c}_{\mb{n}}\mathcal{I}^{-1}= \sigma^z\vec{c}_{-\mb{n}}\ .
\eeq
We see that both time-reversal and inversion symmetry act with opposite signs on the two species of fermion. We note that the 
inversion symmetry also negates the spatial coordinate, and $\mathcal{T}$ is anti-unitary.

The energies of the two bands of this model as a function of $\mb{k}$ are given by
\beq
	E(\mb{k})_{\pm}= \pm \sqrt{ \sin^2(k_x) +  \big(1-m-\cos(k_x)-t_y\cos(k_y)\big)^2     }\ . \label{eq:DSMenergy}
\eeq
When $t_y=0$ and $m=0$, this model has a band touching at $k_x=0$ for any value of $k_y$.
However, for non-zero $m$ and $t_y$ the bands touch only at isolated points along the line $k_x=0$ in the Brillouin zone.
The location of these points is determined by the ratio of $m$ and $t_y$. We focus our attention on the regime of 
$m>0$ but $m\ll 2$ (at $m=2$ the band touching moves to $k_x=\pi$). In this regime the low-energy physics
of this system is completely described by two continuum Dirac Hamiltonians obtained by linearizing around the two band touchings. These band touchings are located at the points $\mb{k}_{\pm}= (0,\pm B_y)$ in the Brillouin zone, where
$B_y$ is the positive solution, in the first Brillouin zone, to the equation $m + t_y\cos(B_y)= 0$.

Performing a $k\cdot P$ expansion around $\mb{k}_{\pm}$  we find the low-energy Dirac Hamiltonians 
$\mathcal{H}_{\pm}(\mb{k})$ 
\beq
	\mathcal{H}_{\pm}(\mb{k})= k_x\sigma^x \pm t_y\sin(B_y)(k_y \mp B_y)\sigma^z\ .
\eeq
We emphasize that these two low-energy Hamiltonians have opposite signs on their $k_y$ terms. This means that
the two Dirac fermions which emerge at low-energy in this model have opposite \emph{helicity}, i.e., the Berry phase for electrons on the two Fermi surfaces when the chemical potential is tuned away from the Dirac points have opposite signs.
This form of the low-energy Hamiltonian for each Dirac point leads us directly to the continuum description of the DSM.

\subsection{Continuum description of the DSM}

For the continuum description of the DSM, we take as our starting point an effective Hamiltonian for two continuum Dirac fermions
$\psi_A$ and $\psi_B$ with
opposite helicity (which is exactly what we found
in the linearized Bloch Hamiltonian for the DSM model).
We then shift the zero in momentum in the $k_y$ direction by $B_y$. This leads to the Hamiltonian
\beq
	\mathcal{H}_{DSM}(\mb{k})= k_x \mathbb{I}\otimes\sigma^x + k_y\sigma^z\otimes\sigma^z - B_y\mathbb{I}\otimes\sigma^z\ ,
\eeq
where for simplicity we have taken $t_y\sin(B_y)= 1$ to make the dispersion of the Dirac cones isotropic. 
In position space the original lattice fermions $\vec{c}_{\mb{n}}$ may be written at low energies in terms of the two 
continuum Dirac fermions $\psi_A$ and $\psi_B$ as 
\beq
	\frac{\vec{c}_{\mb{n}}}{a_0} \sim \psi_A(\mb{x})e^{iB_y y} + \psi_B(\mb{x})e^{-iB_y y}\ , \label{eq:expansion}
\eeq 
where $\mb{x}=(x,y)=(n_x a_0,n_y a_0)$, and we have temporarily restored the lattice spacing. Now we define the multi-
component fermion operator $\Psi= (\psi_A,\psi_B)^T$. One can show that
$\mathcal{I}$ and $\mathcal{T}$ act on
$\Psi$ as
\beq
	\mathcal{I}\Psi(\mb{x})\mathcal{I}^{-1}= \sigma^x\otimes\sigma^z \Psi(-\mb{x})\ ,
\eeq
and
\beq
	\mathcal{T}\Psi \mathcal{T}^{-1}= \sigma^x\otimes\sigma^z \Psi\ .
\eeq
In particular, these operators exchange $\psi_A$ and $\psi_B$, i.e., they each map fermions from one Dirac cone to the other. 

At this point we may
go ahead and generically allow for an offset $B_x$ to the $k_x$ location of the Dirac points, as well as an offset $B_t$ between the energies
of the two Dirac points. This leads to the effective Hamiltonian
\beqa
	\mathcal{H}_{DSM}(\mb{k}) &=& k_x \mathbb{I}\otimes\sigma^x - B_x \sigma^z\otimes\sigma^x + k_y\sigma^z\otimes\sigma^z \nnb \\ 
&-& B_y\mathbb{I}\otimes\sigma^z + B_t \sigma^z\otimes\mathbb{I}\ . \label{eq:DSM-ham}
\eeqa
To more clearly see the final structure we can pass to a Lagrangian formulation of this system. The Lagrangian has the form 
\beq
	\mathcal{L}= \bar{\Psi}\left(i\slashed{\pd} + \slashed{A} + (\sigma^z \otimes\mathbb{I})\slashed{B}\right)\Psi\ ,
\eeq
where we define the gamma matrices $\gamma^0=\mathbb{I}\otimes\sigma^y,\gamma^1=-i\mathbb{I}\otimes\sigma^z$,
and $\gamma^2 = i\sigma^z\otimes\sigma^x$, 
and where $\bar{\Psi}= \Psi^{\dg}\gamma^0$. We have also employed the Feynman slash notation 
$\slashed{\pd}= \gamma^{\mu}\pd_{\mu}$, etc., and 
included minimal coupling to the external electromagnetic field $A_{\mu}$  connected to the $U(1)_c$ symmetry.

\subsection{Electromagnetic response of the DSM}

We now briefly review the electromagnetic response of the DSM to time-reversal and inversion symmetry breaking 
perturbations. The two mass terms
\beqa
	\Sigma_{\mathcal{I}} &=& \mathbb{I}\otimes\sigma^y \\
	\Sigma_{\mathcal{T}} &=& \sigma^z\otimes\sigma^y\ ,
\eeqa
are the only matrices that anti-commute with the kinetic energy terms of the DSM Hamiltonian in Eq.~\eqref{eq:DSM-ham},
\emph{and} preserve translation invariance (i.e., they do not couple $\psi_A$ to $\psi_B$, which are located at different
points in momentum space). The first term $\Sigma_{\mathcal{I}}$ breaks inversion symmetry, while the second term
$\Sigma_{\mathcal{T}}$ breaks time-reversal symmetry.

As shown in Ref.~\onlinecite{Ramamurthy2014}, perturbing the system with a term $-m\Sigma_{\mathcal{T}}$ 
leads to the 2D electromagnetic response $\mathcal{L}_{\mathcal{T}}$ from Eq.~\eqref{eq:DSM-responses}, i.e., it induces a quantum anomalous Hall effect with Chern number/Hall conductance\cite{haldane1988} $\sigma_{xy}=\pm e^2/h$.
On the other hand, perturbing the system with $-m\Sigma_{\mathcal{I}}$ leads to the  
quasi-1D electromagnetic response $\mathcal{L}_{\mathcal{I}}$ from Eq.~\eqref{eq:DSM-responses}. This response indicates that, when starting from the gapped, inversion breaking phase and taking $m\to 0$, the DSM limit will have a charge polarization and/or orbital magnetization. In fact, the response does not depend on the magnitude of $m$ at all, and the dependence on $m$ enters only as a global sign $\text{sgn}(m)$ multiplying the response formula. Interestingly, this second response term depends crucially on the properties of the Dirac nodes, i.e., their relative positions in momentum and energy. 

\subsection{Combined $\mathcal{TI}$ symmetry ensures local stability of the Dirac cones}

We end this section with a quick comment about the stability of the DSM. We saw in the previous section that the only mass
terms that we can add to Eq.~\eqref{eq:DSM-ham} which are allowed by translation symmetry are the terms 
$\Sigma_{\mathcal{T}}$ and $\Sigma_{\mathcal{I}}$. If we add only one of these mass terms to the system then it will
gap out both of the Dirac cones. However, suppose we tried to add a linear combination of these two mass terms. The 
two possible linear combinations are
\beq
	\Sigma_{\pm}= \frac{1}{2} (\Sigma_{\mathcal{I}}\pm\Sigma_{\mathcal{T}})\ .
\eeq
Adding just one of these terms would gap out
either $\psi_A$ (add $\Sigma_{+}$) or $\psi_B$ (add $\Sigma_{-}$). However, both of these terms are forbidden by
the composite symmetry $\mathcal{TI}$. Therefore, the local stability of the Dirac cones is guaranteed by the combined
time-reversal times inversion symmetry $\mathcal{TI}$.\cite{bernevigbook} If we enforce this symmetry, then it 
is impossible to gap out one of the Dirac cones independently of the other cone, and hence they can only be removed if they are perturbed enough to collide with each other in momentum space. This means that with translation, $\mathcal{TI}$, and $U(1)_c$ preserved the DSM is a (perturbatively) stable 2+1-d semi-metal phase.

\section{The surface theory of the bosonic topological insulator}
\label{sec:BTI-surface}

In this section we review, and also clarify some aspects of, the surface theory of the 3+1-d BTI. Since our BSM model
is constructed from two copies of the surface theory of the BTI, it is essential that we discuss this theory
in detail. The surface theory of the BTI
was first derived in Ref.~\onlinecite{VS2013}, where it was obtained from a network model constructed from coupled
edge theories of the BIQH state
(we briefly discuss this network model in  Sec.~\ref{sec:1D-bosonic-wire}). 
The authors of Ref.~\onlinecite{VS2013} then used this theory,
as well as a dual vortex description of the theory, to derive many possible surface phases for the BTI. These different
possible surface phases were further investigated and clarified in Ref.~\onlinecite{MKF2013} which utilized 
monopole configurations of the external gauge field  to probe the properties of the various phases.

In this section we provide a detailed account of the surface theory of the BTI, which is equivalent to an $O(4)$ NLSM with theta term and theta angle $\theta=\pi$.
We first discuss the basic properties of this theory, and also the transformations of the $O(4)$ field under the physical
symmetry group of the BTI. We then give a summary of the dual description of the theory, but from a different point of view
than the one given in Ref.~\onlinecite{VS2013}. We then show how the time-reversal symmetry breaking electromagnetic
response of the BTI surface can be obtained from the dual description. We also describe an alternative method for calculating 
the electromagnetic response of the theory. This method uses a well-known formula derived by Abanov and Wiegmann
in Ref.~\onlinecite{AbanovWiegmann}, which allows one to write the original $O(4)$ NLSM as a path integral over a set
of auxiliary fermions which couple to the $O(4)$ field. Since the $O(4)$ NLSM is such a difficult system to study, having
two different methods for calculating the response which give the same answer is strong corroborating evidence. We next discuss the stability of the gapless theory. In particular, we carefully study the effects of
symmetry-allowed perturbations, some of which were briefly discussed in Ref.~\onlinecite{VS2013}. 
Finally, we end the section with a brief
review of the symmetry-preserving, topologically ordered surface phase for the BTI proposed in Ref.~\onlinecite{VS2013}. After all of this is complete we will be ready to discuss the properties of the bosonic semi-metal state.

\subsection{The $O(4)$ NLSM with theta term}

In this subsection we review the description of the surface of the Bosonic Topological Insulator (BTI) in terms of one 2+1-d $O(4)$ Nonlinear Sigma Model (NLSM) with a topological theta term having $\theta= \pi$. The $O(4)$ NLSM field 
$\mb{N}= (N^1,N^2,N^3,N^4)$ is a real-valued unit vector field (i.e., $\mb{N}\cdot\mb{N}= 1$). The action for 
this theory with a general theta angle takes the form
\beq
	S= \int d^3 x\ \frac{1}{g}(\pd^{\mu}N^a)(\pd_{\mu}N^a) - \theta S_{\theta}[\mb{N}]\ , \label{eq:O4}
\eeq
where we sum over all repeated indices ($\mu= t,x,y$ and $a= 1,2,3,4$), and the theta term is
\beq
	S_{\theta}[\mb{N}]= \frac{1}{12\pi^2}\int d^3 x\ \ep^{\mu\nu\lambda}\ep_{abcd} N^a\pd_{\mu}N^b\pd_{\nu}N^c\pd_{\lam}N^d\ . \label{eq:theta}
\eeq
The coefficient $g$ is a positive coupling constant. 
Small $g$ favors an ordered phase in which $N^a$ is constant everywhere in spacetime,
while large $g$ favors a disordered phase. The theta term only plays a role in the disordered phase, so we assume that 
we are working in the large $g$ regime. 

%(\textbf{An interesting point is that $g$ carries dimensions of length, so one should 
%really say what length $g$ is large compared to. I don't really know the answer to that.})

For the description of the surface of the BTI, it is more convenient to use a formulation of the $O(4)$ NLSM in terms of an 
$SU(2)$ matrix $U$ which is related to the unit vector $\mb{N}$ via $U= N^4\mathbb{I} + \sum_{a= 1}^3 i N^a\sigma^a$.
In terms of $U$ the action takes the form
\beq
	S= \int d^3 x\ \frac{1}{2 g}\text{tr}[\pd^{\mu} U^{\dg} \pd_{\mu} U] - \theta S_{\theta}[U]\ , 
\eeq
where now
\beq
	S_{\theta}[U]=\frac{1}{24 \pi^2} \int d^3 x\ \ep^{\mu\nu\lambda} \text{tr}[(U^{\dg}\pd_{\mu} U)( U^{\dg}\pd_{\nu} U)( U^{\dg}\pd_{\lambda} U)]\ ,\label{eq:thetaPC}
\eeq
and $\text{tr}[\dots]$ denotes the usual trace operation for matrices.
In this form, the $O(4)$ NLSM is also known as the $SU(2)$ Principal Chiral Nonlinear Sigma Model (PCNLSM).

The renormalization group (RG) flows of general $SU(N)$ PCNLSM's in the $(g,\theta)$ plane were 
studied qualitatively in Ref.~\onlinecite{XuLudwig}. In that paper the authors argued that the theory with $\theta=\pi$ 
could either be gapless or have a degenerate ground state. In the gapless case they predicted an RG fixed point at 
$\theta=\pi$ and $g=g^*$ for some finite $g^*$, while for the degenerate case they predicted that
$g$ flows off to positive infinity. In this paper we focus only on the first possibility of 
a gapless theory. We might also suspect that the $O(4)$ NLSM
at $\theta=\pi$ is gapless on the grounds that its lower dimensional cousin, the $O(3)$ NLSM with $\theta=\pi$, was 
also shown to be gapless in Ref. \onlinecite{ShankarRead}.

For the description of the BTI surface, one writes $U$ in terms of two complex fields $b_1$ and $b_2$ as
\beq
	U= \begin{pmatrix}
	b_{1} & -b_{2}^{*} \\
	b_{2} & b_{1}^{*}
\end{pmatrix}\ , \label{eq:b-variables}
\eeq
where $b_1$ and $b_2$ are subject to the constraint $\sum_{I=1}^2 |b_I|^2 = 1$,
which is equivalent to the original constraint $\mb{N}\cdot\mb{N}=1$ of the
$O(4)$ NLSM. We should think of $b_1$ and $b_2$ as representing the physical bosonic degrees of freedom
on the surface of the BTI, and so we will refer to $b_I$, $I=1,2$, as ``bosonic fields" for the rest of the article.  
Using these fields we see that the $O(4)$ NLSM can be viewed as being essentially a theory of two complex scalar fields 
$b_1$ and $b_2$, however, these fields interact with each other due to (i) the constraint $\sum_{I} |b_I|^2 = 1$ , 
and (ii) the theta term $S_{\theta}[U]$.

The BTI is a gapped bosonic phase of matter protected by $U(1)_c$ charge conservation symmetry and $\ZT$ time-reversal
symmetry. Under these symmetries the bosonic fields $b_I$ transform as
\beqa
	U(1)_c: b_I &\to& e^{i\chi}b_I  \label{eq:U1c}\\ 
	\ZT: b_I(t,\mb{x}) &\to& b_I(-t,\mb{x})\ ,
\eeqa
for $I=1,2$, where $\mb{x}= (x,y)$ denotes the spatial coordinates. 
These transformations give the total symmetry group the structure $U(1)_c \rtimes \ZT$, 
where the semi-direct product ``$\rtimes$" indicates that the $U(1)_c$ and $\ZT$ transformations do not commute with each 
other. As we explain in the next few paragraphs, 
the $O(4)$ NLSM theory with a theta term only possesses this time-reversal symmetry when 
$\theta$ is an integer multiple of $\pi$. 

%It is interesting to examine the transformation of the theta term $S_{\theta}[U]$ under the time-reversal
%transformation $b_I(t,\mb{x}) \to b_I(-t,\mb{x})$, $t \to -t$. We find that under time-reversal (\textbf{MFL: actually
%this may only be correct in Euclidean signature, check it})
%\beq
%	S_{\theta}[U] \to -S_{\theta}[U] \ .
%\eeq
%A well-known argument (\textbf{citation?}), which we review in the next few paragraphs, then tells us that the only
%time-reversal symmetric values of $\theta$ are integer multiples of $\pi$. 

To see why the only time-reversal symmetric values of $\theta$ are $\theta= n\pi,\ n \in \mathbb{Z}$, we first 
make a transformation to Euclidean spacetime. Euclidean time is defined by $\tau= it$, and
the theta term in Euclidean spacetime has the form
\beq
	S_{\theta,E}[\mb{N}]= -\frac{i}{12\pi^2}\int d^3 x_E\ \ep^{\mu\nu\lambda}\ep_{abcd} N^a\pd_{\mu}N^b\pd_{\nu}N^c\pd_{\lam}N^d\ ,
\eeq
where $d^3 x_E = d\tau d^2\mb{x}$ is the integration measure for Euclidean spacetime, and now 
$\mu,\nu,\lambda = \tau, x,y$. The theta term is now imaginary, which means that 
$e^{-\theta S_{\theta,E}[\mb{N}]}$ appears as a phase factor in the Euclidean path integral. 
Under a time-reversal transformation we send $t \to -t$, $i \to -i$ (since this symmetry is anti-unitary),
and $b_I(t,\mb{x}) \to b_I(-t,\mb{x})$. Since $\tau = it$, $\tau$ is invariant under this transformation. Therefore we find
that under time-reversal $S_{\theta,E}[\mb{N}] \to - S_{\theta,E}[\mb{N}]$ \cite{CenkeClass1}.

If we impose boundary conditions on
$\mb{N}$ such that $\mb{N}$ tends to a fixed configuration $\mb{N}_0$ at infinity in all directions of Euclidean spacetime,
then we may identify Euclidean spacetime with the sphere $S^3$. The sphere $S^3$ is also the configuration space for the 
$O(4)$ NLSM, so in this situation the theta term becomes quantized,
\beq
	\frac{1}{12\pi^2}\int d^3 x_E\ \ep^{\mu\nu\lambda}\ep_{abcd} N^a\pd_{\mu}N^b\pd_{\nu}N^c\pd_{\lam}N^d = n_I \in \mathbb{Z}\ , \label{eq:inst-no}
\eeq
where $n_I$ is the \emph{instanton} number of the field configuration $\mb{N}$. The quantization of this integral follows
from the homotopy group $\pi_3 (S^3)= \mathbb{Z}$.
In fact, the theta term is just the pull-back to spacetime of the volume form on $S^3$. Since Euclidean spacetime (with the 
boundary conditions discussed above) is just another copy of $S^3$, the integral is required to be an integer, which just
counts the number of times that the spacetime $S^3$ wraps around the configuration space (also $S^3$) of 
the $O(4)$ NLSM field $\mb{N}$.

In the Euclidean path integral, the theta term appears in an exponential, $e^{-\theta S_{\theta,E}[\mb{N}]} = e^{i\theta n_I}$, which
shows that the parameter $\theta$ is only defined modulo $2\pi$. We have already seen that a time-reversal transformation
sends $\theta \to -\theta$. It is then immediate to see that the only time-reversal symmetric values of $\theta$ are
$\theta=n\pi$, $n\in\mathbb{Z}$, since it is only these values of $\theta$ which satisfy $\theta \equiv -\theta$ mod $2\pi$.
 It was shown in Ref.~\onlinecite{VS2013} that the gapless surface termination
of the BTI is described by the $O(4)$ NLSM with $\theta=\pi,$ and hence preserves time-reversal symmetry.

Another comment can be made about the interpretation of the theta term in Euclidean spacetime. 
It was shown in Ref.~\onlinecite{SenthilFisher} that the one-instanton configuration of the $O(4)$ field $\mb{N}$ can be 
re-interpreted in terms of vortex configurations of the bosonic fields $b_1$ and $b_2$. Recall that a \emph{vortex} of
the field $b_I$ is a point in space around which the phase of $b_I$ 
winds by $2\pi$. In 2+1-d the spacetime trajectory, or \emph{worldline},
of a vortex is just a line (or curve) in spacetime. 
In Euclidean spacetime (compactified to the sphere $S^3$ via appropriate boundary conditions), the worldlines
of vortices become closed loops. In Ref.~\onlinecite{SenthilFisher} it was shown that the one-instanton configuration of the 
field $\mb{N}$ is equivalent to a linking configuration in which the worldline of a vortex in the phase of $b_1$ links exactly once 
with a worldline of a vortex in the phase of $b_2$. Since this configuration contributes a phase of $e^{i\theta}$ to the 
Euclidean path integral, the authors of Ref.~\onlinecite{SenthilFisher} interpreted this to mean that a vortex in $b_1$
and a vortex in $b_2$ have a mutual statistical angle of $\theta$. This means that a braiding process in which a vortex in $b_1$
makes a complete circuit around a vortex in $b_2$ should result in an overall phase factor $e^{i\theta}$ for the wave 
functional of the quantum field theory. This result was then used in Ref.~\onlinecite{VS2013} to deduce a topologically
ordered surface phase for the BTI. We will review this topologically ordered phase at the end of this section.
We remark in passing that similar arguments were also used in Ref.~\onlinecite{CenkeBraiding} to deduce the braiding statistics
of particle and loop-like excitations in gauged SPT phases from their description in terms of NLSM's with theta term.

It is clear from the discussion in the preceding paragraphs that the theta term plays an important role in the physics of the 
$O(4)$ NLSM. However, in this section we relied extensively on the interpretation of the theta term in Euclidean spacetime 
to understand its special properties. To better understand the quantum mechanics of the $O(4)$ NLSM with theta term, it is
desirable to understand the role the theta term plays in the Minkowksi spacetime path integral. In 
Appendix~\ref{app:path-integral} we explain the precise interpretation of the theta term in Minkowksi spacetime, and 
show that the theta term does indeed contribute a phase $e^{i\theta}$ to the path integral for configurations
of the $O(4)$ field in which a vortex in the boson $b_2$ makes a complete circuit around a vortex in $b_1$. This result
confirms the interpretation of the theta term given by Senthil and Fisher in Ref.~\onlinecite{SenthilFisher}, which 
was based on an analysis of the theory in Euclidean spacetime. In addition, following an argument from
Ref.~\onlinecite{WilczekZee}, this result implies that a bound state of a vortex in $b_1$ and a vortex in $b_2$
carries intrinsic angular momentum $J=\frac{\theta}{2\pi}$. At $\theta=\pi$ we have $J=\frac{1}{2}$, which means
that the vortex bound state is a fermion, as was discussed in Ref.~\onlinecite{VS2013}.

\subsection{Time-reversal breaking response}

In this section we discuss the calculation of the time-reversal breaking electromagnetic response of the $O(4)$ NLSM
with $\theta=\pi$. This response is in principle obtained by coupling the NLSM to the external electromagnetic field 
$A_{\mu}$, turning on a small time-reversal breaking perturbation, integrating out the matter fields $b_1$ and $b_2$,
and then setting the time-reversal breaking perturbation to zero. 
In practice, however,
it is very difficult to integrate out the NLSM field directly, and so we make use of two alternative and completely different
methods for calculating the time-reversal breaking response of the theory. The fact that these two methods give the
same answer strongly suggests that the answer is the correct one, even though it has not, as yet, been checked with a direct calculation in 
the $O(4)$ NLSM.

%The first method for calculating the time-reversal breaking response  uses
%the dual vortex description of the $O(4)$ NLSM at $\theta=\pi$ and was used extensively in
%Ref.~\onlinecite{VS2013} to analyze the surface physics of the BTI.
%The second method uses a formula derived in Ref.~\onlinecite{AbanovWiegmann} by Abanov and Wiegmann, which allows one 
%to write the $O(4)$ NLSM at $\theta=\pi$ as a path integral over a set of auxiliary fermions which couple to the $O(4)$ field 
%$\mb{N}$. The fermions in this construction must also carry charge under the physical $U(1)_c$ symmetry, so we can directly 
%couple the fermions to the $U(1)_c$ gauge field $A_{\mu}$ and then integrate out the fermions to deduce the electromagnetic 
%response of the system. A similar approach was used recently in Ref.~\onlinecite{CenkeDual} to calculate the electromagnetic
%response of a Bosonic Integer Quantum Hall state in $4+1$ dimensions.
%We now describe both of these methods since they are the methods we will use to calculate the 
%response of our BSM model later in the paper. 

\subsubsection{Method 1: Dual Vortex Description}

The first method for calculating the time-reversal breaking response of the BTI surface is to use the dual vortex description
of the $O(4)$ NLSM which becomes possible at the special value of $\theta=\pi$. 
This dual vortex description was first obtained in Ref.~\onlinecite{SenthilFisher} using a lattice
formulation of the theory. The continuum version of this dual vortex theory was then used extensively in 
Ref.~\onlinecite{VS2013} to study the possible surface phases of the BTI. In this section we give a review of this dual
description from an alternative perspective that is complementary to that given in Refs.~\onlinecite{SenthilFisher,VS2013}.

We have already explained how the $O(4)$ NLSM can be regarded as a theory of two complex scalar fields 
$b_1$ and $b_2$ subject to the constraint $\sum_I |b_I|^2=1$. This constraint has a strong effect on the physics
of vortices in the fields $b_1$ and $b_2$. Recall that a vortex in the 
field $b_1$ is a point in space around which the phase of $b_1$ winds by $2\pi$. At such a point the phase of $b_1$ is 
undefined, and so the amplitude of $b_1$ must vanish at that point. However, since the fields $b_1$ and $b_2$ are subject
to the constraint discussed above, this means that in the core of a vortex in $b_1$ we have $|b_2|= 1$. This indicates
that vortices in $b_1$ can trap charge of $b_2$ and vice-versa. In fact, in Minkowski spacetime the main effect of the
theta term is to attach charge $\frac{\theta}{2\pi}$ of boson $b_1$ to vortices in $b_2$ and vice-versa. A heuristic
argument for this effect was given in Ref.~\onlinecite{VS2013}. In Appendix~\ref{app:vortices} we prove this
result explicitly by computing the charges of global 
excitations on the background of certain exact vortex solutions of the NLSM equations of motion. 

%In Minkowski spacetime (as opposed to Euclidean space time, where the effect of the theta term is to give a phase
%$e^{i n_I \theta}$ in the path integral to configurations of the $O(4)$ field with instanton number $n_I$), the effect of
%the theta term is to attach charge $\frac{\theta}{2\pi}$ of boson $b_1$ to vortices in $b_2$ and vice-versa 
%\cite{VS2013}. It turns out that this result can 
%be demonstrated explicitly by computing the charge \emph{operator} for the $O(4)$ NLSM in 
%the canonical formalism, and we carry out this analysis in Appendix~\ref{app:vortex-charge}. 
%The analysis is quite similar to the case of the $O(3)$ NLSM with Hopf term, studied  in 
%Ref.~\onlinecite{Karabali}
%using canonical quantization, where it was shown that the angular momentum operator for the 
%quantum theory acquires a correction $\frac{\theta}{2\pi}Q^2$, where $Q$ is the soliton number and $\theta$ is the 
%coefficient of the Hopf term.

%\textbf{Could use a footnote about $\kappa$. When you do boson-vortex duality for a condensed complex
%scalar field, $\kappa$ is given in terms of the absolute value of the $vev$ of the scalar field. It is more complicated in the
%case of the $O(4)$ model because $|vev|$ is always 1. So we can just say that $\kappa$ is some constant to be general.}

We first give a short review of the dual vortex description of the theory of an ordinary charged scalar field
in 2+1-d, and refer the reader to Ref.~\onlinecite{LeeFisher} for a more detailed description of this technique. 
Consider first an ordinary complex scalar field $b$, with a Lagrangian of the form
\beq
	\mathcal{L}= |(\pd_{\mu}- i A_{\mu})b|^2 - \frac{\mu}{2} |b|^2 - \frac{\lambda}{4} |b|^4 + \dots
\eeq
For later convenience we write $b$ in a density phase representation as $b= \rho e^{i\vth}$.
When $\mu<0$ this system has a symmetry-broken ground state in which $\rho = \bar{\rho}= \sqrt{-\tfrac{\mu}{\lambda}}$ and 
the phase of $b$ is locked to a particular value (thus spontaneously breaking the original $U(1)_c$ symmetry under 
$b \to e^{i\chi}b$). The low-energy excitations about this ground state are the gapless fluctuations of the phase $\vth$ of 
$b$ (the Goldstone modes), which are described by
\beq
	\mathcal{L}= \bar{\rho}^2 (\pd_{\mu}\vth - A_{\mu})^2\ + \dots. \label{eq:goldstone}
\eeq
The fluctuations $\vth$ consist of two parts, $\vth = \vth^s + \vth^v$. The smooth part $\vth^s$ consists of small
fluctuations around the fixed vacuum value of $\vth$. The second part $\vth^v$ consists of vortices in which 
the phase winds by some multiple of $2\pi$ around the vacuum manifold (i.e., the circle defined by $|b|= \bar{\rho}$).

In the usual boson-vortex duality a sequence of transformations is now applied to the Lagrangian 
Eq.~\eqref{eq:goldstone}  (more precisely, these transformations are applied to the path integral) to obtain a final Lagrangian
of the form
\beqa
	\mathcal{L} &=& |(\pd_{\mu} - i\al_{\mu})\phi|^2 - \frac{\tilde{\mu}}{2} |\phi|^2 - \frac{\tilde{\lam}}{4} |\phi|^4 + \dots  \nnb \\
&-& \frac{1}{4\bar{\rho}^2}\left(\frac{1}{2\pi}\ep^{\mu\nu\lam}\pd_{\nu}\al_{\lambda}\right)^2 - \frac{1}{2\pi}\ep^{\mu\nu\lam}A_{\mu}\pd_{\nu}\al_{\lam}\ .
\eeqa
This expression features two new fields: the gauge field $\al_{\mu}$ and the complex scalar field $\phi$.
The field $\al_{\mu}$ is a non-compact gauge field which is introduced to represent the conserved number current 
$J^{\mu}$ of the original bosons $b$ via the equation $J^{\mu}= \frac{1}{2\pi}\ep^{\mu\nu\lam}\pd_{\nu}\al_{\lambda}$.
Non-compactness of $\al_{\mu}$ is just the statement that $\ep^{\mu\nu\lam}\pd_{\mu}\pd_{\nu}\al_{\lambda}=0$, 
which guarantees the conservation of $J^{\mu}$. 
The excitations of the new complex scalar field $\phi$ represent vortices in the phase of the original boson $b$. 
The vortex current of $b$,
defined by $K^{\mu}= \frac{1}{2\pi}\ep^{\mu\nu\lam}\pd_{\nu}\pd_{\lam}\vth^v$, is given in this representation by the
number current of $\phi$ as $K^{\mu} = i(\phi\ \pd^{\mu}\phi^* - \phi^* \pd^{\mu}\phi)$ (in other words, the 
$U(1)$ charge of $\phi$ is the vortex number). We have also included a number
of potential energy terms which could appear in the action for the vortex field $\phi$.

We now apply this technique to the boson $b_2$ in the $O(4)$ NLSM while leaving $b_1$ untransformed (a nearly identical discussion can be had if
one chooses to dualize $b_1$ and leave $b_2$ fixed instead).
We therefore define a new complex scalar field $\phi_{2,+}$ which creates
a vortex in the phase of $b_2$. From the discussion earlier in this section, and the results of 
Appendix~\ref{app:vortices}, $\phi_{2,+}$ 
carries charge $\frac{\theta}{2\pi}$ under the $U(1)_c$ symmetry. We represent the conserved number current 
$J^{\mu}_2$ of $b_2$ using the non-compact gauge field $\al_{2,\mu}$. 
%
%In terms of $b_2$, the current is given explicitly by
%the expression  
%\beq
%	J^{\mu}_2 = \frac{i}{g}( b_2\ \pd^{\mu} b_2^{*} - b_2^*\pd^{\mu} b_2)\ .
%\eeq
%It is represented in terms of $\al_{2,\mu}$ by $J^{\mu}_2= \frac{1}{2\pi}\ep^{\mu\nu\lam}\pd_{\nu}\al_{2,\lam}$ in the dual 
%vortex description.

At this point, the dual vortex description of the $O(4)$ NLSM with general angle $\theta$ takes the form
\beqa
	\mathcal{L}&=& \frac{1}{g}|(\pd_{\mu}  -i A_{\mu} ) b_1|^2 +  |(\pd_{\mu} - i\al_{2,\mu} -i\frac{\theta}{2\pi}A_{\mu} )\phi_{2,+}|^2 +\dots  \nnb \\
	&-&  \frac{1}{\kappa_2}\left(\frac{1}{2\pi}\ep^{\mu\nu\lam}\pd_{\nu}\al_{2,\lam}\right)^2  - \frac{1}{2\pi}\ep^{\mu\nu\lam}A_{\mu}\pd_{\nu}\al_{2,\lam}\ ,
\eeqa
where the ellipses stand for possible potential energy terms.
The field $\phi_{2,+}$ carries charge of both the dual gauge field $\al_{2,\mu}$ \emph{and} the external field $A_{\mu}$.
The theta term is entirely responsible for the coupling of $\phi_{2,+}$ to $A_{\mu}$. Finally, the constant $\kappa_2$
is given by $\kappa_2= \frac{g}{4\bar{\rho}^2_2}$, where $\bar{\rho}_2$ is the absolute value of $b_2$
in the condensed phase.

%~\footnote{Due to the complex interplay between vortices and the constraint $\sum_I |b_I|^2=1$,
%we prefer not to speculate on the precise value of $\bar{\rho}_2$. In any case, the value of this constant is not
%needed for the long-wavelength topological responses which we calculate from the dual vortex theory.}.

Interestingly, exactly at the special value $\theta=\pi$, it becomes possible to replace our description of the theory in terms 
of $b_1$ and $\phi_{2,+}$ with a much more symmetric dual description in terms of two types of vortices $\phi_{2,+}$ and 
$\phi_{2,-}$, as we now explain. At $\theta=\pi$,
the vortex $\phi_{2,+}$ carries charge $\frac{1}{2}$ of boson $b_1$. In this case the composite field
\beq
	\phi_{2,-} = \phi_{2,+}b^*_1\ ,
\eeq
carries charge  $-\frac{1}{2}$ of boson $b_1$ (note that we are using $^{\ast}$ to represent anti-particles). The field 
$\phi_{2,-}$ can be understood as a vortex-anti-boson bound state, and
at $\theta=\pi$ it is a natural object to consider because of the fact that it carries the same magnitude of charge as the
original vortex $\phi_{2,+}$ 
(note that we can always define $\phi_{2,-}$ in this way for \emph{any} value of $\theta$, 
but this field only transforms nicely under the symmetries of the theory when $\theta=\pi$).

Further justification for the introduction of the field $\phi_{2,-}$ can be obtained by recalling that at the special
value $\theta=\pi$, the time-reversal symmetry of the $O(4)$ NLSM is restored. It follows that the vortex
$\phi_{2,+}$ should have a well-defined transformation under time-reversal when $\theta=\pi$. Vortices should
transform into anti-vortices under the action of time-reversal, since time-reversal is an anti-unitary symmetry. On the other 
hand, the time-reversal partner of $\phi_{2,+}$ should have the same $U(1)_c$ charge as $\phi_{2,+}$ in order to preserve
the structure of the symmetry group of the BTI. It turns out that $\phi^*_{2,-}$ has just the right properties to be the 
partner of $\phi_{2,+}$ under the time-reversal operation. 

We see then that at the special value $\theta=\pi$, the dual description of the $O(4)$ NLSM is given most naturally in terms
of the two-component vortex field $\Phi_2= (\phi_{2,+},\phi_{2,-})^T$, which transforms under the $U(1)_c$ and 
$\ZT$ symmetries according to 
\beqa
	U(1)_c: \Phi_2 &\to& e^{i\tfrac{\chi}{2}\sigma^z}\Phi_2 \\
	\ZT: \Phi_2(t,\mb{x}) &\to& \sigma^x\Phi_2^*(-t,\mb{x})\ .
\eeqa
In terms of the pair of vortex fields making up $\Phi_2$, the final dual action takes the form
\beqa
	\mathcal{L}&=& \sum_{s=\pm}|(\pd_{\mu} - i\al_{2,\mu} -i\frac{s}{2}A_{\mu} )\phi_{2,s}|^2 +\dots \label{eq:O4dual}  \\
	&-&  \frac{1}{\kappa_2}\left(\frac{1}{2\pi}\ep^{\mu\nu\lam}\pd_{\nu}\al_{2,\lam}\right)^2  - \frac{1}{2\pi}\ep^{\mu\nu\lam}A_{\mu}\pd_{\nu}\al_{2,\lam}\ , \nnb 
\eeqa
where again the ellipses stand for possible potential energy terms. Note that the two species of vortex carry
the same charge of $\al_{2,\mu}$, but opposite charge of $A_{\mu}$.

As stated in Ref.~\onlinecite{VS2013}, the original boson field $b_1$ is now represented approximately by
\beq
	b_1 = \phi_{2,+}\phi_{2,-}^*\ ,
\eeq
i.e., it is a bound state of a vortex ($\phi_{2,+}$) and an anti-vortex ($\phi_{2,-}^*$).
We should, however, take a moment to consider this equation carefully. Interestingly, the two sides of this equation
do not have the same dimensions. The field $b_1$ is dimensionless, while the complex scalar fields $\phi_{2,+}$
and $\phi_{2,-}$ carry dimensions of $(\text{length})^{-\tfrac{1}{2}}$ (this is  true because the vortex current 
$K^{\mu}$ has dimensions of $(\text{length})^{-2}$). 
A more precise version of this equation would be to write
\beq
	b_1 \sim g\ \phi_{2,+}\phi_{2,-}^*\ , \label{eq:composite}
\eeq
where $g$ is the NLSM coupling which has units of length, and where an arbitrary dimensionless constant could be included
on the right-hand side of this equation.

The time-reversal breaking response of the $O(4)$ NLSM at $\theta=\pi$ can now be explored using the dual description
in Eq.~\eqref{eq:O4dual}. A gapped, time-reversal breaking phase is realized when, for example, $\phi_{2,+}$ condenses
and $\phi_{2,-}$ becomes gapped, or vice-versa. In order to induce this phase, one needs to include in Eq.~\eqref{eq:O4dual}
a potential energy of the form
\beq
	V(\Phi_2) = \mu \Phi_2^{\dg}\sigma^z\Phi + \lambda_{+}|\phi_{2,+}|^4 + \lambda_{-}|\phi_{2,-}|^4\ ,
\eeq
where $\lambda_{\pm}$ are both positive. 
The choice of which vortex condenses and which is gapped depends on the sign of $\mu$. Note
that the term $\Phi_2^{\dg}\sigma^z\Phi$ explicitly breaks time-reversal symmetry.

When $\phi_{2,+}$ condenses and $\phi_{2,-}$ is gapped, we may (at low energies) set $\phi_{2,-}=0$ and 
$\phi_{2,+}= \text{const.}$ to find that the minimum energy configuration is realized when 
$\al_{2,\mu}= -\frac{1}{2}A_{\mu}$,
which yields the response
\beq
	\mathcal{L}_{eff}= \frac{e^2}{4\pi}e^{\mu\nu\lam}A_{\mu}\pd_{\nu}A_{\lam}\ ,
\eeq
where we have restored the charge $e$. This response yields a ``half" bosonic quantum Hall effect with $\sigma_{xy}=\frac{1}{2}\frac{2e^2}{h}.$
If we had instead condensed $\phi_{2,-}$ and gapped out $\phi_{2,+}$, we would have found the same response but
with the opposite sign.

\subsubsection{Method 2: Abanov-Wiegmann integration over fermions}

The second method for calculating the time-reversal breaking response of the BTI surface uses a formula due to 
Abanov and Wiegmann\cite{AbanovWiegmann} which allows one to express the $O(4)$ NLSM with 
theta term as a path integral over a set of auxiliary fermions.
The fermions in this construction must also carry charge under the physical $U(1)_c$ symmetry, so we can directly 
couple the fermions to the $U(1)_c$ gauge field $A_{\mu}$ and then integrate out the fermions to deduce the electromagnetic 
response of the system. A similar approach was used recently in Ref.~\onlinecite{CenkeDual} to calculate the electromagnetic
response of a Bosonic Integer Quantum Hall state in 4+1-d.

The starting point for this construction is a multi-component fermionic field $\Psi= (\psi_1,\psi_2,\psi_3,\psi_4)^T$, 
where each of $\psi_a$, $a=1,2,3,4,$ is a 
two-component Dirac fermion in 2+1-d. In what follows we use tensor product notation in order to treat spinor
and ``isospace" indices on equal footing. All indices are traced over in the evaluation of the fermion path integral.
The rightmost $2\times2$ matrix in the tensor products acts on the spinor indices of $\psi_a$, while the left and middle 
matrices in the tensor products act on the isospace indices.

We define two sets of gamma matrices $\gamma^{\mu}$ and $\Gamma^a$ by
\begin{subequations}
\label{eq:lc-gamma}
\beqa
	\gamma^0 &=& \mathbb{I}\otimes\mathbb{I}\otimes\sigma^y \\ 
	\gamma^1 &=& -i\mathbb{I}\otimes\mathbb{I}\otimes\sigma^z \\
	\gamma^2 &=& i\mathbb{I}\otimes\mathbb{I}\otimes\sigma^x\ ,
\eeqa
\end{subequations}
and 
\begin{subequations}
\label{eq:uc-gamma}
\beqa
	\Gamma^1 &=& \sigma^x\otimes\sigma^x\otimes \mathbb{I} \\
	\Gamma^2 &=& \sigma^y\otimes\sigma^x\otimes \mathbb{I} \\
	\Gamma^3 &=& \sigma^z\otimes\sigma^x\otimes \mathbb{I} \\
	\Gamma^4 &=& \mathbb{I}\otimes\sigma^y\otimes \mathbb{I} \ ,
\eeqa
\end{subequations}
where $\mathbb{I}$ is the $2\times 2$ identity matrix.
In this case we can also define a fifth matrix for the second set, 
\beq
	\Gamma^5 = \mathbb{I}\otimes\sigma^z\otimes \mathbb{I}\ .
\eeq
The first set of gamma matrices obey a Clifford Algebra in Lorentz signature, 
$\{\gamma^{\mu},\gamma^{\nu}\}= 2\eta^{\mu\nu}\mathbb{I}_{8\times 8}$, and are used
to construct the derivative operator for the Dirac action. The second set obeys a Euclidean Clifford Algebra, 
$\{\Gamma^{a},\Gamma^{b}\}= 2\delta^{ab}\mathbb{I}_{8\times 8}$, and is used to construct the mass terms
which couple $\Psi$ to the $O(4)$ field $\mb{N}$. 

According to Ref.~\onlinecite{AbanovWiegmann}, a fermionic action of the form
\beq
	\mathcal{L}_{f}= \bar{\Psi}\left(i\slashed{\pd} -\cos(\nu)M\Gamma^5- \sin(\nu)M\sum_{a=1}^4 N^a\Gamma^a\right)\Psi\ ,
\eeq
with large mass $M>0$ will produce, after integration over the fermions, an $O(4)$ NLSM of the form of 
Eq.~\eqref{eq:O4} with the theta angle given by
\beq
	\theta= \pi\left( 1-\frac{9}{8}\cos(\nu) + \frac{1}{8}\cos(3\nu)   \right)\ .\label{eq:theta-angle}
\eeq
Taking $\nu= \frac{\pi}{2}$ gives $\theta=\pi$. The evaluation of this fermion path integral is not completely
straightforward, and so we refer the reader to Ref.~\onlinecite{AbanovWiegmann} as well as Ref.~\onlinecite{TanakaHu2}
for explanations of this calculation. 

If we set $\nu= \frac{\pi}{2} - \delta$ for small $\delta$, then the action takes the form
\beq
	\mathcal{L}_{f}= \bar{\Psi}\left(i\slashed{\pd} - (M\delta)  \Gamma^5 - M\sum_{a=1}^4 N^a\Gamma^a \right)\Psi\ .
\eeq
Since the only time-reversal invariant values of $\theta$ are multiples of $\pi$, this corresponds to adding a small
time-reversal breaking perturbation to the action (we would now get $\theta\approx \pi(1-\tfrac{3}{2}\delta)$ after
integrating out the fermions).
We now calculate the response of the theory in the presence of this perturbation, and then take the limit 
$\delta \to 0$.

Before we proceed with the calculation, we mention the following puzzle.
The calculation in Ref.~\onlinecite{AbanovWiegmann} is controlled by an expansion in powers of $M^{-1}$, so we
must take $M$ to be large for this expansion to make sense. On the other hand, the coupling constant $g$ of the $O(4)$ NLSM
is related to $M$ via a formula of the form $M\propto 1/g$. For $M$ large we seem to obtain an $O(4)$ NLSM
in the ordered (small $g$) phase, whereas we are interested in studying the disordered (large $g$) phase. 
It is therefore not immediately clear why the calculation in this subsection agrees with the response calculation
of Ref.~\onlinecite{VS2013} using the dual vortex theory, which we reviewed in the previous subsection. 
We resolve this puzzle in Appendix~\ref{app:fermion-puzzle}, where we use the Abanov-Wiegmann formula to argue that the 
theory $S_f=  \int d^3 x\ i\bar{\Psi}\slashed{\pd}\Psi$ of four massless fermions $\psi_a$ must possess exactly
the same topological response as the original $O(4)$ NLSM at $\theta=\pi$. In the rest of this section we will
therefore calculate the response of the fermions $\psi_a$ to the time-reversal breaking mass term 
$- (M\delta) \bar{\Psi}\Gamma^5\Psi$. According to the arguments in Appendix~\ref{app:fermion-puzzle}, this
response (or at least its topological part), should be identical to the response of the $O(4)$ NLSM at $\theta=\pi$.

%Recall, however, that $g$ is a parameter with dimensions of length. Therefore we cannot say what values of
%$g$ are large or small without comparing $g$ to some fixed reference scale, say $\bar{g}$. 
%As far as we know, the identity and physical 
%meaning of any reference scale $\bar{g}$ for the $O(4)$ NLSM with theta term are completely unknown, which leaves open the 
%possibility that $M$ can be taken
%large enough for the expansion of Ref.~\onlinecite{AbanovWiegmann} to work, but still give a value of $g$ which is larger
%than the mystery scale $\bar{g}$. The fact that the response calculation in this section agrees with the calculation
%in the previous section (which was performed originally in Ref.~\onlinecite{VS2013}) 
%offers some evidence that this possibility is indeed realized.

%\textbf{MFL: What do you think of this explanation for the paradox that $M\sim 1/g$?}
%
%Let us briefly mention one puzzling aspect of this formula. The coupling constant $g$ of the $O(4)$ NLSM
%is related to $M$ via a formula of the form $M\sim 1/g$. Since the derivative expansion of Ref.~\onlinecite{AbanovWiegmann}  
%relies on the fact that $M$ is large, it seems that this produces an $O(4)$ NLSM in the ordered (small $g$) phase, whereas we 
%are interested in studying the disordered phase (large $g$) phase.

Before we can do this we need to determine the charges $q_a$ of the four Dirac fermions $\psi_a$. These charges should
be chosen so that the coupling term $\sum_{a=1}^4 N^a\bar{\Psi}\Gamma^a\Psi$ is invariant under the $U(1)_c$ symmetry.
Each fermion $\psi_a$ is assumed to transform as
\beq
	U(1)_c: \psi_a \to e^{iq_a \chi} \psi_a\ .
\eeq
The transformation of the $O(4)$ field under the $U(1)_c$ symmetry was described in Eq.~\eqref{eq:U1c}. Using the
relation
\beqa
	b_1 &=& N^4 + i N^3 \\
	b_2 &=& -N^2 + iN^1\ ,
\eeqa
and the explicit form of the matrices $\Gamma^a$, we find that in order for the 
term  $\sum_{a=1}^4 N^a\bar{\Psi}\Gamma^a\Psi$ to be invariant under $U(1)_c$, the charges $q_a$ must
satisfy the matrix equation 
\beq
	\begin{pmatrix}
		-1 &0&0&1 \\
		0&-1&1&0 \\
		-1&1&0&0 \\
		0&0&-1&1	
\end{pmatrix}\begin{pmatrix}
q_1 \\
q_2 \\
q_3 \\
q_4 
\end{pmatrix}= \begin{pmatrix}
	1 \\
	1 \\
	-1 \\
	1
\end{pmatrix}\ .
\eeq
This matrix has a null vector $(1,1,1,1)^T$, so the solution of the system is not 
unique. One possible way to parameterize a general solution is
\beq
	\begin{pmatrix}
q_1 \\
q_2 \\
q_3 \\
q_4 
\end{pmatrix} = \begin{pmatrix}
	\bar{q} \\
	-1+\bar{q} \\
	\bar{q} \\
	1+\bar{q}
\end{pmatrix}\ ,
\eeq
where the parameter $\bar{q}$ is completely arbitrary because of the non-uniqueness of the solution. 
In what follows, we keep $\bar{q}$ to be some arbitrary number. Importantly, when we calculate a physical quantity 
pertaining to the $O(4)$ NLSM we will see that the answer is independent of $\bar{q}$.

We now define the diagonal matrix of charges $Q= \text{diag}(q_1,q_2,q_3,q_4)\otimes \mathbb{I}$, given explicitly by, 
\beq
	Q= \bar{q}\II\otimes\II\otimes \mathbb{I} + \frac{1}{2}(\sigma^z\otimes\sigma^z\otimes \mathbb{I} - \sigma^z\otimes\II\otimes \mathbb{I})\ , \label{eq:charge-matrix}
\eeq
(note that it acts as the identity on the spinor indices of the fermions)
and then use this matrix to couple $\Psi$ to $A_{\mu}$ to obtain the action
\beqa
	\mathcal{L}_{f,gauge} &=& \bar{\Psi}\left(i\slashed{\pd}  - (M\delta) \Gamma^5\right.  \\
&-& \left. M\sum_{a=1}^4 N^a\Gamma^a + Q\slashed{A}\right)\Psi\ .\nnb
\eeqa
We now integrate out the fermions and collect the lowest order terms in derivatives involving only $A_{\mu}$, because
those terms will give the dominant contribution to the electromagnetic response. For completeness we give a basic
outline of this calculation below. 

Since we are currently only interested in the electromagnetic response of the fermions, we set $N^a=0$ for the response 
calculation. Integrating out $\Psi$ then gives
\beqa
	S_{eff}[A_{\mu}] &=& -i\ln \text{det}\left(i\slashed{\pd}  - (M\delta) \Gamma^5 + Q\slashed{A} \right) \nnb \\
	&=&  -i\text{Tr}\ln\left(i\slashed{\pd}  - (M\delta) \Gamma^5  + Q\slashed{A} \right)\ ,
\eeqa
where $\text{Tr}[\dots]$ indicates a trace over spacetime, spinor, and isospace indices. We now write
\beqa
	S_{eff}[A_{\mu}] &=& -i\text{Tr}\ln(i\slashed{\pd}  - (M\delta) \Gamma^5)  \\
	&-&  i\text{Tr}\ln\left[ 1  +  (i\slashed{\pd}  - (M\delta) \Gamma^5)^{-1}(Q\slashed{A})\right]\ , \nnb
\eeqa
and expand the second term using $\ln(1+x)= \sum_{n=1}^{\infty} (-1)^{n+1}\frac{x^n}{n}$. 

Here is a technical point. The effective action we wrote down is divergent for $\delta \to 0$. 
Therefore a procedure is needed to define the effective action for $\delta=0$. 
Let us indicate the dependence of the effective action on $\delta$ by writing it as $S_{eff}[A_{\mu},\delta]$. 
We follow Ref.~\onlinecite{Redlich} and define
the renormalized effective action at $\delta=0$ by
\beq
	S^R_{eff}[A_{\mu},0] = S_{eff}[A_{\mu},0] - \lim_{\delta\to\infty} S_{eff}[A_{\mu},\delta]\ .
\eeq
The second term in this expression also has a divergent term which cancels the divergence from the first term. Now
as $\delta\to\infty$, the second term gives a finite contribution, which is a Chern-Simons term. We find that
%The first interesting contribution to the electromagnetic response comes from the second order 
%term in the expansion, 
%\beq
%	S^{(2)}_{eff}[A_{\mu}] = \frac{i}{2}\text{Tr}[(i\slashed{\pd}  - \delta M \Gamma^5)^{-1}(iQ\slashed{A}) (i\slashed{\pd}  - \delta M \Gamma^5)^{-1}(iQ\slashed{A})]\ .
%\eeq
\beq
	S^R_{eff}[A_{\mu},0]= \frac{1}{2}\text{tr}_I[Q^2 \Gamma^5]\frac{\text{sgn}(\delta)}{4\pi}\int d^3x\ e^{\mu\nu\lambda}A_{\mu}\pd_{\nu}A_{\lambda} \ ,
\eeq
where $\text{tr}_I[\dots]$ denotes a trace over isospace indices only (the trace over spacetime and spinor indices has 
already been performed). Since $\frac{1}{2}\text{tr}_I[Q^2 \Gamma^5] = -1$,
the final response is given by
\beq
	\mathcal{L}^R_{eff}= -\text{sgn}(\delta)\frac{e^2}{4\pi}\ep^{\mu\nu\lambda}A_{\mu}\pd_{\nu}A_{\lambda}\ ,
\eeq
where we have restored the charge $e$. The answer depends on $\text{sgn}(\delta)$ and not $\text{sgn}(\delta M)$ because 
the mass $M$ is assumed positive in the Abanov-Wiegmann method.
Note that the result is independent of $\bar{q}$ (the arbitrary offset to the charges of the four fermions $\psi_a$), which
is expected because we have calculated a physical quantity related to the $O(4)$ NLSM at $\theta=\pi$.

%In this case the sign of the electromagnetic response depends on the sign of $\delta$, and the sign of $\delta$ indicates 
%whether we have shifted $\theta$ to a value slightly greater than or slightly less than $\pi$ when we added the time-reversal
%breaking term to the Lagrangian.

%\subsection{Connection to the dual description of the BTI surface in terms of $QED_3$ with two flavors of 
%fermions}
%
%\textbf{MFL: I will try to see if I can make more sense of this connection in the next few days.}
%
%Recently there has been an exciting discussion of dual descriptions of the surfaces of the 3D ETI and the 3D BTI
%in terms of $2+1$ dimensional Quantum Electrodynamics ($QED_3$) with one and two flavors of fermions, respectively 
%\textbf{Cite many papers}. These developments began with the proposal of a new particle-hole symmetric
%description of the $\nu =\frac{1}{2}$ fractional Quantum Hall state \textbf{Cite D. Son}. Our study of the 
%electromagnetic response of the BTI surface using the Abanov-Wiegmann integration over fermions allows us to 
%get an interesting perspective on the dual description of the BTI surface in terms of $QED_3$ with two flavors
%of fermions \textbf{Cite new Cenke paper}. 

\subsection{Connection to the dual description of the BTI surface in terms of $N=2$ QED$_3$}

In this section we briefly comment on the relationship between the descriptions of the BTI surface theory discussed above: (i)
the dual vortex description, (ii) the description in terms of Abanov-Wiegmann fermions, and (iii) the recently 
proposed dual description of the BTI surface in terms of $2+1$-d Quantum Electrodynamics with 
two flavors of Dirac fermion, also known as $N=2$ QED$_3$ \cite{xu2015} (a quasi-1D derivation
of this dual description was later given in Ref.~\onlinecite{mross2015}). 

Before writing down the dual description of Ref.~\onlinecite{xu2015}, we first remind the reader that in their original study of 
the BTI in Ref.~\onlinecite{VS2013}, Vishwanath and Senthil assigned an additional ``pseudo-spin" quantum number to
the bosons $b_1$ and $b_2$ for convenience, with $b_1$ carrying spin $1$ and $b_2$ carrying spin $-1$. We refer to the $U(1)$ symmetry
associated with pseudo-spin conservation as $U(1)_s$. Under this symmetry the bosons transform as
\begin{subequations}
\label{eq:spin-sym}
\beqa
	U(1)_s: b_1 &\to& e^{i\xi}b_1, \\
	b_2 &\to& e^{-i\xi}b_2\ .
\eeqa
\end{subequations}
The fermions in the $N=2$ QED$_3$ description of the BTI surface are charged under this $U(1)_s$ symmetry. 

The $N=2$ QED$_3$ description of the BTI surface consists of two flavors of Dirac fermions, $\chi_1$ and $\chi_2$,
which can be combined into one multi-component spinor $X= (\chi_1,\chi_2)^T$. 
These fermions do not carry any $U(1)_c$ charge, but $\chi_1$
carries spin $1$ while $\chi_2$ carries spin $-1$. The time-reversal symmetry of the BTI acts on $X$ like a particle-hole
transformation. Both fermions also carry charge $1$ of a dual non-compact gauge field
$\al_{\mu}$, whose curl represents the total number current $J^{\mu}_{tot}$ of the bosons on the BTI surface via
$J^{\mu}_{tot}= \frac{1}{2\pi}\ep^{\mu\nu\lam}\pd_{\nu}\al_{\lam}$. The dual Lagrangian takes the form
\begin{align}
	\mathcal{L}=& \bar{X}(i(\mathbb{I}\otimes\gamma^{\mu})\pd_{\mu} + (\sigma^z\otimes\gamma^{\mu})A^s_{\mu} + (\mathbb{I}\otimes\gamma^{\mu})\al_{\mu} )X \nnb \\ 
 &- \frac{1}{2\pi}\ep^{\mu\nu\lam}A^c_{\mu}\pd_{\nu}\al_{\lam}\ , 
\end{align}
where $\gamma^{\mu}$, are the usual $2\times 2$ gamma matrices for $2+1$ dimensional Dirac fermions
(e.g. the matrices from Eq.~\eqref{eq:lc-gamma} without the additional identity matrices in the tensor product), 
$\bar{X}= X^{\dg}(\mathbb{I}\otimes\gamma^0)$, $A^c_{\mu}$ is the external electromagnetic field (denoted simply by
``$A_{\mu}$" in the other sections of this paper), and $A^s_{\mu}$ is a new external $U(1)$ gauge field which probes the 
$U(1)_s$ symmetry.

We now speculate on the relation between the fermions $\chi_1$ and $\chi_2$, the vortices $\phi_{1,\pm}$ and 
$\phi_{2,\pm}$ from the dual vortex description of the BTI surface, and the four
Abanov-Wiegmann fermions $\psi_a$. Out of the four vortices $\phi_{1,\pm}$ and 
$\phi_{2,\pm}$, we may form four composite vortices $\phi_{1,\pm}\phi_{2,\pm}$ by taking every possible combination
of ``$+$" and ``$-$" vortices of species $1$ and $2$. As discussed in Ref.~\onlinecite{VS2013}, and as we show
in Appendix~\ref{app:path-integral}, when $\theta=\pi$ a bound state of a vortex in $b_1$ and a vortex in $b_2$ is
a fermion. This means that the four composite vortices $\phi_{1,\pm}\phi_{2,\pm}$ are in fact fermions. The
charges and spins of these four composite vortices can be easily determined and they are shown in 
Table~\ref{tab:charge-spin1}. The spins $s_a$ of the four Abanov-Wiegmann fermions $\psi_a$ can also be calculated,
using the same method used to determine their charges $q_a$ (just like the charges, the spins are also determined only
up to an arbitrary offset $\bar{s}$, which we ignore here). The charges and spins of the four Abanov-Wiegmann fermions
are shown in Table~\ref{tab:charge-spin2}. Interestingly, each composite
vortex has precisely the same charge and spin as one of the Abanov-Wiegmann fermions.

Since the composite vortices have precisely the same charges and spins as the Abanov-Wiegmann fermions, and
since the composite vortices in the $O(4)$ NLSM at $\theta=\pi$ are known to be fermions, 
we conjecture that these objects should be identified with each other.
Furthermore, we propose that the fermions $\chi_1$ and $\chi_2$ from the $N=2$ QED$_3$ description can be identified
with $\psi_1$ and $\psi_3$, respectively, which in turn correspond to the composite vortices $\phi_{1,-}\phi_{2,+}$
and $\phi_{1,+}\phi_{2,-}$. The particle-hole-like transformation of $\chi_1$ and $\chi_2$ under time-reversal
then follows immediately from the transformations of the vortices under time-reversal.
Also, since the individual vortices $\phi_{I,\pm}$ couple
to the non-compact gauge fields $\al_{I,\mu}$, where the conserved current of boson $b_I$ is given by 
$J^{\mu}_I = \frac{1}{2\pi}\ep^{\mu\nu\lam}\pd_{\nu}\al_{I,\lam}$, the composite vortices are coupled to the 
total gauge field $\al_{\mu}=\al_{1,\mu}+\al_{2,\mu}$, whose curl represents the total boson number current. This
is the exact same gauge field which $\chi_1$ and $\chi_2$ couple to in the $N=2$ QED$_3$ description. It would be 
an interesting challenge for future investigations to provide a derivation of the $N=2$ QED$_3$ description of the BTI
surface directly from the description in terms of an $O(4)$ NLSM at $\theta=\pi$. Such a derivation would provide the 
details necessary to support the picture we have presented here.

\begin{table}
\centering
\begin{tabular}{|c|c|c|c| c|}
\hline
  & $\phi_{1,+}\phi_{2,+}$ & $\phi_{1,+}\phi_{2,-}$ & $\phi_{1,-}\phi_{2,+}$ & $\phi_{1,-}\phi_{2,-}$ \\ \hline
$q$ & 1 & 0 & 0 & -1 \\ \hline
$s$ & 0 & -1 & 1 & 0 \\ \hline 
\end{tabular}
\caption{Charges and spins of the composite vortices of the form $\phi_{1,\pm}\phi_{2,\pm}$. As
was discussed in Ref.~\onlinecite{VS2013}, and as we show in Appendix~\ref{app:path-integral}, a 
bound state of a vortex in $b_1$ and a vortex in $b_2$ is a fermion.}
\label{tab:charge-spin1}
\end{table}

\begin{table}
\centering
\begin{tabular}{|c|c|c|c| c|}
\hline
  & $\psi_1$ & $\psi_2$ & $\psi_3$ & $\psi_4$ \\ \hline
$q$ & 0 & -1 & 0 & 1 \\ \hline
$s$ & 1 & 0 & -1 & 0 \\ \hline 
\end{tabular}
\caption{Charges and spins of the fermions $\psi_a$ used in the Abanov-Wiegmann formula for the $O(4)$ NLSM (ignoring
the arbitrary offsets $\bar{q}$ and $\bar{s}$ discussed in the main text).}
\label{tab:charge-spin2}
\end{table}

\subsection{Symmetry-breaking phases accessible from the dual theory}

In this section we will complete our  discussion, following Ref.~\onlinecite{VS2013}, of the symmetry-breaking phases
of the surface of the BTI which are accessible from the dual vortex description of the $O(4)$ NLSM at $\theta=\pi$. We have 
already seen that condensing just one vortex, say $\phi_{2,+}$,
and gapping out the other one leads to a phase which breaks time-reversal symmetry. In that case it was
necessary to add the time-reversal breaking term $\Phi_2^{\dg}\sigma^z\Phi_2$ to the Lagrangian to simultaneously gap out 
one vortex and force the other vortex to condense.

There are two other basic options for condensing and/or gapping out the vortices in the dual theory. These 
options are: (i) condense both vortices, and (ii) gap both vortices. Both options lead to a superfluid phase which
can be understood as a phase in which one of the original fields $b_1$ or $b_2$ condenses. 
To identify which boson is condensing in each case, it is convenient to separately gauge the $U(1)$ symmetries
corresponding to $b_1 \to e^{i\chi}b_1$ and $b_2\to e^{i\chi}b_2$. We couple $b_1$ to the external field 
$A_{1,\mu}$ and $b_2$ to the external field $A_{2,\mu}$. In this case the dual theory takes the form
(recall that $\phi_{2,\pm}$ carry charge $\pm \frac{1}{2}$ of the boson $b_1$)
\beqa
	\mathcal{L}&=& \sum_{s=\pm}|(\pd_{\mu} - i\al_{2,\mu} -i\frac{s}{2}A_{1,\mu} )\phi_{2,s}|^2 +\dots  \nnb \\
	&-&  \frac{1}{\kappa_2}(\ep^{\mu\nu\lam}\pd_{\nu}\al_{2,\lam})^2  - \frac{1}{2\pi}\ep^{\mu\nu\lam}A_{2,\mu}\pd_{\nu}\al_{2,\lam}\ . \label{eq:O4dual2}
\eeqa

Consider first the phase obtained by condensing both $\phi_{2,+}$ and $\phi_{2,-}$. To be 
precise, we consider the condensation $\lan \phi_{2,+} \ran = \lan \phi_{2,-} \ran^*= v$, which
does not break $\ZT$. In this case we get a Higgs term
for the gauge field $\al_{2,\mu}$ and the external field $A_{1,\mu}$. The gauge field $\al_{2,\mu}$, which represents
the Goldstone boson of a condensate of $b_2$, is therefore gapped and can be safely integrated out. The resulting action
contains only a Higgs term for $A_{1,\mu}$, and so the phase where both $\phi_{2,+}$ and $\phi_{2,-}$ condense
can be identified with the phase where $b_1$ condenses.

Next consider the second case in which $\phi_{2,+}$ and $\phi_{2,-}$ are both gapped. We can then set 
$\phi_{2,+}$ and $\phi_{2,-}$ equal to zero to study the low energy properties of this phase. 
At this point the gauge field $\al_{2,\mu}$ can be integrated out to give a Higgs term for $A_{2,\mu}$, and so the phase
where both $\phi_{2,+}$ and $\phi_{2,-}$ are gapped can be identified with the phase in which $b_2$ condenses.

Finally, we note that the dual vortex theory can be driven into either of these two phases
by a potential that does not break the $U(1)_c$
or $\ZT$ symmetries, which means that the superfluid phase of the BTI surface spontaneously breaks
the $U(1)_c$ symmetry (and it does not break the time-reversal symmetry).

\subsection{Symmetry-allowed perturbations}

In this section we carefully investigate the effects of symmetry-allowed perturbations on the BTI surface. This is important as 
we want to understand the stability of the gapless phase of the surface, and hence the related 2+1-d semi-metal, as explicitly 
as possible.
In Ref.~\onlinecite{VS2013} Vishwanath and Senthil initially studied the $O(4)$ NLSM at $\theta=\pi$ assuming a larger
symmetry group consisting not only of $U(1)_c$ charge conservation and $\ZT$ time-reversal,
 but also an additional $U(1)_s$ ``pseudo-spin"
conservation symmetry and a $Z_2^s$ ``spin-flip" symmetry. The action of the $U(1)_s$ symmetry on the
bosons $b_I$ was already given in Eq.~\eqref{eq:spin-sym}. The $Z_2^s$ spin-flip symmetry acts as
\beq
	Z_2^s: b_1 \leftrightarrow b_2 \ .
\eeq
In the presence of these additional symmetries, interspecies tunneling terms of the form 
$b^{*}_1 b_2 + b^{*}_2 b_1$, as well as chemical potential terms of the form $\mu_1 |b_1|^2 + \mu_2 |b_2|^2$
(with $\mu_1 \neq \mu_2$), cannot be added to the Lagrangian. However,
the BTI is supposed to be protected by $U(1)_c$ and $\ZT$ symmetry alone. It is therefore 
essential to understand the effects
that such terms can have on the $O(4)$ NLSM at $\theta=\pi$, since we are allowed to add these terms to the 
Lagrangian in the generic case when the extra $U(1)_s$ and $Z_2^s$ symmetries are broken.

Interspecies tunneling and chemical potential terms can have a drastic effect on the physics of 
the $O(4)$ NLSM with theta term. 
However, we will show that these terms always drive the system into a symmetry-breaking
phase. To show this we make use of the commutation relations of the $O(4)$ NLSM fields in the
canonical formalism. Because of the constraint between the bosonic fields $b_{I}$, these commutation relations must be derived
using the Dirac Bracket formalism, and we review their derivation in Appendix~\ref{app:canonical}.

There is a simple way to understand why interspecies tunneling and chemical potential terms can have a strong effect on the
physics of the $O(4)$ NLSM with theta term. When these terms are strong, they can drive the fields into a configuration
in which the theta term vanishes identically. This is easiest to see when the theta term is written in Hopf coordinates
on the the sphere $S^3$. In Hopf coordinates the fields $b_1$ and $b_2$ are parameterized as
$b_1= \sin(\eta)e^{i\vth_1}$, $b_2= \cos(\eta)e^{i\vth_2}$ with $\eta \in [0,\pi/2]$, and $\vth_1,\vth_2 \in [0,2\pi)$,
and the theta term takes the form
\beq
	S_{\theta}[U]= \frac{1}{2\pi^2}\int d^3 x\ \cos(\eta)\sin(\eta)\ep^{\mu\nu\lam}\pd_{\mu}\eta\pd_{\nu}\vth_1\pd_{\lam}\vth_2\ . \label{eq:theta-hopf}
\eeq
The interspecies tunneling and chemical potential terms take the form
\beq
	b^{*}_1 b_2 + b^{*}_2 b_1 = 2\cos(\eta)\sin(\eta)\cos(\vth_1 - \vth_2)\ ,
\eeq
and 
\beqa
	\mu_1|b_1|^2 + \mu_2 |b_2|^2 &=& \mu_1 \cos^2(\eta) + \mu_2 \sin^2(\eta) \nnb \\
	&=& \mu_1  + (\mu_2 -\mu_1) \sin^2(\eta)\ .
\eeqa

Consider the interspecies tunneling term. When it is strong, the lowest energy configurations of the $O(4)$ field are those
configurations which have $\vth_1=\vth_2 + n\pi$ for some integer $n$ which is even or odd depending on the sign of the 
coefficient of this term. It is easy to see that the theta term vanishes identically on this kind of field configuration. The analysis
of the chemical potential term is even simpler.  Depending on the sign of $\mu_2 - \mu_1$, the lowest energy configurations are
those with $\sin(\eta)=0$ or $\sin(\eta) = 1$. In either case $\eta$ is a constant and so the theta term completely 
vanishes. This analysis makes it clear that a more thorough understanding of the effects of symmetry-allowed perturbations
on the BTI surface is needed.

\subsubsection{Interspecies tunneling}

%Naively, a term like $b^{*}_1 b_2 + b^{*}_2 b_1$ should have a drastic effect on the $O(4)$ NLSM at $\theta=\pi$. One
%way to see this is to use an amplitude-phase representation for the bosonic fields $b_I$ by writing them as 
%$b_I= \rho_I e^{i\theta_I}$, where now the constraint is $\sum_I \rho_I^2 = 1$. Then we have
%\beq
%	b^{*}_1 b_2 + b^{*}_2 b_1 = 2\rho_1\rho_2\cos(\theta_1 - \theta_2)\ .
%\eeq
%It seems that if this term is added to the Lagrangian and becomes dominant, then it will drive the system into a phase where
%$\theta_1=\theta_2$ (up to a possible multiple of $2\pi$). One can then easily check that the theta term $S_{\theta}[U]$
%vanishes identically on the configuration
%\beq
%	U= \begin{pmatrix}
%	\rho_1 e^{i\theta_1} & -\rho_2e^{-i\theta_1} \\
%	\rho_2 e^{i\theta_1} & \rho_1 e^{-i\theta_1}
%\end{pmatrix}\ ,
%\eeq
%where we have set $\theta_2 = \theta_1$. So it seems that strong inter-species tunneling can force the theta term to vanish 
%completely, destroying the topological properties of the model.
%
%
%\textbf{``are not important" is bad wording. Fix it.}

We now show that interspecies tunneling terms such as  $b^{*}_1 b_2 + b^{*}_2 b_1$, 
and even interaction terms such as $(b^{*}_1 b_2)^n + (b^{*}_2 b_1)^n$ for $n \geq 1$, 
do not condense (i.e., have zero expectation value)
in \emph{any} time-reversal invariant state $|\Psi\ran$. This means that interspecies tunneling terms can only condense in
the ground state of the system if that ground state breaks time-reversal symmetry. It also means that weak interspecies
tunneling terms should have a negligible effect on the gapless time-reversal invariant ground state of the $O(4)$ NLSM
with $\theta=\pi$.

To show that these expectation values vanish, we canonically quantize the theory and study the
(equal-time) commutation relations of 
the \emph{operators} $b_I(\mb{x})$, their hermitian conjugates $b^{\dg}_I(\mb{x})$, and their canonically conjugate 
momenta. We discuss the canonical quantization of this system in Appendix~\ref{app:canonical}.
The only commutation relation we will need here is 
\beq
	[b_I(\mb{x}),\pi_J(\mb{y})] = i\left(\delta_{IJ} - \frac{1}{2}b_I(\mb{x})b^{\dg}_J(\mb{y})\right)\delta^{(2)}(\mb{x}-\mb{y})\ , \label{eq:CR2}
\eeq
where $\pi_I = \pd \mathcal{L}/\pd (\pd_t b_I)$ is the momentum conjugate to $b_I$. Consider this commutation relation
first in the case where $I=J$, say for $I=J=1$. We have
\beq
	[b_1(\mb{x}),\pi_1(\mb{y})] = i\left(1 - \frac{1}{2}b_1(\mb{x})b^{\dg}_1(\mb{y})\right)\delta^{(2)}(\mb{x}-\mb{y})\ .
\eeq 
In the Hilbert space the action of the time-reversal symmetry $\ZT$ is represented by an anti-unitary operator 
$\mathcal{T}$, obeying $\T^2=1$, which acts on the boson operators $b_I(\mb{x})$ as
\beq
	\mathcal{T}b_I(\mb{x})\mathcal{T}^{-1}= b_I(\mb{x})\ .
\eeq
Then we must have 
\beq
	\mathcal{T}\pi_I(\mb{x})\mathcal{T}^{-1}= -\pi_I(\mb{x})\ ,
\eeq
in order for the commutation relations to be invariant under conjugation by $\mathcal{T}$. Now suppose we have a state
$|\Psi\ran$ of the system which is time-reversal invariant, i.e., $\mathcal{T}|\Psi\ran = |\Psi\ran$. Then the expectation
value $\lan \Psi|\mathcal{O}|\Psi\ran$ of any operator $\mathcal{O}$ which is odd under time-reversal, 
$\mathcal{T}\mathcal{O}\mathcal{T}^{-1}= -\mathcal{O}$, must vanish. 

We now apply this reasoning to the off-diagonal commutation relation
\beq
	[b_1(\mb{x}),\pi_2(\mb{y})] = -i\frac{1}{2}b_1(\mb{x})b^{\dg}_2(\mb{y})\delta^{(2)}(\mb{x}-\mb{y})\ . \label{eq:CRoff-diag}
\eeq
If we take the expectation value of both sides of this equation in the state $|\Psi\ran$, then the expectation value of the 
left-hand side vanishes since all operators on the left-hand side are odd under the action of $\mathcal{T}$. We are
left with the equation
\beq
	0 = -i\frac{1}{2}\lan \Psi| b_1(\mb{x})b^{\dg}_2(\mb{y})|\Psi\ran \delta^{(2)}(\mb{x}-\mb{y})\ ,
\eeq
and integrating both sides of this equation over $\mb{y}$ yields the final result
\beq
	\lan \Psi| b_1(\mb{x})b^{\dg}_2(\mb{x}) |\Psi\ran = 0\ .
\eeq
So we find that the operator $b_1(\mb{x})b^{\dg}_2(\mb{x})$ has zero expectation value in
any time-reversal invariant state $|\Psi\ran$. Going further, we may first multiply both sides of Eq.~\eqref{eq:CRoff-diag}
by \emph{any} time-reversal invariant operator $\tilde{\mathcal{O}}(\mb{x})$, and then take an expectation value in 
$|\Psi\ran$ to find that
\beq
	\lan \Psi| \tilde{\mathcal{O}}(\mb{x})b_1(\mb{x})b^{\dg}_2(\mb{x})|\Psi\ran = 0\ .
\eeq
For example we could take $\tilde{\mathcal{O}}(\mb{x})= (b_1(\mb{x})b^{\dg}_2(\mb{x}))^{n-1}$ to find that
the expectation value of $(b_1(\mb{x})b^{\dg}_2(\mb{x}))^{n}$ vanishes for any $n\geq 1$. 
We can conclude from this analysis that if interspecies tunneling and interaction terms of the form 
$(b^{*}_1 b_2)^n + (b^{*}_2 b_1)^n$ do condense in the ground state of the system, then that ground state
must break time-reversal symmetry. However, our analysis is not limited to just these terms, since there are many
more possible choices for the form of the operator $\tilde{\mathcal{O}}(\mb{x})$ which we are allowed to insert.

\subsubsection{Chemical potential}

We now discuss the effects of the chemical potential term on the quantum theory. In a theory of two independent complex scalar fields, a
chemical potential term, combined with suitable quartic terms in the potential, can have many possible effects on the fields
in the theory. 
For example, both scalar fields could become gapped, or they could both condense, or one scalar field could become
gapped and the other scalar field could condense. But the $O(4)$ NLSM is not a theory of two \emph{independent} complex scalar fields. Instead,
the fields $b_1$ and $b_2$ obey the very important constraint $\sum_I |b_I|^2 = 1$. In fact, 
with the help of the constraint, any chemical potential
term can be re-written as
\beq
	\mu_1|b_1|^2 + \mu_2 |b_2|^2 =\frac{1}{2}(\mu_1+\mu_2) + \frac{1}{2}(\mu_1-\mu_2)(|b_1|^2 - |b_2|^2)\ .
\eeq
This result indicates that for the $O(4)$ NLSM,
the effect of a general quartic potential of the form
\beq
	V(b_1,b_2)= \mu_1|b_1|^2 + \mu_2 |b_2|^2 + \lam_1 |b_1|^4 + \lam_2 |b_2|^4\ ,
\eeq
is to cause one of the fields $b_1$ or $b_2$ to condense and to cause the other field to become gapped. The choice of which
of $b_1$ or $b_2$ is condensed and which is gapped depends only on the sign of $\mu_1-\mu_2$ (assuming that $\lam_1$
and $\lam_2$ are positive). In particular, it seems that it is impossible to write down any potential which could cause both
$b_1$ and $b_2$ to condense. Further evidence for this conclusion can be obtained from an analysis of the 
commutation relations of the theory, as we now show.

Consider a state $|\Phi\ran$ which represents a superfluid ground state of the $O(4)$ NLSM for the boson $b_1$. In
such a state the $U(1)$ symmetry $b_1 \to e^{i\chi}b_1$ is spontaneously broken, and $\lan \Phi |b_1|\Phi\ran \neq 0$.
In general, the state $|\Phi\ran$ is not an eigenstate of the operator $b_1$, or even of the phase of $b_1$ 
(this can be seen from the form of the symmetry broken ground state for the phase excitations of an ordinary complex scalar 
field shown in Chapter 11 of Ref.~\onlinecite{itzykson}, for example). 
Below we show that in the special case where $|\Phi\ran$ \emph{is} an eigenstate of $b_1$, it is possible to prove
that $\lan \Phi|b_2|\Phi\ran =0$. For the general case where $|\Phi\ran$ is not an eigenstate of $b_1$, we must
instead rely on the qualitative argument presented above, and another argument which we present below which is based on 
the expression for $b_2$ in terms of the vortices $\phi_{1,\pm}$ in $b_1$ (Eq.~\eqref{eq:composite} with the indices
$1$ and $2$ swapped).

For now we assume that $|\Phi\ran$ is an eigenstate of the operators 
$b_1(\mb{x})$ and $b^{\dg}_1(\mb{x})$, and that
$b_1(\mb{x})|\Phi\ran = \alpha|\Phi\ran$ and $b^{\dg}_1(\mb{x})|\Phi\ran = \beta|\Phi\ran$, 
where $\alpha$ and $\beta$ are complex numbers which do not depend on $\mb{x}$. The relation 
$\lan\Phi|b_1(\mb{x})|\Phi\ran = \lan\Phi|b^{\dg}_1(\mb{x})|\Phi\ran^*$ implies that $\beta= \alpha^*$. 
Now assume that $\alpha \neq 0$, and take the expectation value of Eq.~\eqref{eq:CRoff-diag} in the state $|\Phi\ran$.
The left-hand side vanishes and we find
\beq
	0 = -i\frac{\alpha^*}{2}\lan\Phi|b^{\dg}_2(\mb{y})|\Phi\ran \delta^{(2)}(\mb{x}-\mb{y})\ .
\eeq
Since we assumed that $\alpha \neq 0$, we are forced to conclude that $\lan\Phi|b^{\dg}_2(\mb{x})|\Phi\ran=0$, which 
shows that $b_2$ cannot condense in an eigenstate of $b_1$ (which we have
argued is a representative ground state of the superfluid phase of $b_1$). 
Similarly, we can show that $b_1$ cannot condense in an eigenstate of $b_2$.

Another intuitive way of seeing that $b_2$ cannot condense in a superfluid ground state of $b_1$ is
to recall that $b_2$ can be expressed in terms of the two kinds of vortices in $b_1$ as $b_2 \sim \phi_{1,+}\phi^*_{1,-}$.
In a superfluid ground state of $b_1$ we expect that the vortices $\phi_{1,\pm}$ in $b_1$ are gapped (i.e., not
condensed), which means that we should also have $\lan b_2 \ran = 0$ in such a state.

We conclude that the main effect of a chemical potential term (combined with suitable quartic terms) is to spontaneously 
break the $U(1)_c$ symmetry, since this term will condense one of $b_1$ or $b_2$ and gap out the other one. Our analysis
of the commutation relations confirms that the vacuum expectation value of one boson always vanishes in a state which 
represents a superfluid ground state of the other boson. 

\subsection{Symmetry-preserving state with topological order}

In Ref.~\onlinecite{VS2013} Vishwanath and Senthil showed that it was possible for the surface phase of the BTI to retain
the full $U(1)_c \rtimes \ZT$ symmetry while gapped, but at the expense of having intrinsic topological order, 
and they went on to derive a specific topologically ordered state for the BTI surface. That
same topologically ordered state was constructed in Ref.~\onlinecite{CDL} using a coupled wires construction
consisting of Bosonic Integer Quantum Hall effect edge modes decorated with Toric Code/$\mathbb{Z}_2$ topological order (Abelian Chern-Simons theory with $K$-matrix given by $\pm 2\sigma^x$) edge
modes. In this section we briefly review
the construction of this topologically ordered state via vortex condensation in the $O(4)$ 
NLSM with $\theta=\pi$ as shown in Ref.~\onlinecite{VS2013}.

Recall the interpretation of the theta term that was derived in Ref.~\onlinecite{SenthilFisher} 
(see also our Appendix~\ref{app:path-integral}). According to Ref.~\onlinecite{SenthilFisher}, in the $O(4)$ NLSM with
$\theta=\pi$, a braiding process in which a vortex in the phase of $b_1$ makes a full circuit around a vortex in
the phase of $b_2$ results in an overall phase of $e^{i\pi}$ in the path integral for the theory. In other words, the vortex
in $b_1$ and the vortex in $b_2$ can be regarded as anyons with a mutual statistical angle of $\pi$. 

In the $O(4)$ NLSM at $\theta=\pi$, all quasi-particles with non-trivial statistics can be built just from the fundamental vortices
$\phi_{1,+}$ and $\phi_{2,+}.$ Indeed, recall that 
the other two vortices  $\phi_{1,-}$ and $\phi_{2,-}$ should really be understood as bound states of a vortex and a boson:
$\phi_{1,-} \sim \phi_{1,+}b^*_2$ and $\phi_{2,-} \sim \phi_{2,+}b^*_1$, and are hence not \emph{topologically} distinguishable from $\phi_{1,+}$ and $\phi_{2,+}.$
This means that the two vortices $\phi_{1,+}$ and $\phi_{2,+}$ should be sufficient building blocks 
to describe any topologically ordered states derived from this system. 
According to the arguments given in the previous paragraph, these two vortices have
a mutual statistical angle of $\pi$, i.e., they are mutual semions. Also, a composite of a vortex with
itself, such as $(\phi_{1,+})^2$, should be regarded as trivial (topologically equivalent to the vacuum quasi-particle), since that object 
has the exact same quantum numbers as the boson $b_2$ and braids trivially with all other quasi-particles. %The same statement holds for the ``$-$" vortices $\phi_{1,-}$ and $\phi_{2,-}$. 
This property is partly responsible for the $\mathbb{Z}_2$ structure
of the topological order discussed below.

We can now consider condensing some field $\mathcal{O}$ which is a composite of the vortices. 
Standard reasoning then tells us
that any quasi-particles that have trivial mutual statistics with $\mathcal{O}$ will survive as anyons in the state obtained
by condensing $\mathcal{O}$. In Ref.~\onlinecite{VS2013} Vishwanath and Senthil construct a topologically ordered phase
for the surface of the BTI by choosing to condense $\mathcal{O}= \phi_{1,+}\phi_{1,-}$ in such a 
way that $\lan \mathcal{O} \ran$ is real. Since $\ZT$  maps $\mathcal{O} \to \mathcal{O}^*$, this 
condensation does not break time-reversal symmetry. In addition, $\mathcal{O}$ is invariant under $U(1)_c$, so the resulting
phase actually retains the full $U(1)_c\rtimes \ZT$ symmetry of the BTI.
%\beqa
%	\mathcal{O} &=& \phi_{1,+}\phi_{1,-} + \text{c.c.}\ . \\ \label{eq:topo-order}
%			&\sim&  (\phi_{1,+})^2 b^*_2 + \text{c.c.}\ ,
%\eeqa
%where we have used the relation $\phi_{1,-} \sim \phi_{1,+}b^*_2$. Note that $\mathcal{O}$ is invariant under the full 
%$U(1)_c\rtimes \ZT$ symmetry, so the phase obtained by condensing
%$\mathcal{O}$ does not break any symmetries of the system. 

We see that both $\phi_{1,+}$ and $\phi_{2,+}$ have trivial mutual statistics with $\mathcal{O}$, so these vortices
both survive as quasi-particles in the condensed state. The condensed state therefore has quasi-particle content
(recall that fusing a vortex with itself gives a trivial excitation)
\beq
	\left\{1,\phi_{1,+},\,\phi_{2,+},\, \phi_{1,+}\phi_{2,+}  \right\}\ ,
\eeq 
where ``$1$" is the trivial (vacuum) quasi-particle and $\phi_{1,+}\phi_{2,+}$ is the composite of the two vortices
$\phi_{1,+}$ and $\phi_{2,+}$. The composite vortex $\phi_{1,+}\phi_{2,+}$ is actually a fermion.
The exchange and mutual statistics of these quasi-particles is shown in 
Table~\ref{tab:anyons}. These quasi-particles with the braiding statistics shown in Table~\ref{tab:anyons} form a 
$\mathbb{Z}_2$ topological order which is characterized by a $K$-matrix $K= 2\sigma^x$ and charge vector
$\vec{t}= (1,1)^T$. In a purely 2D system which admits an edge to the vacuum, such a system would exhibit a charge
Hall conductance of $\vec{t}\cdot (K^{-1}\vec{t}) = 1$ (it is non-zero and therefore breaks time-reversal symmetry), 
which  hints that the topologically-ordered state on the surface of the BTI realizes time-reversal symmetry
in a way which is forbidden in a real 2D system~\cite{VS2013}.

\begin{table}
\begin{tabular}{|c|c|c|c|}
\hline
 & $\phi_{1,+}$ & $\phi_{2,+}$ & $\phi_{1,+}\phi_{2,+}$ \\ \hline
$\phi_{1,+}$ & 0 & $\pi$ & $\pi$ \\ \hline
$\phi_{2,+}$ & $\pi$ & 0 & $\pi$ \\ \hline
$\phi_{1,+}\phi_{2,+}$ & $\pi$ & $\pi$ & $\pi$ \\ \hline
\end{tabular}
\caption{Exchange and mutual statistics for the quasi-particles contained in the topologically ordered surface phase of the 
BTI which is accessed by condensing the composite field $\mathcal{O} = \phi_{1,+}\phi_{1,-}$.  
The diagonal entries in the table (\emph{exchange} statistics) are the phase for a process in which two identical particles exchange
positions (so a phase of $\pi$ represents fermions),
while the off-diagonal entries (\emph{mutual} statistics) are the phase picked up when a particle of one type makes a complete circuit around a particle of 
a different type. }
\label{tab:anyons}
\end{table}

\section{The Bosonic Semi-Metal model: two $O(4)$ NLSM's with $\theta= \pm \pi$}
\label{sec:BSM-model}

In this section we introduce our Bosonic Semi-Metal (BSM) model. The model is constructed from two copies of the $O(4)$
NLSM with theta term, and we take one copy to have $\theta=\pi$ and the other copy to have $\theta=-\pi$. The intuition
behind the construction of our model is as follows. Recall that the surface theory of the 3D ETI 
is a single massless 2+1-d Dirac fermion. The two-cone DSM phase in 2+1-d can then be viewed as being 
constructed from two copies of the surface theory of the 3D ETI, with the two copies separated
in momentum space and having opposite helicity. 
For our BSM model we instead take two copies of the $O(4)$ NLSM with $|\theta|=\pi$,
(i.e., two copies of the surface theory of the BTI), but we take the two copies
to have opposite signs of $\theta$, which is the bosonic analog of the helicity of the 2+1-d Dirac fermion. One way to see this is in the
construction by Abanov and Wiegmann in Ref.~\onlinecite{AbanovWiegmann}, 
where the helicity of the auxiliary fermions directly determines the sign of the theta angle in the resulting $O(4)$ NLSM.

This section is broken up into several subsections as follows. We first define our BSM model and the transformations of the 
fields in the model under $U(1)_c$ charge conservation symmetry,
$U(1)_t$ ``translation" symmetry (to be defined), $\ZT$ time-reversal symmetry, and $\ZI$ inversion symmetry. We 
then discuss the dual description of our BSM model and derive the action of the different symmetries on the vortex fields in the 
dual theory. Finally, we calculate the time-reversal and inversion breaking electromagnetic responses of our BSM 
model (again using two different methods), and compare the result with that of the 2+1-d DSM model discussed in 
Ref.~\onlinecite{Ramamurthy2014}, and reviewed in 
Sec.~\ref{sec:DSM-review}. 
We then discuss the stability of the model and find that the composite $\ZTI$ symmetry again plays an important
role. Finally, we close the section with a discussion of phases with $\mathbb{Z}_2$ and $\mathbb{Z}_2\times\mathbb{Z}_2$ 
topological order which can be accessed from our BSM model by condensing a composite of the vortices appearing in the dual 
description of the model. 
We show that these phases break either the time-reversal or the inversion 
symmetry of the BSM model. This is interesting because the gapped phases which do not have topological order also must break 
one of these two symmetries.

\subsection{BSM model and symmetries}

Our BSM model consists of two copies of an $O(4)$ NLSM with theta term, called ``A" and ``B" copies, with the 
theta angles for the two copies being $\theta_A=\pi$ and $\theta_B=-\pi$. We write the model in terms of $SU(2)$ matrices
$U_A$ and $U_B$, which are each expressed in terms of bosonic fields $b_{I,A}$ and $b_{I,B}$, $I=1,2$, 
as in Eq.~\eqref{eq:b-variables}. The action for the system is
\beqa
	S= \int d^3 x\  \left[\frac{1}{2 g}\text{tr}[\pd^{\mu} U_A^{\dg} \pd_{\mu} U_A + (A\to B)]\right. \nonumber\\ - \left.\pi S_{\theta}[U_A] + \pi S_{\theta}[U_B]\right]\label{eq:O4PC}\ ,
\eeqa
where the explicit form of the theta term was given in Eq.~\eqref{eq:thetaPC}.

The fields in the BSM model transform under $U(1)_c$ charge conservation symmetry, $U(1)_t$ ``translation" symmetry, 
$\ZT$ time-reversal symmetry, and $\ZI$ inversion symmetry. In this section we explain the action of each of these 
symmetries on the fields $b_{I,A}$ and $b_{I,B}$ in the model. Just as for the DSM, the composite symmetry
$\ZTI$, consisting of a time-reversal transformation followed by an inversion transformation, will be important for guaranteeing
the stability of the gapless phase of our model. 

The fields transform under the $U(1)_c$ symmetry as 
\beq
	U(1)_c: b_{I,A/B} \to e^{i\chi}b_{I,A/B}\ . 
\eeq
This just indicates that each bosonic field $b_{I,A/B}$ carries charge $1$ of the external gauge field $A_{\mu}$.
Under the ``translation" $U(1)$ symmetry, $U(1)_t$, the fields transform as
\begin{subequations}
\label{eq:U1t}
\beqa
	U(1)_t : b_{I,A} &\to& e^{i\xi}b_{I,A} \\
			 b_{I,B} &\to& e^{-i\xi}b_{I,B}\ . 
\eeqa
\end{subequations}

To explain the physical meaning of this $U(1)$ translation symmetry, we need to imagine that our BSM model
has been obtained in the low-energy continuum limit of a bosonic lattice model, as we now explain.
Let us assume that the fields $b_{I,A}$ and $b_{I,B}$ arise from the low-energy continuum limit of a 
bosonic lattice model, and that they are related to the lattice boson operators in a way similar to the DSM
case illustrated in Eq.~\eqref{eq:expansion}.  In other words, the combinations $e^{i\mb{k}_{+}}b_{I,A}$
and $e^{i\mb{k}_{-}}b_{I,B}$ appear in the expression for the lattice boson operator, indicating that the 
continuum fields $b_{I,A}$ and $b_{I,B}$ are located at positions $\mb{k}_{\pm}= \pm (B_x,B_y)$ in momentum space. We 
provide an explicit example of such a model using a coupled-wire construction in the next section.

To model this momentum shift, the kinetic term in the Lagrangian for the fields $b_{I,A}$ and $b_{I,B}$ will feature a minimal coupling to the 
vector field $B_{\mu}$ (and just as in the DSM case, we also allow for a relative energy offset given by $B_t$).
The translation properties of the fields $b_{I,A}$ and $b_{I,B}$ can then be thought of in terms of  carrying charges $1$ and 
$-1$, respectively, of the field $B_{\mu}$, and the action is invariant under the $U(1)_t$ gauge transformation in which
the bosonic fields transform according to Eq.~\eqref{eq:U1t} while $B_{\mu} \to B_{\mu} + \pd_{\mu}\xi$.
This is the physical origin of the $U(1)_t$ symmetry
\footnote{We also assume here that $\mb{k}_{\pm}$ are incommensurate with the 
reciprocal lattice vectors of the underlying lattice. If $\mb{k}_{\pm}$ were commensurate with the lattice then 
we will have $n\ \mb{k}_{\pm}\cdot \mb{x} = 2\pi m$ for some integers $m$ and $n$, where $\mb{x}$ is a coordinate
on the lattice. In this case the continuous translation symmetry which we are assuming would be broken down to a 
discrete subgroup.}.

We now discuss the discrete symmetries $\ZI$, $\ZT$ and $\ZTI$. We take $\ZI,$ and $\ZT$ to act on the bosonic
fields as
\beqa
	\ZI: b_{I,A}(t,\mb{x}) &\to& b_{I,B}(t,-\mb{x}) \label{eq:symI} \\  
	\ZT: b_{I,A}(t,\mb{x}) &\to& b_{I,B}(-t,\mb{x})\ , \label{eq:symT}
\eeqa
and vice-versa. The composite symmetry $\ZTI$ then acts as
\beq
	\ZTI: b_{I,A}(t,\mb{x}) \to b_{I,A}(-t,-\mb{x})\ ,
\eeq
with an identical transformation for $b_{I,B}(t,\mb{x})$. 
In the canonical formalism, the action of time-reversal is represented by the anti-unitary operator
$\mathcal{T}$, and the action of inversion is represented by the unitary operator $\mathcal{I}$. From the symmetry
transformations defined above we can see that these operators satisfy the identities $\mathcal{T}^2 = 1$, 
$\mathcal{I}^2=1$, and $[\mathcal{T},\mathcal{I}]=0$, which implies that $(\mathcal{TI})^2=1$ as well. 

Just as in the fermionic DSM case, the composite $\ZTI$ symmetry is important for ensuring the local (in momentum space) stability of each $O(4)$ NLSM copy in our BSM model.
We will have more to say on this subject later in this section, but for now we note the following important property
of the $\ZTI$ symmetry for the BSM model. In the BSM model it is actually the $\ZTI$ symmetry which fixes the theta angles $\theta_A$ 
and $\theta_B$ to be multiples of $\pi$, just as the $\ZT$ symmetry guaranteed this property for the BTI surface theory. 
Therefore, from this general argument, the gaplessness of the BSM model (which can occur only when the theta angles are odd multiples of $\pi$) depends
crucially on this symmetry.

%Note that the two $U(1)$ symmetries forbid all possible bilinear terms $b^{\dg}_{I,A}b_{J,B}$ and $b_{I,A}b_{J,B}$
%that could couple the two copies of the $O(4)$ model together. Later we will show that the combined $\mathcal{TI}$
%symmetry prevents gapping out just one of the two $O(4)$ NLSM's (either the $A$ or $B$ copy).
%
%We will probe the $U(1)_c$ symmetry by coupling the theory to the background gauge field $A_{\mu}$, and we will
%probe the $U(1)_t$ symmetry by coupling the theory to the background gauge field $B_{\mu}$
%(whose physical meaning we have already discussed). In the following sections
%we investigate electromagnetic response of this system to time-reversal and inversion breaking perturbations using
%two completely different approaches. 

\subsection{Dual vortex description of the BSM model}

We now turn to the dual vortex description of our BSM model, using the dual description of one $O(4)$ model which 
we reviewed in Sec.~\ref{sec:BTI-surface}.
We choose to employ the dual vortex description in terms of vortices in $b_{2,A}$ and $b_{2,B}$,
although a description starting in terms of vortices in $b_{1,A}$ and $b_{1,B}$ is also possible.
For the ``$A$" NLSM, vortices in $b_{2,A}$ are represented by the two-component field
$\Phi^{(A)}_2= (\phi^{(A)}_{2,+},\phi^{(A)}_{2,-})^T$. For the ``$B$" copy of the NLSM, vortices in 
$b_{2,B}$ are represented by the two-component field $\Phi^{(B)}_2= (\phi^{(B)}_{2,+},\phi^{(B)}_{2,-})^T$.

As discussed in Sec.~\ref{sec:BTI-surface}, and as we explicitly prove in Appendix~\ref{app:vortices},  a vortex in the phase of one boson binds a charge of $\frac{\theta}{2\pi}$ of the other boson. This
result holds for any $U(1)$ symmetry under which the bosons are charged; for example, the $U(1)_c$ and $U(1)_t$ symmetries
in our case. 
This means that under the $U(1)_c$ and $U(1)_t$ 
symmetries, the field $\Phi^{(A)}_2$ transforms as 
\beqa
	U(1)_c : \Phi^{(A)}_2 \to e^{i\tfrac{\chi}{2}\sigma^z}\Phi^{(A)}_2 \\
	U(1)_t : \Phi^{(A)}_2 \to e^{i\tfrac{\xi}{2}\sigma^z}\Phi^{(A)}_2\ .
\eeqa
On the other hand, the ``$B$" copy of the $O(4)$ NLSM in our BSM model has theta angle $\theta_B = -\pi$. 
The elementary vortices $\phi^{(B)}_{1,+}$ and $\phi^{(B)}_{2,+}$ now both carry charges $-\frac{1}{2}$ and 
$\frac{1}{2}$ under the $U(1)_c$ and $U(1)_t$ symmetries, respectively. The ``$-$" vortices must now be defined
as $\phi^{(B)}_{2,-} = \phi^{(B)}_{2,+}b_{1,B}$ and $\phi^{(B)}_{1,-} = \phi^{(B)}_{1,+}b_{2,B}$. Unlike for
the ``A" copy, these relations involve the bosons $b_{I,B}$ and not the anti-bosons $b_{I,B}^*$ since $\theta_B=-\pi$ 
and not $+\pi.$
We then find that under the $U(1)_c$ and $U(1)_t$ 
symmetries, the field $\Phi^{(B)}_2$ transforms as
\beqa
	U(1)_c : \Phi^{(B)}_2 \to e^{-i\tfrac{\chi}{2}\sigma^z}\Phi^{(B)}_2 \\
	U(1)_t : \Phi^{(B)}_2 \to e^{i\tfrac{\xi}{2}\sigma^z}\Phi^{(B)}_2\ .
\eeqa

In terms of the fields $\Phi^{(A)}_2$ and $\Phi^{(B)}_2$, the dual description of the BSM model has the Lagrangian
$\mathcal{L}= \mathcal{L}^{(A)} + \mathcal{L}^{(B)}$, with
\beqa
	\mathcal{L}^{(A)} &=& \sum_{s=\pm } |[\pd_{\mu} - i\al^{(A)}_{2,\mu} - i\frac{s}{2}(A_{\mu} + B_{\mu})]\phi^{(A)}_{2,s}|^2  \\ 
	&-& \frac{1}{\kappa_{2,A}}\left(\frac{1}{2\pi}\ep^{\mu\nu\lam}\pd_{\nu}\al^{(A)}_{2,\lam}\right)^2 - \frac{1}{2\pi}\ep^{\mu\nu\lam}(A_{\mu} + B_{\mu})\pd_{\nu}\al^{(A)}_{2,\lam}\ , \nnb
\eeqa
and
\beqa
	\mathcal{L}^{(B)} &=& \sum_{s=\pm } |[\pd_{\mu} - i\al^{(B)}_{2,\mu} + i\frac{s}{2}(A_{\mu} - B_{\mu})]\phi^{(B)}_{2,s}|^2 \\ 
	&-& \frac{1}{\kappa_{2,B}}\left(\frac{1}{2\pi}\ep^{\mu\nu\lam}\pd_{\nu}\al^{(B)}_{2,\lam}\right)^2 - \frac{1}{2\pi}\ep^{\mu\nu\lam}(A_{\mu} - B_{\mu})\pd_{\nu}\al^{(B)}_{2,\lam}\ . \nnb
\eeqa
In these expressions, $\frac{1}{2\pi}\ep^{\mu\nu\lam}\pd_{\nu}\al^{(A)}_{2,\lam}$ and  
$\frac{1}{2\pi}\ep^{\mu\nu\lam}\pd_{\nu}\al^{(B)}_{2,\lam}$ represent the number currents of the bosons $b_{2,A}$
and $b_{2,B}$, respectively. We have
included coupling to the external probe fields $A_{\mu}$ and $B_{\mu}$ associated with the two $U(1)$
symmetries $U(1)_c$ and $U(1)_t$. It is also possible to add various potential energy terms to these dual Lagrangians. 

\subsection{Transformation of vortices under $\mathcal{T}$ and $\mathcal{I}$ symmetries} 

In this section we deduce the transformations of the vortices under the $\ZT$ and $\ZI$ symmetries. 
First we note that because of the quantum numbers carried by the vortex fields, we have the approximate  
relations
\beqa
	b_{1,A} &\sim& \phi^{(A),*}_{2,-}\phi^{(A)}_{2,+} \\
	b_{1,B} &\sim& \phi^{(B),*}_{2,+}\phi^{(B)}_{2,-}\ ,
\eeqa
which are just Eq.~\eqref{eq:composite} written for the two copies of the $O(4)$ NLSM, and taking into account the fact
that the ``B" copy of the $O(4)$ NLSM has $\theta_B=-\pi$. 
Also, recall that a dimensionful quantity like $g$, the NLSM coupling
constant, is needed to balance the units in this equation, but we ignore that subtlety here.
We now deduce the transformations of the vortices under $\ZT$ and $\ZI$ by requiring that
the transformations of the vortices under these symmetries reproduce the transformations of $b_{1,A}$ and $b_{1,B}$
under the symmetries, and that the action of the symmetries on the vortices is consistent with the general structure of the 
symmetry group.

Consider first the inversion symmetry. Inversion commutes with the $U(1)_c$ symmetry, whereas it negates
the $U(1)_t$ charge. In addition, since inversion is a unitary symmetry, it should take vortices to vortices, not anti-vortices 
(conjugation by the operator $\mathcal{I}$ \emph{does not} negate the phase of $b_{I,A/B}$).
We have only two options: either $\Phi^{(A)}_2(t,\mb{x}) \to \sigma^x\Phi^{(B)}_2(t,-\mb{x})$ or 
$\Phi^{(A)}_2(t,\mb{x}) \to i\sigma^y\Phi^{(B)}_2(t,-\mb{x})$. Only the first option is consistent with
Eq.~\eqref{eq:symI}. We hence find that
\beq
	\ZI: \Phi^{(A)}_2(t,\mb{x}) \to \sigma^x\Phi^{(B)}_2(t,-\mb{x})\ ,
\eeq
and vice-versa.

Next consider time-reversal symmetry. Time-reversal is anti-unitary, so it should take vortices to anti-vortices
(conjugation by $\mathcal{T}$ \emph{does} negate the phase of $b_{I,A/B}$).
In addition, time-reversal preserves the $U(1)_c$ charge and negates the $U(1)_t$ charge. 
The only two possibilities are then $\Phi^{(A)}_2(t,\mb{x}) \to  \Phi^{(B),*}_2(-t,\mb{x})$
or $\Phi^{(A)}_2(t,\mb{x}) \to  \sigma^z\Phi^{(B),*}_2(-t,\mb{x})$. Only the first option is consistent with
Eq.~\eqref{eq:symT}, so we find that
\beqa
	\ZT: \Phi^{(A)}_2(t,\mb{x}) \to  \Phi^{(B),*}_2(-t,\mb{x})\ ,
\eeqa
and vice-versa. 

We see that the time-reversal and inversion symmetry continue to commute with each other when acting
on the vortices. The combined $\ZTI$ symmetry then acts on the vortices as
\beq
	\ZTI:  \Phi^{(A)}_2(t,\mb{x}) \to  \sigma^x \Phi^{(A),*}_2(-t,-\mb{x})\ ,
\eeq
and similarly for $\Phi^{(B)}_2$.

\subsection{Time-reversal and inversion breaking mass terms , electromagnetic response, and a bosonic Chern insulator}

Now that we know how the vortex fields transform under the various symmetries, we can use the dual vortex 
theory to calculate the responses of our BSM model to time-reversal and inversion breaking perturbations. 
Analogous to the fermion DSM, we can define a time-reversal breaking mass term for the vortices,
\beq
	\Sigma_{\mathcal{T}}= \Phi^{(A),*}_2\sigma^z\Phi^{(A)}_2 - \Phi^{(B),*}_2\sigma^z\Phi^{(B)}_2\ ,
\eeq
and also an inversion breaking mass term
\beq
	\Sigma_{\mathcal{I}}= \Phi^{(A),*}_2\sigma^z\Phi^{(A)}_2 + \Phi^{(B),*}_2\sigma^z\Phi^{(B)}_2\ .
\eeq
The term $\Sigma_{\mathcal{T}}$ is odd under $\ZT$ but even under $\ZI$. On the other hand, $\Sigma_{\mathcal{I}}$
is even under $\ZT$ but odd under $\ZI$.

%It is odd under the action of $\mathcal{I}$ but even under the action of $\mathcal{T}$. Note that any mass
%term with $\sigma^x$ or $\sigma^y$ in place of $\sigma^z$ in the above expressions would explicitly break the 
%$U(1)_c$ symmetry, and is therefore forbidden.

Now let us consider the electromagnetic response in these two gapped phases. Suppose we add the time-reversal breaking 
mass term $\mu \Sigma_{\mathcal{T}}$ to the vortex potential energy.
If $\mu <0$ (and in the presence of suitable quartic terms in the vortex action), this
will cause $\phi^{(A)}_{2,-}$
and $\phi^{(B)}_{2,+}$ to become gapped, and $\phi^{(A)}_{2,+}$
and $\phi^{(B)}_{2,-}$ to condense. 
In this case we can then integrate out $\phi^{(A)}_{2,-}$
and $\phi^{(B)}_{2,+}$. A mean-field treatment of the remaining terms in the action then gives 
$\al^{(A)}_{2,\mu} = -\frac{1}{2}(A_{\mu}+B_{\mu})$ and $\al^{(B)}_{2,\mu} = -\frac{1}{2}(A_{\mu}-B_{\mu})$,
which gives the 2D time-reversal breaking response
\beq
	\mathcal{L}_{\mathcal{T}}=  \frac{e^2}{2\pi}\ep^{\mu\nu\lam}A_{\mu}\pd_{\nu}A_{\lam} + \frac{1}{2\pi}\ep^{\mu\nu\lam}B_{\mu}\pd_{\nu}B_{\lam}\ . \label{eq:2D-BSM-response}
\eeq
The first term in this expression is a Quantum Hall response with Hall conductivity $\sigma_{xy}=2e^2/h$,
exactly the same as one finds for the Bosonic Integer Quantum Hall effect \cite{VL2012,SenthilLevin}. If we took $\mu >0$
we would get the same response but with the opposite sign. We note that we cannot add a simple mass term to find a $\sigma_{xy}$ quantized as an odd multiple of $e^2/h.$ This gapped phase represents a Bosonic Chern insulator. 

On the other hand, we can add the inversion breaking mass term $\mu \Sigma_{\mathcal{I}}$ to the vortex
potential energy instead. If $\mu <0$ (and again, assuming suitable quartic terms), this will cause $\phi^{(A)}_{2,-}$
and $\phi^{(B)}_{2,-}$ to become gapped and $\phi^{(A)}_{2,+}$
and $\phi^{(B)}_{2,+}$ to condense. In a mean-field treatment this gives 
$\al^{(A)}_{2,\mu} = -\frac{1}{2}(A_{\mu}+B_{\mu})$ and $\al^{(B)}_{2,\mu} = \frac{1}{2}(A_{\mu}-B_{\mu})$,
which yields the quasi-1D inversion breaking response
\beq
	\mathcal{L}_{\mathcal{I}}= \frac{e}{\pi}\ep^{\mu\nu\lam}B_{\mu}\pd_{\nu}A_{\lam}\ . \label{eq:1D-BSM-response}
\eeq
Again, if we took $\mu>0$ then we would get the same response but with the opposite sign. This response encodes a charge 
polarization $P^{i}=\frac{e}{\pi}\epsilon^{ij}B_{j}$ ($i,j=x,y$) and an orbital magnetization $M=\frac{e}{\pi}B_t.$

We see that both the time-reversal breaking and inversion breaking electromagnetic responses of the BSM are \emph{twice}
as large as the responses for the free fermion DSM shown in Eq.~\eqref{eq:DSM-responses}. Let us now provide an alternate 
derivation of these responses. 

%\subsection{$\mathcal{TI}$ symmetry forbids gapping out one copy of the theory}
%
%Next we consider all possible mass terms that one can add for just one copy of the theory, and then show that the 
%combined $\mathcal{TI}$ symmetry (together with the $U(1)$ symmetry) forbids all such terms. The general mass
%term for theory $A$ takes the form $\Phi^{(A),\dg}_2 M \Phi^{(A)}_2$, where $M= \mathbb{I},\sigma^x,\sigma^y$ or 
%$\sigma^z$. The $U(1)_c$ symmetry forbids $M= \sigma^x$ or $\sigma^y$. The choice $M= \sigma^z$ is forbidden by
%the combined $\mathcal{TI}$ symmetry.
%\textbf{We need to give a clear discussion of the identity matrix mass term. $\Phi_2$ is actually supposed to be a CP1 field,
%so we have $\Phi_2^{\dg}\Phi_2= 1$. This means that the identity matrix term does nothing/does not exist. The original
%claim by Senthil and Fisher is that $\Phi_2$ is a CP1 field, so I think we are ok assuming that.}
%
%So we have shown that the combined $\mathcal{TI}$ symmetry prevents us from just gapping out only one 
%of the theories (either $A$ or $B$).

\subsection{Electromagnetic Responses from Abanov-Wiegmann Method}

We now briefly show how the time-reversal and inversion breaking responses of our BSM model can be computed
using the Abanov-Wiegmann method of integration over auxiliary fermions which we discussed in Sec.~\ref{sec:BTI-surface}. 
We first rewrite our BSM model in terms of two four-component unit vector fields $\mb{N}_A$ and $\mb{N}_B$.
Now introduce the multi-component complex fermion
$\Psi= (\psi_{1,A},..,\psi_{4,A},\psi_{1,B},..,\psi_{4,B})^T$, where each of $\psi_{a,A/B}$ is a two-component Dirac 
fermion in $2+1$ dimensions. The fermion $\Psi$ has a total of $16$ components. 
In terms of the two sets of gamma matrices introduced in Eqs.~\eqref{eq:lc-gamma} and \eqref{eq:uc-gamma},
our BSM model can be obtained from the fermionic Lagrangian
\beqa
	\tilde{\mathcal{L}}_{f} &=& \bar{\Psi}\left(i\tilde{\gamma}^{\mu}\pd_{\mu} - \frac{M}{2}\sum_{a=1}^4 N^a_A (\mathbb{I}+\sigma^z)\otimes\Gamma^a \right. \nnb \\
&-& \left. \frac{M}{2}\sum_{a=1}^4 N^a_B (\mathbb{I}-\sigma^z)\otimes\Gamma^a \right)\Psi\ ,
\eeqa
where we have defined
\begin{subequations}
\beqa
	\tilde{\gamma}^0 &=& \mathbb{I}\otimes\gamma^0 \\
	\tilde{\gamma}^1 &=& \mathbb{I}\otimes\gamma^1 \\ 
	\tilde{\gamma}^2 &=& \sigma^z\otimes\gamma^2\ ,
\eeqa
\end{subequations}
and $\bar{\Psi} = \Psi^{\dg}\tilde{\gamma}^0$ now.
The extra $\sigma^z$ on $\tilde{\gamma}^2$ means that the fermions $\psi_{a,B}$ have opposite helicity to the 
fermions $\psi_{a,A}$. This change directly accounts for the opposite signs of the theta angle for the ``A" and ``B"
copies of the $O(4)$ NLSM that we get when we integrate out $\Psi$. This is the reason why we stated earlier that the
sign of $\theta$ is the analogue in the BSM of the helicity of the Dirac fermions in the DSM. Indeed, we see that the helicity
of the Abanov-Wiegmann auxiliary fermions directly translates into the sign of $\theta$ in the $O(4)$ NLSM.

By the same reasoning used in  Sec.~\ref{sec:BTI-surface} 
to deduce the charges of the fermions used to generate one $O(4)$ NLSM
on the surface of the BTI, we now find that the field $\Psi$ transforms under the $U(1)_c$ and $U(1)_t$ symmetries as
\beqa
	U(1)_c: \Psi &\to& e^{i\chi \II\otimes Q}\Psi \\
	U(1)_t: \Psi &\to& e^{i\xi \sigma^z\otimes Q}\Psi\ ,
\eeqa
where $Q$ is the $8\times 8$ charge matrix introduced in Eq.~\eqref{eq:charge-matrix}. We can now couple $\Psi$ to the
background gauge fields $A_{\mu}$ and $B_{\mu}$ and calculate the response of the system to various perturbations.
%\beqa
%	\tilde{\mathcal{L}}_{f,gauge} &=& \bar{\Psi}\Biggl(i\tilde{\gamma}^{\mu}\pd_{\mu} - \frac{M}{2}\sum_{a=1}^4 N^a_A (\mathbb{I}+\sigma^z)\otimes\Gamma^a  \nnb \\
%&-&  \frac{M}{2}\sum_{a=1}^4 N^a_B (\mathbb{I}-\sigma^z)\otimes\Gamma^a -  i( \II\otimes Q)\tilde{\gamma}^{\mu}A_{\mu}  \nnb \\
%&-& i(\sigma^z\otimes Q)\tilde{\gamma}^{\mu}B_{\mu} \Biggr)\Psi \ .
%\eeqa

The time-reversal breaking response is obtained by adding the term $- (M\delta) \bar{\Psi}(\sigma^z\otimes \Gamma^5)\Psi$
to the Lagrangian. According to Eq.~\eqref{eq:theta-angle}, this will give $\theta_A\approx \pi(1-\tfrac{3}{2}\delta)$
and $\theta_B \approx -\pi(1+\tfrac{3}{2}\delta)$, 
 so this breaks $\ZT$ (which requires
$\theta_A \equiv -\theta_B$ mod $2\pi$). The inversion breaking 
response is obtained by adding the term $ -( M\delta) \bar{\Psi}(\mathbb{I}\otimes \Gamma^5)\Psi$.
This will give  $\theta_A\approx \pi(1-\tfrac{3}{2}\delta)$and $\theta_B \approx -\pi(1-\tfrac{3}{2}\delta)$, 
so this breaks $\ZI$ (which requires $\theta_A \equiv \theta_B$ mod $2\pi$).

%After integrating out the fermions, these terms have the effect of shifting the theta angles of the ``A" and ``B" copies of the
%$O(4)$ NLSM's in such a way that either the time-reversal or the inversion symmetry of the BSM action is broken 
%(\textbf{MFL: I can say something more precise here if needed. Just need to work out the signs in the shifts of the
%theta angles.}). 

In the limit that $\delta \to 0$, the time-reversal breaking perturbation generates the 2+1-d response
\beq
	\tilde{\mathcal{L}}_{\mathcal{T}} =  -\text{sgn}(\delta)\frac{1}{2\pi}\ep^{\mu\nu\lam}\left( e^2 A_{\mu}\pd_{\nu}A_{\lam} + B_{\mu}\pd_{\nu}B_{\lam}\right)\ , \label{eq:below-theta-discussion}
\eeq
coming from the contributions of each of the Dirac fermions (and according to their charge), while the inversion breaking perturbation gives the quasi-1D response
\beq
	\tilde{\mathcal{L}}_{\mathcal{I}} = -\text{sgn}(\delta)\frac{ e}{\pi}\ep^{\mu\nu\lam} B_{\mu}\pd_{\nu}A_{\lam}\ .
\eeq
These are the same responses which we derived in the previous subsection using the dual vortex formulation of the BSM model.

\subsection{Stability of the BSM Effective Theory}
We have provided an effective theory for a gapless bosonic semi-metal in 2+1-d and we now want to evaluate the perturbative stability of this theory to see under what conditions the semi-metal is a stable phase.
In discussing the stability of the BSM model, there are a few expected properties which we would like to verify. First, the translation
symmetry of the model should prevent us from trivially gapping out the model by coupling the ``A" copy of the $O(4)$ NLSM
to the ``B" copy (for our purposes, by a trivially gapped phase, we mean a gapped phase which retains all the symmetries
of the original gapless system and has no interesting electromagnetic response). And second, the composite $\ZTI$ symmetry 
should guarantee the \emph{local} stability of
the $O(4)$ NLSM's which make up our BSM model (recall that local stability means that it should be impossible to gap out one 
copy of the $O(4)$ NLSM independently of the other copy without breaking required symmetries). We claim that analogous to the 2+1-d fermionic DSM, these symmetries are enough to provide perturbative stability to the BSM. However, just as with any symmetry-protected phase (gapped or gapless) it is also important to keep in mind the possibility that even
symmetry-allowed perturbations may spontaneously break one or more of the symmetries of the system if those perturbations 
are strong enough. 

Also, as a caveat, the $O(4)$ NLSM with $\theta=\pi$ is a difficult
interacting theory to study in general. Since many of its properties are still unknown, 
it is impossible for us to give a complete characterization of the stability
of our BSM effective theory. We do provide a thorough analysis of the the effects of many important perturbations on the BSM model, but there are still
many other symmetry-allowed perturbations that we have not been able to completely understand: for example, a quartic 
coupling of the form $|b_{I,A}|^2|b_{J,B}|^2$ between bosons in the ``A" and ``B" copies of the $O(4)$ NLSM. Our discussion in this section gives strong evidence for the stability of the semi-metal phase so we will leave a possible discussion of these untreated terms to future work. 

Let us begin by addressing the issue of trivially gapping out the system by coupling the ``A" copy of the $O(4)$ NLSM to the 
``B" copy. Since the two copies of the $O(4)$ NLSM have opposite theta angles, an interaction which could enforce
$\mb{N}_A = \pm \mb{N}_B$ would have the effect of canceling the theta terms and leaving us with just a single $O(4)$ NLSM
without theta term. According to Ref.~\onlinecite{CenkeClass1}, an $O(4)$ NLSM with $\theta=0$ represents a trivial 
gapped phase of charged bosons in $2+1$ dimensions (i.e.,
this phase has no topological term in its electromagnetic response). 
Hence, as one consideration, we should make sure that it is not possible to get 
$\mb{N}_A = \pm \mb{N}_B$  in
our model.
To show that it is impossible to drive our system into a phase where $\mb{N}_A = \pm \mb{N}_B$, we should examine the
term $\mb{N}_A \cdot \mb{N}_B$, as any interaction which could set $\mb{N}_A = \pm \mb{N}_B$ should be a function
of $\mb{N}_A \cdot \mb{N}_B$. In terms of $U_A$ and $U_B$ we have
\beq
	\mb{N}_A \cdot \mb{N}_B= \frac{1}{2}\text{tr}[U^{\dg}_A U_B]\ .
\eeq
This term is invariant under the $U(1)_c$ symmetry, but under $U(1)_t$ we have
\beq
	\text{tr}[U^{\dg}_A U_B] \to \text{tr}[U^{\dg}_A U_B e^{-i2\xi\sigma^z} ]\ ,
\eeq	
where we used the fact that the $U(1)_t$ transformation of the bosons from Eq.~\eqref{eq:U1t}
 is equivalent to $U_A \to U_A e^{i\xi\sigma^z}$ and $U_B \to U_B e^{-i\xi\sigma^z}$. Since this term is
not invariant under $U(1)_t$, we see that translation symmetry forbids terms which could drive our BSM model
into a trivial gapped phase.

As we mentioned earlier in this subsection, there are symmetry-allowed quartic terms which
can couple the two copies of the $O(4)$ NLSM in the BSM model, for example the term $|b_{I,A}|^2|b_{J,B}|^2$. Another
possibility would be a current-current interaction of the form $\eta^{\mu\nu}J^{\mu}_{I,A}J^{\nu}_{J,B}$ where
$J^{\mu}_{I,A}=\frac{i}{g}(\pd^{\mu} b_{I,A}^{*} b_{I,A} - b_{I,A}^*\pd^{\mu} b_{I,A})$ is the conserved number 
current for boson $I$ in the ``A" NLSM,  similarly for $J^{\nu}_{J,B},$ and $\eta^{\mu\nu}= \text{diag}(1,-1,-1)$ is 
the Minkowksi metric. A precise analysis of these terms is very difficult and beyond the scope of this paper. To address them what
is really needed is the scaling dimension of the $O(4)$ field at the RG fixed point at $\theta=\pi$ discussed in 
Ref.~\onlinecite{XuLudwig}. Despite this, we
expect the BSM model to be perturbatively stable to these interactions since, at least when treated in a mean-field limit, these
terms do not cause the theta terms for the ``A" and ``B" copies of the $O(4)$ NLSM to cancel each other.

We see that translation symmetry prevents us from coupling the two NLSM copies (if they are not at the same momentum point), so it remains to discuss the local stability of each NLSM copy . Recall that in the dual description of the BSM model
we added mass terms of the form $\Phi^{(A),\dg}_2 \sigma^z \Phi^{(A)}_2 \pm \Phi^{(B),\dg}_2 \sigma^z \Phi^{(B)}_2$
to gap out the system and induce an interesting electromagnetic response. Suppose instead that we tried to add
just a single term $\Phi^{(A),\dg}_2 \sigma^z \Phi^{(A)}_2$ or $\Phi^{(B),\dg}_2 \sigma^z \Phi^{(B)}_2$ to the dual
theory in order to gap out just one of the ``A" or ``B" copies of the model. It turns out that adding one of these terms alone
is actually forbidden by the composite $\ZTI$ symmetry. Indeed, under $\ZTI$ we have 
$\Phi^{(A),\dg}_2(t,\mb{x}) \sigma^z \Phi^{(A)}_2(t,\mb{x}) \to - \Phi^{(A),\dg}_2(-t,-\mb{x}) \sigma^z \Phi^{(A)}_2(-t,-\mb{x})$, and likewise for the ``B" copy. Thus, if we require our system to obey $\ZTI$ then these terms are forbidden, and some measure of stability is provided for the BSM phase. 

While the requirement of $\ZTI$ forbids the conventional mass terms listed above, we should also consider the local stability of each $O(4)$ NLSM in the presence of symmetry-\emph{allowed} perturbations.
The discussion here closely parallels the discussion in Sec.~\ref{sec:BTI-surface} of the effects of symmetry-allowed
perturbations on the surface theory of the BTI. We start by considering interspecies tunneling terms of the
form $b^{*}_{1,A}b_{2,A} + \text{c.c.}$ for the ``A" copy of the $O(4)$ NLSM. In the canonical formalism
the operators $b_{I,A}(\mb{x})$ and their conjugate momenta $\pi_{I,A}(\mb{x})$ also obey the commutation 
relation of Eq.~\eqref{eq:CR2}. Since the composite $\ZTI$ symmetry acts on the bosons as
$(\mathcal{TI})b_{I,A}(\mb{x})(\mathcal{TI})^{-1} = b_{I,A}(-\mb{x})$, we deduce from the diagonal commutator that
$(\mathcal{TI})\pi_{I,A}(\mb{x})(\mathcal{TI})^{-1} = -\pi_{I,A}(-\mb{x})$. Now consider a state $|\Psi\ran$ which
is $\ZTI$-symmetric, i.e., $(\mathcal{TI})|\Psi\ran = |\Psi\ran$. Then in such a state we find that
\beq
	\lan \Psi | [b_{1,A}(\mb{x}),\pi_{2,A}(\mb{y}) ] |\Psi \ran = - \lan \Psi | [b_{1,A}(-\mb{x}),\pi_{2,A}(-\mb{y}) ] |\Psi \ran\ .
\eeq
If we now plug in for the commutators on both sides of this equation using Eq.~\eqref{eq:CR2}, then we find (again, after
 integration over the $\mb{y}$ coordinate) that
\beq
	\lan \Psi | b_{1,A}(\mb{x}) b^{\dg}_{2,A}(\mb{x}) | \Psi \ran = - \lan \Psi | b_{1,A}(-\mb{x}) b^{\dg}_{2,A}(-\mb{x}) | \Psi \ran \ .
\eeq
On the other hand, if the state $|\Psi\ran$ is really invariant under the action of $\ZTI$, then we should have
\beq
	\lan \Psi | b_{1,A}(\mb{x}) b^{\dg}_{2,A}(\mb{x}) | \Psi \ran =  \lan \Psi | b_{1,A}(-\mb{x}) b^{\dg}_{2,A}(-\mb{x}) | \Psi \ran \ .
\eeq
Therefore we find that $\lan \Psi | b_{1,A}(\mb{x}) b^{\dg}_{2,A}(\mb{x}) | \Psi \ran=0$ in any
state $|\Psi\ran$ which is invariant under the combined $\ZTI$ symmetry. Just as in Sec.~\ref{sec:BTI-surface}, 
we may conclude that weak interspecies tunneling terms should have a negligible effect on the BSM model 
(which has $\ZTI$ symmetry), but strong interspecies tunneling can drive the system into 
a phase which spontaneously breaks $\ZTI$ symmetry. The same conclusion holds for interspecies tunneling
terms in the ``B" copy of the $O(4)$ NLSM.

Also, in close analogy to the case in Sec.~\ref{sec:BTI-surface}, 
this result may be generalized to include insertions of any operator
$\tilde{\mathcal{O}}(\mb{x})$ which transforms nicely under the action of $\mathcal{TI}$ (recall that in 
Sec.~\ref{sec:BTI-surface} the result was generalized to include operators $\tilde{\mathcal{O}}(\mb{x})$ invariant under 
$\mathcal{T}$).  Suppose $\tilde{\mathcal{O}}(\mb{x})$ transforms under the action of $\mathcal{TI}$ as
$(\mathcal{TI})\tilde{\mathcal{O}}(\mb{x})(\mathcal{TI})^{-1} = \tilde{\mathcal{O}}(-\mb{x})$. Then we find
that 
\begin{align}
	\lan \Psi | \tilde{\mathcal{O}}(\mb{x}) [b_{1,A}(\mb{x}),\pi_{2,A}(\mb{y}) ] |\Psi \ran &= \nnb \\
- \lan \Psi | \tilde{\mathcal{O}}(-\mb{x})& [b_{1,A}(-\mb{x}),\pi_{2,A}(-\mb{y}) ] |\Psi \ran\ ,
\end{align}
and following the same steps as above gives the result that
$\lan \Psi | \tilde{\mathcal{O}}(\mb{x}) b_{1,A}(\mb{x}) b^{\dg}_{2,A}(\mb{x}) | \Psi \ran=0$ in any state 
$|\Psi\ran$ which is invariant under $\ZTI$. Note that $\tilde{\mathcal{O}}(\mb{x})$ could in principle contain operators
from both the ``A" and ``B" copies of the NLSM, as long as it transforms under $\mathcal{TI}$ as specified above.

Finally, we can again consider chemical potential terms;  the discussion of these terms is nearly identical to that
in Sec.~\ref{sec:BTI-surface} since the discussion of the terms in that section did not involve the time-reversal symmetry
at all. For the BSM model we can add chemical potential terms of the form $\mu_1 |b_{1,A}|^2 + \mu_2 |b_{2,A}|^2$
for just one copy of the $O(4)$ NLSM. As in Sec.~\ref{sec:BTI-surface} we again find that this term (combined with
suitable quartic terms) will in general cause one of $b_{1,A}$ or $b_{2,A}$ to condense and the other to be come gapped
(with the choice depending on the sign of $\mu_1 - \mu_2$). The only new feature in this context is that if a boson from
one of the $O(4)$ NLSM's were to condense, then both $U(1)_c$ \emph{and} $U(1)_t$ symmetries would be spontaneously
broken (i.e., condensing a boson from just one of the $O(4)$ NLSM's also spontaneously breaks translation symmetry).

\subsection{Topologically ordered phases accessible from the BSM theory}

In this section we briefly discuss the possibility of generating topologically ordered states from the BSM model
by condensing composite vortices. As in Sec.~\ref{sec:BTI-surface}, a basis for describing any possible topological
orders generated from the BSM model is provided by the ``$+$" vortices 
$\phi^{(A)}_{1,+},\phi^{(A)}_{2,+},\phi^{(B)}_{1,+}$, and $\phi^{(B)}_{2,+}$, since the ``$-$" vortices may be obtained
by binding a ``$+$" vortex with a trivial boson excitation. 

In exploring different composite vortices to condense, we note first that if we condense
a composite vortex of the form $\phi^{(A)}_{I,\pm}\phi^{(B)}_{J,\pm}$, then the only ``$+$" vortices which braid 
trivially with this object are $\phi^{(A)}_{I,+}$ and $\phi^{(B)}_{J,+}$, and these two vortices braid trivially with each other.
The resulting state is therefore trivial.
This means that it is impossible to generate
any topologically ordered states by condensing a product of one vortex from the ``A" NLSM and one vortex
from the ``B" NLSM. We must therefore consider composites which have at least two vortices from the same
copy of the $O(4)$ NLSM. In this section we discuss one particular phase with $\mathbb{Z}_2\times\mathbb{Z}_2$
topological order which is generated by condensing two fields which are themselves quadratic in the vortex fields from a 
single $O(4)$ NLSM. We then show that this same phase can be constructed by condensing a single field which
is quartic in the vortex fields. {We also show how to construct phases with $\mathbb{Z}_2$ topological order by 
condensing a composite vortex in one copy of the $O(4)$ NLSM and in the other copy simultaneously condensing a single vortex of one species and gapping 
out the other one. 
%In principle, more exotic topologically ordered phases should be accessible by condensing composite operators of
%four, six, or more vortices (e.g. $\phi^{(A)}_{1,+}\phi^{(A)}_{1,-}\phi^{(B)}_{1,+}\phi^{(B)}_{1,-}$), but we
%have not explored those possibilities in detail.

We now show how to construct a phase with $\mathbb{Z}_2\times\mathbb{Z}_2$ topological order
by condensing the composite vortices
$\mathcal{O}_A= \phi^{(A)}_{1,+}\phi^{(A)}_{1,-}$ and $\mathcal{O}_B= \phi^{(B)}_{1,+}\phi^{(B)}_{1,-}$
in such a way that $\lan \mathcal{O}_A\ran=\lan \mathcal{O}_B\ran \equiv \bar{\mathcal{O}}$ with $ \bar{\mathcal{O}}$
real. The vortices $\phi^{(A)}_{1,+}, \phi^{(A)}_{2,+}, \phi^{(B)}_{1,+}$, and $\phi^{(B)}_{2,+}$ all braid trivially with
$\mathcal{O}_A$  and $\mathcal{O}_B$ and so they survive as quasi-particles in the resulting topologically ordered state. 
The particular condensation shown here, with $\bar{\mathcal{O}}$ real, appears to respect all symmetries of the system
($U(1)_c$, $U(1)_t$, $\ZT$, and $\ZI$), however, we show below that this state must break either the 
time-reversal ($\ZT$) or the inversion ($\ZI$) symmetry. 

Since $\{ \phi^{(A)}_{1,+}, \phi^{(A)}_{2,+} \}$ braid trivially with $\{ \phi^{(B)}_{1,+}, \phi^{(B)}_{2,+} \}$ the 
resulting state is nearly identical to two copies of the $\mathbb{Z}_2$ topological order shown in Table~\ref{tab:anyons}.
The first factor of $\mathbb{Z}_2$ is represented exactly by Table~\ref{tab:anyons}. 
This part of the topological order is
generated by $\{ \phi^{(A)}_{1,+}, \phi^{(A)}_{2,+} \}$ and is described in the $K$-matrix formalism by
$K^{(A)}= 2\sigma^x$, $\vec{t}^{(A)}= (1,1)^T$ and $\vec{u}^{(A)}= (1,1)^T$, where $\vec{u}^{(A)}$
is a $U(1)_t$ charge vector which describes the coupling of the vortices to the external field $B_{\mu}$. 
Based on this data, the contribution of the ``A" vortices to the electromagnetic responses of this state are
\beq
	\mathcal{L}^{(A)}_{\mathcal{T}}= \frac{e^2}{4\pi}\ep^{\mu\nu\lam}A_{\mu}\pd_{\nu}A_{\lam} \ ,
\eeq
and 
\beq
	\mathcal{L}^{(A)}_{\mathcal{I}}= \frac{e}{2\pi}\ep^{\mu\nu\lam}B_{\mu}\pd_{\nu}A_{\lam} \ .
\eeq

The second factor of $\mathbb{Z}_2$ is generated by 
$\{ \phi^{(B)}_{1,+}, \phi^{(B)}_{2,+} \}$. 
For the ``B" copy, since we actually have $\theta_B=-\pi$, it seems that we should choose $K^{(B)}= -2\sigma^x$, however,
there is some ambiguity here because a statistical phase of $\pi$ is equivalent to a phase of $-\pi$. So let us
consider both possibilities $K^{(B)}= \pm 2\sigma^x$. On the other hand, there is no ambiguity in the charges of the ``B" 
vortices under the $U(1)_c$ and $U(1)_t$ symmetries: the coupling of 
$\{ \phi^{(B)}_{1,+}, \phi^{(B)}_{2,+} \}$ to $A_{\mu}$ and $B_{\mu}$ is described by the charge vectors 
$\vec{t}^{(B)}= (-1,-1)^T$ and $\vec{u}^{(B)}= (1,1)^T$, respectively.
Based on this, the contribution of the ``B" copy to the responses of this state are given by
\beq
	\mathcal{L}^{(B)}_{\mathcal{T}}= \pm\frac{e^2}{4\pi}\ep^{\mu\nu\lam}A_{\mu}\pd_{\nu}A_{\lam} \ ,
\eeq
and 
\beq
	\mathcal{L}^{(B)}_{\mathcal{I}}= \mp\frac{e}{2\pi}\ep^{\mu\nu\lam}B_{\mu}\pd_{\nu}A_{\lam} \ ,
\eeq
where the signs out front correspond to the choice of $K^{(B)}= \pm 2\sigma^x$.

We see that if we choose $K^{(B)}=  2\sigma^x$, then the entire system will break time-reversal (we get the full
2D time-reversal breaking response of the BSM), but if we choose
$K^{(B)}= - 2\sigma^x$, the entire system breaks inversion (we get the full quasi-1D inversion breaking response of the BSM). 
In particular, it seems like one cannot construct a topological order consisting of the quasi-particles 
$\phi^{(A)}_{1,+},\phi^{(A)}_{2,+},\phi^{(B)}_{1,+}$, and $\phi^{(B)}_{2,+}$, which also preserves all of the
symmetries of the BSM model.

The topologically ordered phase which we constructed above can also be accessed by condensing the single
quartic vortex field $\mathcal{O}' = \phi^{(A)}_{1,+}\phi^{(A)}_{1,-}\phi^{(B)}_{1,+}\phi^{(B)}_{1,-}$ in such a way 
that the expectation value $\lan \mathcal{O}' \ran$ is real. The field $\mathcal{O}'$ does not
carry any charge under the $U(1)_c$ or $U(1)_t$ symmetries, is invariant under inversion, and is complex conjugated
by time-reversal (so we should take $\lan \mathcal{O}' \ran$ real in an attempt to preserve time-reversal). 

In analyzing the resulting topological order, we first note that all four fundamental vortices
$\phi^{(A)}_{1,+}, \phi^{(A)}_{2,+}, \phi^{(B)}_{1,+}$, and $\phi^{(B)}_{2,+}$ braid trivially with $\mathcal{O}'$,
so they all survive as quasi-particles in the resulting topologically ordered state. The composite quasi-particles 
that can be constructed from these four fundamental vortices have the form 
$(\phi^{(A)}_{1,+})^{n_1}(\phi^{(A)}_{2,+})^{n_2}(\phi^{(B)}_{1,+})^{n_3}(\phi^{(B)}_{2,+})^{n_4}$, 
where the integers $n_j$ are either $0$ or $1$ (since the fusion of a vortex with itself is topologically trivial). A total of
16 possible quasi-particles can be constructed by letting all $n_j$ range over their values $0$ and $1$. 
To see whether the resulting topologically ordered state actually supports all of these quasi-particles as distinct excitations, we need to check
whether any quasi-particle can be obtained from another one by fusing with the condensate $\mathcal{O}'$, which is 
equivalent to the vacuum  (in the phase where $\mathcal{O}'$ is condensed). We find that each of the 16 quasiparticles is topologically distinct and that this set is sufficient to label all of the anyon sectors. %above should befusing any
%of these 16 quasi-particles with $\mathcal{O}'$ results in no new quasi-particles %the same quasi-particle, up to further fusions with
%$\mathcal{O}'$ or with topologically trivial bosons. For example 
%$\phi^{(A)}_{1,+}\mathcal{O}' \sim \phi^{(A)}_{1,-}\phi^{(B)}_{1,+}\phi^{(B)}_{1,-}$, and 
%$\phi^{(A)}_{1,-}\phi^{(B)}_{1,+}\phi^{(B)}_{1,-}$ can be fused again with $\mathcal{O'}$ to give back 
%$\phi^{(A)}_{1,+}$. Another example is 
%$\phi^{(A)}_{1,+}\phi^{(A)}_{2,+}\phi^{(B)}_{1,+}\mathcal{O}' \sim \phi^{(A)}_{2,+}\phi^{(A)}_{1,-}\phi^{(B)}_{1,-}$,
%which is again equivalent to the original composite $\phi^{(A)}_{1,+}\phi^{(A)}_{2,+}\phi^{(B)}_{1,+}$ up to trivial
%bosons since $\phi^{(A)}_{1,-} = \phi^{(A)}_{1,+}b^*_{2,A}$ and $\phi^{(B)}_{1,-} = \phi^{(B)}_{1,+}b_{2,B}$.
%Since all 16 composites are preserved under fusion with $\mathcal{O}'$, all of these
%composites will survive as quasi-particles in the topologically ordered state (we do not need 
%to quotient the set of 16 quasi-particles by identifying two or more of them), 
Hence,  the resulting
state is actually identical to the state obtained earlier from simultaneously condensing $\mathcal{O}_A$ and $\mathcal{O}_B$. 
This result could
have been anticipated since $\mathcal{O}' =\mathcal{O}_A\mathcal{O}_B$, and the vortices from the ``A" copy of the 
NLSM braid trivially with the vortices from the ``B" copy.

Another way to see that condensing $\mathcal{O}'$ leads to 
$\mathbb{Z}_2\times\mathbb{Z}_2$ topological order, and not, for example, $\mathbb{Z}_4$ topological order, is
as follows. First, note that $\phi^{(A)}_{1,\pm}$ carry charge $1$ of the dual gauge field $\al^{(A)}_{1,\mu}$ 
(whose curl is the number current of $b_{1,A}$),
while $\phi^{(B)}_{1,\pm}$ carry charge $1$ of the dual gauge field $\al^{(B)}_{1,\mu}$ (whose curl is the number
current of $b_{1,B}$). So the composite field $\mathcal{O}'$ carries charge $2$ of $\al^{(A)}_{1,\mu}$ and charge $2$ of 
$\al^{(B)}_{1,\mu}$. Therefore, condensing $\mathcal{O}'$ will break the $U(1)$ symmetries associated with
$\al^{(A)}_{1,\mu}$ and $\al^{(B)}_{1,\mu}$ down to a $\mathbb{Z}_2$ subgroup, i.e., the symmetry-breaking
associated with this condensation is $U(1)\times U(1) \to \mathbb{Z}_2\times\mathbb{Z}_2$. 
If instead it were the case that the four vortices $\phi^{(A)}_{1,\pm}$ and $\phi^{(B)}_{1,\pm}$ 
all carried charge $1$ of the \emph{same} $U(1)$ gauge field, then we would expect the condensation of 
$\mathcal{O}'$ to break that $U(1)$ symmetry down to a $\mathbb{Z}_4$ subgroup, 
leading to a $\mathbb{Z}_4$ topological order.
This does not happen in our case since the vortices $\phi^{(A)}_{1,\pm}$ and $\phi^{(B)}_{1,\pm}$ couple
to different $U(1)$ gauge fields.

% and is instead described by $K= -2\sigma^x$ and $\vec{t}_c= (-1,-1)^T$. 
%The reason for the change of sign of $K$ is that the theta angle of the ``B" copy of the $O(4)$ NLSM is $\theta=-\pi$,
%so strictly speaking, the mutual statistical angle of  $\phi^{(B)}_{1,+}$ and  $\phi^{(B)}_{2,+}$ is actually $-\pi$. The
%charge vector $\vec{t}_c$ has the opposite sign for the second copy since the vortices $\phi^{(B)}_{1,+}$ and  
%$\phi^{(B)}_{2,+}$ carry charge $-\frac{1}{2}$ of $U(1)_c$. 
%
%The full topologically ordered state is then described by $\mathcal{K}= 2\sigma^x\oplus (-2\sigma^x)$ and
%$\vec{t}_c= (1,1,-1,-1)^T$. This state has zero Hall conductance, and 
%the exchange and mutual statistical phases of the quasi-particles are all real, which means that this state
%preserves time-reversal symmetry. Since
%time-reversal and inversion symmetry map vortices in the ``A" copy of the $O(4)$ NLSM to vortices in the ``B" copy,
%these symmetries simply act to exchange the two factors of $\mathbb{Z}_2$ that generate
%the topological order. 

As we mentioned above, 
it is also possible to generate a phase with $\mathbb{Z}_2$ topological order by condensing
composite vortices in one copy of the $O(4)$ NLSM, and in the other copy simply gapping out one vortex species and condensing the
other. We show that such a phase will break either the time-reversal or the inversion symmetry of the BSM model. As an
example, consider condensing the composite vortex $\mathcal{O}_A$ in the ``A" copy of the NLSM, while in the 
``B" copy  condensing the single vortex $\phi^{(B)}_{2,+},$ and gapping out the vortex $\phi^{(B)}_{2,-}$. The 
resulting phase has a $\mathbb{Z}_2$ topological order generated by $\phi^{(A)}_{1,+}$ and  $\phi^{(A)}_{2,+}$. Note that 
the ``B" copy does not contribute to the topological order since $\phi^{(B)}_{2,+}$ has been condensed 
(i.e., it is now topologically equivalent to the vacuum quasi-particle) and $\phi^{(B)}_{1,+}$ is confined (it has non-trivial
braiding with $\phi^{(B)}_{2,+}$, which is condensed). The electromagnetic response of this phase can be easily calculated
using the results contained in this section, and we find that this phase has no time-reversal breaking response, but
it does possess the full inversion breaking response of the BSM,
as shown in Eq.~\eqref{eq:1D-BSM-response}. If for the ``B" copy we instead chose to condense $\phi^{(B)}_{2,-}$
and gap out $\phi^{(B)}_{2,+}$ (while still condensing $\mathcal{O}_A$ for the ``A" copy), we would get a phase with
$\mathbb{Z}_2$ topological order which has no inversion breaking response, but the full time-reversal breaking response of
the BSM, as in Eq.~\eqref{eq:2D-BSM-response}.

\section{Quantization of polarization in gapped 2D phases and a criterion for semi-metal behavior}
\label{sec:pol}

In this section we give a general discussion of the quantization of the charge polarization in gapped phases of 2D 
quantum many-body systems with translation, inversion, and $U(1)_c$ charge conservation symmetries,
with the goal of establishing a criterion for detecting whether a given system is in a semi-metallic phase by
measuring its polarization response. The systems in question can be either bosonic or fermionic, and we assume they are made 
of up some fundamental particles of charge $e$. For simplicity we focus on systems on a square lattice with lattice spacing 
$a_0$, but the result can be easily extended to any Bravais lattice. We consider three broad classes (to be described below) of 
gapped phases of 2D systems in which one can define a charge polarization, and we show that in these three classes the 
polarization in (say) the $x$-direction is quantized in units of
\beq
	P^{(min)}_x= r\frac{e}{2 a_0}\ ,
\eeq
where $r \in \mathbb{Q}$ is a rational number. This result then implies
that \emph{if} a 2D quantum many-body system is found to have a continuously tunable polarization of 
the form $\al\frac{e}{2 a_0}$ for a generic real number $\al$, \emph{then} this system cannot be in one of the three classes
of gapped phases mentioned above. If these three classes of gapped phases exhaust all possible gapped phases with
translation symmetry which can support a polarization response, and from their definitions below it is clear that they do, 
then this implies that a polarization of the form $\al\frac{e}{2 a_0}$ for generic $\al \in\mathbb{R}$ is indicative of
a gapless semi-metal phase. Therefore our argument in this section provides
a direct relation between the gaplessness of a semi-metal and the tunability of its polarization response. As we mentioned in 
Sec.~\ref{sec:overview}, since the polarization response is expected to be reasonably robust, this provides additional 
evidence for the stability of the semi-metal phase to perturbations which do not destroy its polarization response.

As we show below, the rational number $r$ mentioned above can be related to specific measurable properties of the three 
types of gapped phases that we consider, so we do not need to worry about the difficulty of ``measuring an irrational 
number", as the number $r$ can be readily obtained for these gapped phases in other ways. Thus, the charge 
polarization response of a 2D system can be used as a criterion for detecting a gapless semi-metal phase. The reason for
focusing on gapped phases \emph{with} translation symmetry is that we know that a semi-metal requires translation 
symmetry for its stability. Since we are looking for a way to distinguish a semi-metal phase from other phases with a polarization 
response, we need to compare to other systems with translation symmetry as we know that without translation symmetry the 
semi-metal phase is not even a possibility. We now give the details of our argument.

The three classes of gapped systems
which we consider are (i) systems with a unique ground state and translation symmetry by one site, (ii) systems with a 
ground state which spontaneously breaks translation symmetry by one site down to translation symmetry
by $q$ sites (so $q$ is a positive integer), and (iii) systems with intrinsic topological order as well as translation symmetry 
by one site.  To calculate the charge polarization of these systems we use a many-body formula for the polarization introduced
by Resta in Ref.~\onlinecite{resta1998}, which we now review. We focus on an analysis of the polarization in the 
$x$-direction, and so we assume periodic boundary conditions in that direction. This assumption of periodic boundary
conditions in at least one direction will also allow us to invoke certain theorems~\cite{oshikawa2000,hastings2005} which
will be crucial for our results in this section.

%For the systems in class (iii) we make one additional assumption, which is that the braiding statistics for the
%Abelian anyons in the topologically ordered system can be described by Abelian Chern-Simons theory (the ``K-matrix" 
%formalism). The reason for making this extra assumption is that in this case the mutual statistical angle 
%$\theta_{a,b}$ for a process in which an anyon $a$ makes a complete circuit around an anyon $b$ can be shown to have
%the form $2\pi\times(\text{rational number})$, which will be important for our calculation of the polarization in gapped phases
%in class (iii).

\subsection{Polarization in 2D, ambiguity with translation invariance, and quantization with inversion symmetry}

%\textbf{MFL: Does this expression in 2D look correct to you both? It certainly has the right units and the correct ambiguity. 
%Do we need to take the limit $L_y \to \infty$ as well if we are only calculating the polarization in the $x$ direction?}

Consider a quantum many-body system defined on a square lattice with lattice spacing $a_0$ and lengths $L_x$ and 
$L_y$ in the $x$ and $y$ directions. Let $N_s$ be the number of sites so that $N_s a_0^2 = L_x L_y$. We label
sites on the square lattice by the vector of integers $\mb{j}=(j_x,j_y)$, $j_x,j_y\in\mathbb{Z}$. Finally, let
$|\Psi_0\ran$ be the ground state of the system. We assume $|\Psi_0\ran$ is an eigenstate of the number operator
with eigenvalue $N_p$ so that the filling factor in the ground state is $\nu=\frac{N_p}{N_s}$. The total number operator
can be expressed as $\hat{N}= \sum_{\mb{j}} \hat{n}_{\mb{j}}$ where $\hat{n}_{\mb{j}}$ is the number operator for
site $\mb{j}$. Then, assuming that
$|\Psi_0\ran$ is the ground state of a \emph{gapped} system, Resta's formula tell us that the polarization in the 
$x$-direction is given by
\beq
	P_x= \lim_{L_x\to\infty} \frac{e}{2\pi L_y} \text{Im}[ \ln \lan \Psi_0 | e^{i\frac{2\pi}{L_x}\hat{X}}|\Psi_0\ran]\ ,
\eeq
where the position operator $\hat{X}$ is given by
\beq
	\hat{X}= \sum_{\mb{j}} (j_x a_0) \hat{n}_{\mb{j}}\ .  
\eeq
The polarization in the $y$-direction has a similar definition. 

Let us suppose that the state $|\Psi_0\ran$ has translation invariance by one site in the $x$-direction, i.e., 
$|\Psi_0\ran$ is an eigenstate of the translation operator $\hat{T}_x$ with some eigenvalue $e^{ik^{(0)}_x}$ (the
precise value of $k^{(0)}_x$ will not be important in what follows). Concretely, $\hat{T}_x$ acts as 
$\hat{T}_x^{\dg}\hat{\mathcal{O}}_{\mb{j}}\hat{T}_x = \hat{\mathcal{O}}_{\mb{j}+(1,0)}$ on operators 
$\hat{\mathcal{O}}_{\mb{j}}$ carrying a position index. If this is the case then one can show that $P_x$ is only
well-defined modulo $\frac{e\nu}{a_0}$. To see it we compute the polarization $P'_x$ of 
$|\Psi'_0\ran \equiv \hat{T}_x|\Psi_0\ran$ in two ways. On one hand we can just write 
$|\Psi'_0\ran \equiv e^{ik^{(0)}_x}|\Psi_0\ran$ to find that $P'_x= P_x$. However, we can also use
\beqa
	\lan\Psi_0|\hat{T}^{\dg}_x e^{i\frac{2\pi}{L_x}\hat{X}} \hat{T}_x|\Psi_0\ran &=&\lan\Psi_0| e^{i\frac{2\pi}{L_x}\hat{X}} e^{-i\frac{2\pi a_0}{L_x}\sum_{\mb{j}} \hat{n}_{\mb{j}}}|\Psi_0\ran \nnb \\
	&=&  \lan\Psi_0|e^{i\frac{2\pi}{L_x}\hat{X}}|\Psi_0\ran e^{-i\frac{2\pi a_0 N_p}{L_x}}\ ,
\eeqa
to show that
\beq
	P'_x = P_x - \frac{e\nu}{a_0}\ .
\eeq
So we conclude that $P_x$ is defined only modulo $\frac{e\nu}{a_0}$ in the presence of translation symmetry by one
site. 

The last ingredient in the polarization calculation is to enforce inversion symmetry in the system. We consider inversion
which acts simply as $\mb{j} \to -\mb{j}$ for the coordinates on the square lattice. It is clear that under inversion we have
$P_x \to - P_x$ and similarly for the polarization in the $y$-direction. So the polarization in the inversion symmetric system
must obey the relation 
\beq
	P_x \equiv -P_x\ \text{mod}\ \frac{e\nu}{a_0}\ .
\eeq
The solutions to this relation are
\beq
	P_x \equiv 0\ \text{or}\ \frac{e\nu}{2a_0}\ \text{mod}\ \frac{e\nu}{a_0}\ ,
\eeq
with a similar result for $P_y$. So in a gapped 2D system with translation and inversion symmetry and filling factor $\nu$, the 
polarization is quantized in units of 
\beq
	P^{(min)}_x = \frac{e\nu}{2a_0}\ . \label{eq:polarization-one-species}
\eeq

Before moving on let us make a few general comments about this formula for the polarization. First,
in a band insulator made out of free fermions the filling $\nu$ must be an integer in order for the system to be 
gapped (i.e., in order to have a completely filled band). This is why the filling $\nu$ usually does not appear explicitly in 
discussions of the polarization in band insulators. Also, in the discussion above we have assumed that there is only
one type of particle. More generally, our system could have several different species of particles, for example spin up and
spin down electrons, and in this case one can separately consider the polarization for each species. 
If we label different particles species by $\sigma$ then we can compute $P_{x,\sigma}$, the polarization
from particles of species $\sigma$, by modifying the
position operator $\hat{X}$ to
\beq
	\hat{X}_{\sigma}= \sum_{\mb{j}} (j_x a_0) \hat{n}_{\mb{j},\sigma}\ , 
\eeq
where $ \hat{n}_{\mb{j},\sigma}$ is the number of particles of species $\sigma$ on site $\mb{j}$. The
total polarization is then given by $P_x= \sum_{\sigma} P_{x,\sigma}$. The importance of computing the polarization in this
way is demonstrated by the following example. Suppose we have a band insulator of spinful electrons (so $\sigma=\up,\down$)
and we have a completely filled band of up and down electrons. Then we have $\nu_{\up}=1$ and $\nu_{\down}=1$
and so the total filling is $\nu=2$. However, in the absence of time-reversal symmetry both bands do not have to have
the same polarization. Since each individual band is at filling $\nu_{\sigma}=1$ we could have $P_{x,\up}= \frac{e}{2 a_0}$
but $P_{x,\down}=0$, and so $P_{x}= \frac{e}{2a_0}$. This result could not have been predicted from
Eq.~\eqref{eq:polarization-one-species}, since that formula does not distinguish between different particle species.

We now discuss the specific values that $\nu$ can take in the three classes of gapped systems discussed above, 
and in this way constrain the possible values of $P^{(min)}_x$ in such phases.

\subsection{The filling factor $\nu$ in the three classes of gapped phases}

%\textbf{MFL: We probably need a few general refs on SET phases.}
%
%\textbf{MFL: OK I am not confused anymore on Oshikawa vs. Hastings. The paper ``Topological Enrichment of Luttinger's 
%Theorem"  just cited the wrong Hastings paper. He proved Oshikawa's theorem in a 2005 paper in Europhysics
%letters (cited below), not in his famous 2004 PRB on the higher-dimensional generalization of Lieb-Schultz-Mattis. I have
%tried to write this section in way which correctly states what is known on this topic.}

Now we discuss the possible values of the filling factor $\nu$ in the three classes of gapped phases,  which will in turn give us 
the minimum value $\frac{e\nu}{2a_0}$ of the polarization in these systems. To start we go back to a theorem of
Oshikawa~\cite{oshikawa2000} which was later proven rigorously (under slightly more restrictive assumptions) by 
Hastings~\cite{hastings2005}. What Oshikawa/Hastings showed is that if the
filling $\nu$ of a \emph{gapped} system is a rational number, say $\nu=\frac{p}{q}$ with $p$ and $q$ coprime, then the 
system will in general have $q$ degenerate ground states (in the thermodynamic limit), each with a different momentum in the 
(for example) $x$-direction. For integer $\nu$ the ground state is unique. 
On the other hand, irrational values of $\nu$ in the ground state generally imply a gapless system. 
In Hastings' rigorous proof the condition is actually that 
$\nu\left(\frac{N_s a_0}{L_x}\right)= \frac{N_p a_0}{L_x}= \frac{p}{q}$, where $L_x$ is the length of the system in the 
$x$-direction~\cite{hastings2005}. In what follows we assume that this result holds for the condition 
$\nu=\frac{p}{q}$, as is expected on general physical grounds, 
although the reader should be aware that there is no rigorous proof available in this case 
(and there are even counterexamples in 2D systems which are long in one direction but short in the other, see e.g., 
Ref.~\onlinecite{hastings2010locality}). 

Using this theorem we can immediately conclude that in the case of integer filling the minimum value of the 
polarization in the ground state of a gapped, translation-invariant 2D system with a unique ground state is
\beq
	P^{(\text{min})}_{x} = \frac{e}{2a_0}\ ,
\eeq
which corresponds to the filling $\nu=1$. This gives the answer for the minimum value of the polarization in a gapped 
system in class (i) discussed above. 

Next we discuss the case of rational filling factor $\nu=\frac{p}{q}$, which will turn out to include gapped systems in 
classes (ii) and (iii). For rational filling factor $\nu= \frac{p}{q}$ there are two possible physical explanations
for the $q$ degenerate ground states~\cite{oshikawa2000,hastings2005}. The first possibility is that
the $q$ degenerate states correspond to a spontaneous breaking of the translation symmetry by one lattice site
down to translation symmetry by $q$ lattice sites. In this case the actual ground state in the thermodynamic limit is 
expected to be a particular linear combination of the $q$ ground states (which each have different momenta
in the $x$-direction) which is an eigenstate of $(\hat{T}_x)^q$ but not of $\hat{T}_x$, thus breaking the
symmetry of translation by one site. This corresponds to our class (ii) of gapped phases. If we repeat the analysis from above 
of the ambiguity of the polarization in the presence of translation symmetry, but replace $\hat{T}_x$ with 
$(\hat{T}_x)^q$, then we find that the polarization is only well-defined modulo $\frac{q e \nu}{a_0}$. Then in 
the presence of inversion symmetry the minimum value of the polarization in this case is also
\beq
	P^{(\text{min})}_{x} = \frac{e}{2a_0}\ ,
\eeq
corresponding to the choice $\nu=\frac{1}{q}$. 

The final possibility is that the system at filling factor $\nu=\frac{p}{q}$ does not break translation symmetry but instead
has intrinsic topological order, which can also explain the $q$-fold ground state degeneracy in the thermodynamic
limit. This corresponds to our class (iii) of gapped systems. In this case the filling factor $\nu$ can be related to the data
describing a 2D symmetry-enriched topological (SET) phase with $U(1)_c$ and translation 
symmetry~\cite{cheng2015translational,bonderson2016topological} and so we now give a brief overview of the physical
properties of 2D SET phases with $U(1)_c$ and translation symmetry. For more details see 
Ref.~\onlinecite{cheng2015translational}.

An SET phase in 2D is a gapped phase possessing intrinsic topological order, but which also has global
symmetry of a group $G$ (see Ref.~\onlinecite{barkeshli2014symmetry} for an in-depth discussion of these phases). 
The group $G$ can act in various non-trivial ways on the anyons which are present in the
topologically ordered system. For example if $G=U(1)_c$ then an anyon can carry a fractional charge under 
$G$ (i.e., the anyon transforms in a projective representation of $G$). A more exotic possibility is that the
action of $G$ can \emph{exchange}, or permute, two different kinds of anyons. In the case where the symmetry does 
not permute the anyons it is known that 2D SET phases with symmetry group $G$ are classified by the cohomology group
$H^2(G,\mathcal{A})$, where $\mathcal{A}$ is the group of Abelian anyons in the topologically ordered system.

In a 2D SET phase with $U(1)_c$ symmetry, each anyon $a$ can carry a particular fractional charge $e_a= Q_a e$ 
under the $U(1)$ symmetry, where $Q_a$ is a dimensionless number. The number $Q_a$ can also be expressed in terms of
the mutual braiding statistics $M_{a,v}$ of $a$ with the anyon $v$, which is the excitation created in the 
system by threading $2\pi$ delta function flux of the $U(1)_c$ gauge field at a point in the system (this excitation was
referred to as a \emph{vison} in Ref.~\onlinecite{bonderson2016topological}). Here 
$M_{a,a'}= e^{i\theta_{a,a'}}$ is the $U(1)$ phase accumulated during a process in which the anyon $a$ makes
a complete circuit around the anyon $a'$. This essentially calculates the Aharonov-Bohm phase of the $U(1)_c$ charge carried by $a$ when dragged around the fundamental flux of $U(1)_c$ carried by $v.$ Hence, we have the relation
\beq
	e^{i2\pi Q_a}= M_{a,v}\ , 
\eeq
or
\beq
	Q_a = \frac{\theta_{a,v}}{2\pi}\ . \label{eq:anyon-charge}
\eeq
A 2D SET phase with translation symmetry is characterized by one additional property. This is the
\emph{anyonic flux} $b$ per unit cell, where $b$ is an Abelian anyon in the topologically ordered system under consideration. 
The physical meaning of the anyonic flux $b$ is that if an anyon $a$ is translated around a unit cell, then the state of the 
system picks up the phase $M_{a,b}$. 

We see that we can characterize a 2D SET phase with $U(1)_c$ and translation symmetry by the data $(\{e_a\},b)$, which 
includes the set of charges $\{e_a\}$ of the anyons under the $U(1)_c$ symmetry, and the particular anyon $b$ which 
provides the anyonic flux per unit cell in the system.

The authors of Refs.~\onlinecite{cheng2015translational,bonderson2016topological} showed that the filling factor $\nu$ in a 
2D SET phase with translation and $U(1)_c$ symmetry can be expressed in terms of the data of the SET phase as
\beq
	\nu \equiv Q_b\ \text{mod}\ 1\ ,
\eeq
or, using Eq.~\eqref{eq:anyon-charge}, 
\beq
	\nu \equiv \frac{\theta_{b,v}}{2\pi}\ \text{mod}\ 1\ .
\eeq
So the filling factor of the 2D SET phase is equal to the $U(1)_c$ charge of the anyon $b$ characterizing the anyonic flux
per unit cell in the system, and this is in turn related to the mutual statistical angle $\theta_{b,v}$ 
between $b$ and the excitation $v$. The derivation of this equation essentially uses Oshikawa's original flux threading 
argument and the fact that threading a flux through the hole of the torus is equivalent to wrapping a string operator for 
$v$ around the cycle of the torus which does not enclose the hole~\cite{bonderson2016topological}. Note that
$Q_b$ must be a rational number since if it were not then the relation between $Q_b$ and $\nu$, combined with the
Oshikawa/Hastings argument, would imply that the phase was gapless and not a gapped SET phase. 
From this relation between $Q_b$ and $\nu$ we find that the minimum value of the polarization for systems in our class (iii) is 
\beq
	P^{(\text{min})}_{x} = \frac{e_b}{2a_0}= \frac{e Q_b}{2 a_0}\ .
\eeq

We have succeeding in showing that for all three classes of gapped phases considered in this section, 
the polarization is quantized to some rational multiple of $\frac{e}{2a_0}$ and,
in particular, is not continuously variable since tuning $\nu$ away from a rational value leads to gapless phase
according to the Oshikawa/Hastings argument. Thus, we see that generic non-rational values of the polarization are 
indicative of a gapless semi-metal phase. Furthermore, we have shown how the polarization in these gapped phases
can be simply related to various physical data describing those phases, which means that it should be simple to diagnose
whether a given value of the polarization implies a gapped or gapless phase.

\section{Coupled Wires Construction of the Bosonic Semi-metal}
\label{sec:1D-bosonic-wire}
%So far we have provided an effective theory for a 2+1-d bosonic semi-metal, and discussed its electromagnetic response properties and stability criteria. In this Section we provide a foundation for a more explicit construction of this phase using a coupled wires approach. We are able to find a suitable wire building-block, and to determine the appropriate interactions to generate the gapless BSM phase, as well as the neighboring insulator phases with non-trivial electromagnetic response. However, although we are able to solve the model in the two gapped phases, it remains open how to deal with the interactions we propose to generate the gapless phase. This is an important open question that we carefully discuss below.

So far we have provided an effective theory for a 2+1-d bosonic semi-metal, and discussed its electromagnetic response 
properties and stability criteria. In this Section we provide an explicit construction of this phase using a coupled wires approach
which is modeled after the coupled wires construction of a single $O(4)$ NLSM with $\theta=\pi$ derived in 
Ref.~\onlinecite{SenthilFisher} (see also Refs.~\onlinecite{TanakaHu,VS2013}).
We are able to find a suitable wire building-block, as well as suitable inter-wire tunneling terms, which together generate our 2D 
BSM model after taking the continuum limit in the wire stacking direction. 

The rationale for a coupled wires construction of the BSM model is provided
by the general demonstration in Ref.~\onlinecite{Ramamurthy2014} that free fermion DSMs admit a coupled wires 
construction in terms of 1+1-d topological insulator wires, each with a charge $\frac{e}{2}$ polarization response.
Indeed, one of the most important aspects of the coupled wires construction of the free fermion DSM is the intuitive explanation 
it provides for the quasi-1D inversion-breaking electromagnetic response of the DSM model, shown in 
Eq.~\eqref{eq:DSM-responses}.

This Section is organized as follows. We begin by reviewing the coupled wires construction of the free fermion DSM. 
We then construct a wire building block for the 2D BSM phase using two copies of the Bosonic Integer Quantum Hall (BIQH) 
edge theory. For our purposes, we require the description of the BIQH edge in terms of an $SU(2)_1$ Wess-Zumino-Witten 
(WZW) theory, as discussed in Refs.~\onlinecite{SenthilLevin,VS2013}. For completeness, we briefly review this description
of the BIQH edge, and carefully discuss how this edge theory couples to an external electromagnetic field. We then review
the derivation of Ref.~\onlinecite{SenthilFisher} of a single $O(4)$ NLSM with $\theta=\pi$ from coupled wires consisting
of a single copy of an $SU(2)_1$ WZW theory. 

With all of this information in hand, we go on to present a coupled wires
construction of the 2D BSM model, and we include a careful discussion of how to define the action of time-reversal and inversion 
symmetries in the coupled wires model so that the correct action of these symmetries on the continuum fields is recovered
in the continuum limit. Finally, we conclude this section by contrasting the coupled wires constructions of the DSM
and BSM phases, and we briefly comment on how the symmetry breaking phases of the BSM model can be accessed within
the coupled wires description.

\subsection{Coupled Wire Construction of a Fermionic Dirac Semi-metal}

We begin by reviewing the construction of the free-fermion DSM via a stacking of 1D gapped topological
free fermion wires, each with charge $\frac{e}{2}$ polarization. This construction was introduced in Ref.~\onlinecite{Ramamurthy2014}, and 
it provides a clear physical interpretation of the quasi-1D inversion-breaking response of the 
DSM in terms of the polarization response of the individual wires in the stacking
construction. Just as in Sec.~\ref{sec:DSM-review}, the degrees of freedom are two-component spinless fermions $\vec{c}_{n}$ living on a 1D lattice with site index $n.$ The Bloch Hamiltonian for the 1+1-d free fermion topological wire model has the form 
\beq
	\mathcal{H}_{1D}(k_x)= \sin(k_x)\sigma^x + (1-m-\cos(k_x))\sigma^z\ . \label{eq:fermion-wire}
\eeq
This model is in a topological phase for $0< m < 2$, and one can show that charge $\pm \frac{e}{2}$ is
trapped at a domain wall between a state with $m\lesssim 0$ and $m \gtrsim 0$\cite{jackiw1976,ssh1979,QHZ2008,Mulligan}. For our interest, we consider the topological
phase of this model to be protected by inversion symmetry\cite{turner2012,hughes2011inv}, where the inversion operator $\mathcal{I}$ acts on the
lattice fermions as
\beq
	\mathcal{I}\vec{c}_{m}\mathcal{I}^{-1}= \sigma^z\vec{c}_{-m}\ .
\eeq 

To obtain the DSM model we now stack these 1+1-d fermion wires into two dimensions and introduce a hopping term:
$t_y ( \vec{c}^{\ \dg}_{\mb{n}+\mb{\hat{y}}}\sigma^z \vec{c}_{\mb{n}} + \text{h.c.} )$ between 
fermions on adjacent wires. The Bloch Hamiltonian for the resulting 2D system has exactly the form of 
Eq.~\eqref{eq:DSM-lattice}.
Now we note that the 2+1-d model Eq.~\eqref{eq:DSM-lattice} 
looks like many copies of the 1+1-d model in Eq.~\eqref{eq:fermion-wire}
where the different copies of the 1+1-d wire are labeled by $k_y$, and with each having a $k_y$-dependent mass
\beq
	m_{k_y}= m + t_y\cos(k_y)\ .
\eeq Essentially, the Bloch Hamiltonian for each value of $k_y$ represents a 1+1-d insulator of the type Eq.~\ref{eq:fermion-wire}, but with a $k_y$-dependent mass parameter.

Consider the parameter range $m, t_y >0$, and recall the definition of $B_y$ from Sec.~\ref{sec:DSM-review} 
(it is the positive
solution to $m+t_y\cos(B_y)=0$ with $B_y \in [0,\pi)$). We see that the 1+1-d systems labeled by $k_y$ have
$m_{k_y} > 0$ for $k_y \in (-B_y, B_y)$, but $m_{k_y} < 0$ for  $k_y \in  (B_y,\pi)$ or $k_y \in  (-\pi,-B_y)$.
So the 1+1-d systems in the range $-B_y < k_y < B_y$ are in the topological phase, while the rest  are
in the trivial phase. 

As we now review, this observation immediately leads to a microscopic description of the quasi-1D response of the DSM.
First, note that each \emph{topological} wire contributes a factor $\frac{e}{2}\int dx dt\ F_{tx}$ to the electromagnetic response 
of the system. Here $F_{tx}= \pd_t A_x - \pd_x A_t = -E_x$ (the electric field in the $x$-direction), so this 
response represents a charge polarization of magnitude $\frac{e}{2}$ in the $x$-direction. 
The total number of 1+1-d systems in the range $-B_y < k_y < B_y$ is $\frac{2 B_y}{\left(\frac{2\pi}{N_y a_0}\right)},$
where $N_y$ is the number of wires that we stack to construct the 2+1-d system, and $a_0$ is the lattice 
spacing in the $y$-direction. So the total electromagnetic response
from all of the topological wires in the range $-B_y < k_y < B_y$ is
\beq
	S_{eff,1D}= \frac{2 B_y}{\left(\frac{2\pi}{N_y a_0}\right)} \frac{e}{2}\int dx dt\ F_{tx}\ .
\eeq
Using $N_y a_0 = \int dy$ and the fact that $B_y$ is uniform in this case, we get
\beq
	S_{eff,1D}= \frac{e}{2\pi}\int d^3 x\  B_y F_{tx}\ ,
\eeq
which is exactly the response from Eq.~\eqref{eq:DSM-responses} for the case where only $B_y \neq 0$.

We would like to make one more comment about the free fermion topological wire model of Eq.~\eqref{eq:fermion-wire}.
If one linearizes this model near the $m=0$ critical point, and then takes a continuum limit, the resulting model
is a $1+1$-d Dirac fermion with a Dirac mass (back-scattering) term acting between the left and right-moving
fermions that make up the Dirac fermion. For the BSM construction it will be useful to make the following analogy.  We note that 
the edge theory of the $\nu=1$
Integer Quantum Hall Effect (IQHE) (for fermions) is a single right-moving fermion. Hence, the fermion topological wire
model used to construct the DSM can then be interpreted as being built from the edge theory of a $\nu=1$ IQH state and a 
$\nu=-1$ IQH state,
with an additional back-scattering mass term introduced to gap out the entire system. Alternatively, we could think of the wire 
as just a thin strip of $\nu=1$ IQHE where the opposing edges are close enough to interact with each other. Similarly, in our coupled 
wires construction of the BSM model, each individual wire will contain the two counter-propagating edge modes of a thin strip of the BIQH system.

%This observation will
%motivate our construction of a bosonic wire model using a thin strip of the BIQH system, with
%suitable back-scattering terms between the two edge theories.

\subsection{Edge Theory of the Bosonic Integer Quantum Hall System}

In this section we briefly discuss the edge theory of the BIQH system, paying close attention to how the edge theory 
couples to an external electromagnetic field. 
This edge theory will help form the basic building block for the 1D bosonic wires we will use to construct our 2D BSM model, 
just as the edge theory for the fermion IQH system forms
the basic building block for the 1+1-d fermionic topological wire considered in the previous subsection. 
We expect the edge theory for the BIQH system to satisfy 
(at least) two requirements: (i) the basic fields in the model are bosonic, and (ii) the $U(1)_c$ charge conservation symmetry
is realized in an anomalous way so that the variation of the boundary action under a gauge transformation cancels the 
contribution from the bulk Chern-Simons action for the BIQH system.

The edge theory for the BIQH state can be described using the $K$-matrix formalism familiar
from Abelian quantum Hall systems (in which case it is described
by $K=\sigma^x$, c.f. Ref~\onlinecite{VL2012}), however, we will use the description of the
edge theory in terms of an $SU(2)_1$ Wess-Zumino-Witten (WZW) theory, which was proposed in 
Refs.~\onlinecite{SenthilLevin,VS2013}.
Here we review some details of this theory and explicitly show that the anomaly of the edge theory with the correct charge 
assignment exactly cancels the boundary term we obtain when we perform a gauge transformation on the bulk Chern-Simons 
effective action for the BIQH system. Indeed, this clearly shows that the BIQH state can be terminated with an $SU(2)_1$ 
WZW edge theory.

The bulk Chern-Simons effective action for the BIQH system can be written in differential form notation as
\beq
	S_{BIQH}= \frac{1}{2\pi}\int_M A\wedge dA\ ,
\eeq 
where $M$ is the space-time manifold and $A= A_{\mu}dx^{\mu}$.
Under a gauge transformation $A\to A+ d\chi$ we have $S_{BIQH} \to S_{BIQH} + \delta S_{BIQH}$ with
\beq
	\delta S_{BIQH}= \frac{1}{2\pi}\int_{\pd M} A\wedge d\chi\ .
\eeq
Therefore, in order for the system as a whole to be gauge invariant, we should expect that the edge theory has an anomaly when we couple to an electromagnetic field, in order to
cancel this term coming from the gauge-variation of the bulk action.

The $SU(2)_1$ WZW theory takes the form (see Ref.~\onlinecite{CFTbook} for an introduction)
\beq
	S= \frac{1}{8\pi}\int d^2x\ \text{tr}[(\pd^{\mu}U^{\dg})(\pd_{\mu}U)] - S_{WZ}[U]\ ,
\eeq
where $U$ is an $SU(2)$ matrix field, and the Wess-Zumino (WZ) term is
\beqa
	&&S_{WZ}[U]=  \\
	&&\frac{1}{12\pi}\int_0^1 ds\int d^2 x\  \ep^{\mu\nu\lambda} \text{tr}[(\tilde{U}^{\dg}\pd_{\mu} \tilde{U})( \tilde{U}^{\dg}\pd_{\nu} \tilde{U})( \tilde{U}^{\dg}\pd_{\lambda} \tilde{U})].\nonumber
\eeqa
As usual, the WZ term involves integration over an auxiliary direction of spacetime. In this expression 
$\tilde{U}(s,t,x)$ denotes an extension of $U(t,x)$ into the $s$-direction and $\mu,\nu,\lam = s,t,x$ in
the sum  (we take $\ep^{stx}=1$) . 
By convention, one typically chooses boundary conditions $\tilde{U}(0,t,x) = \mathbb{I}$ (i.e., a trivial configuration) and 
$\tilde{U}(1,t,x) = U(t,x)$, so that the physical spacetime is located at $s=1$.

The $SU(2)_1$ WZW theory has an $SU(2)\times SU(2)$ symmetry: the action is invariant under the replacement
$U\to g^{\dg} U h$, for $g,h \in SU(2)$. The transformation with $g= \mathbb{I}$ is referred to as the \emph{right}
$SU(2)$ symmetry, while the transformation with $h=\mathbb{I}$ is referred to as the \emph{left} $SU(2)$ symmetry.
The case $g=h$ is called the \emph{diagonal} $SU(2)$ symmetry. Just as in Sec.~\ref{sec:BTI-surface},  the matrix $U$ can 
be written in terms of bosonic fields $b_I$, $I=1,2$, and the physical $U(1)_c$ symmetry $b_I \to e^{i\chi}b_I$ is
realized on $U$ as $U \to U e^{i\chi\sigma^z}$. Hence, the $U(1)_c$ symmetry is a $U(1)$ subgroup of the right $SU(2)$
symmetry of the $SU(2)_1$ WZW theory.

It is known that one cannot obtain a gauge invariant action by only gauging
the right or left $SU(2)$ symmetry of the WZW theory (or a subgroup of one of these symmetry groups) \cite{HullSpence}.
However, in the case where one chooses to gauge a left or right symmetry of the theory, there is a ``best possible" action that 
one can obtain, in which the gauge transformation produces a term that only depends on the gauge field itself and the 
element of the Lie algebra involved in the gauge transformation (instead of a more complicated expression involving the actual
field $U$) \cite{WittenHolo}. In our case, this ``best possible"
action takes the form
\beqa
	S_{gauged} &=& \frac{1}{4\pi}\int d^2x\ \sum_I (D^{\mu}b_I^{*})(D_{\mu}b_I) - S_{WZ}[U] \nnb \\
&+& \frac{1}{4\pi}\int d^2 x\ \ep^{\mu\nu}\text{tr}[iA_{\mu}\sigma^z U^{\dg}\pd_{\nu}U]\ , \label{eq:WZWgauged}
\eeqa
where $D_{\mu}= \pd_{\mu} - iA_{\mu}$. In the kinetic term we applied the usual minimal coupling procedure
$\pd_{\mu} \to D_{\mu}$. The last term, however, is more mysterious. Its purpose is to make the gauge-variation
of this action as nice as possible (the WZ term is not gauge invariant). Indeed, under a gauge transformation
$U \to U e^{i\chi(x)\sigma^z}$, $A_{\mu} \to A_{\mu} + \pd_{\mu}\chi(x)$, we have
\beq
	\delta S_{gauged} = -\frac{1}{2\pi}\int d^2 x\ \ep^{\mu\nu}A_{\mu}\pd_{\nu}\chi = -\frac{1}{2\pi}\int_{\pd M} A\wedge d\chi\ .
\eeq
This precisely cancels the gauge variation of the bulk Chern-Simons term, which shows that the $SU(2)_1$ WZW
theory with gauged right $U(1)$ symmetry is an appropriate description of the edge of the BIQH system.

In our coupled wires construction of the BSM we will take each wire to consist of two copies of the $SU(2)_1$ WZW
theory, with fields $U_{+}$ and $U_{-}$, but with the two copies having opposite signs on their WZ terms. 
Based on the form of the gauged action Eq.~\eqref{eq:WZWgauged} for one WZW theory, it is clear that this doubled
system can be gauged in such a way that the total action is completely gauge-invariant. This 1D wire model, which
can be interpreted as consisting of two counter-propagating BIQH edge modes, is a completely consistent 1D system and
is therefore an appropriate building block for a coupled wires construction of the BSM model.

\subsection{Review: Coupled wires model for one $O(4)$ NLSM with $\theta=\pi$}

Before presenting the coupled wires construction of the BSM model, we first review the coupled wires construction of 
a single $O(4)$ NLSM with $\theta=\pi$, which was first derived in Ref.~\onlinecite{SenthilFisher} 
(see also Refs.~\onlinecite{TanakaHu,VS2013}). In this construction each 1D wire consists of just one copy of the 
$SU(2)_1$ WZW theory. 
We note briefly that in accordance with the discussion in the previous subsection, if each wire contains only one 
copy of the $SU(2)_1$ WZW theory, then the left or right $SU(2)$ symmetry of each wire cannot be consistently gauged. 
This was not a problem in the physical context of Refs.~\onlinecite{SenthilFisher,TanakaHu}, where the  $SU(2)_1$ WZW 
theory was considered in connection with 1D spin chains. In that case the $SU(2)$ subgroup of the theory which one might
consider gauging is actually the diagonal subgroup ($U \to h^{\dg}U h$), and this subgroup can be consistently 
gauged\cite{HullSpence}.

We label the individual wires in the wire model by the discrete coordinate $j= 0,\dots,N-1$. 
The lattice spacing in the stacking direction is $a_0$, and
the continuum coordinate for the stacking direction will be $y= j a_0$. The unperturbed action for the collection of 
wires is
\beq
	S_0= \sum_j \left\{ \frac{1}{8\pi}\int d^2 x\ \text{tr}[(\pd^{\mu}U^{\dg}_j) (\pd_{\mu} U_j)] + (-1)^j S_{WZ}[U_j]\right\}\ .
\eeq
We see that the sign of the WZ term alternates between adjacent wires. 
The coupling between the wires takes the form
\beq
	S_{\perp}= \frac{t_{\perp}}{2} \sum_j \int d^2 x\ \sum_{I=1}^2 (b^*_{I,j}b_{I,j+1} + \text{c.c.})\ \ ,
\eeq
where $t_{\perp} >0$, and $b_1$ and $b_2$ are the matrix elements of $U$. This term is proportional to 
$\text{tr}[U^{\dg}_j U_{j+1} + \text{h.c.}]$. We now Fourier transform in the stacking direction
\beq
	b_{I,j}= \frac{1}{\sqrt{N}}\sum_k b_{I,k} e^{ikja_0}\ , \label{eq:FT}
\eeq
to get
\beq
	S_{\perp} = t_{\perp}\sum_k \cos(k a_0) \int d^2 x\ \sum_I b^*_{I,k}b_{I,k} \ .
\eeq

The key point now is that we should expand this term near its lowest energy point. This should be contrasted
with the free fermion case, where the correct procedure was to expand the dispersion near
the band touchings at zero energy (which is where the low energy excitations are located when the lattice is at half-filling). 
The potential energy associated with $S_{\perp}$ is
\beq
	H_{\perp} = -t_{\perp}\sum_k \cos(k a_0) \int d x\ \sum_I b^*_{I,k}b_{I,k}\ ,
\eeq
which has its minimum value at $k=0$ for $t_{\perp} >0$. Expanding around $k=0$ gives 
\beq
	S_{\perp} \approx \text{const.} - \frac{t_{\perp}}{2}\sum_k (k a_0)^2 \int d^2 x\ \sum_I b^*_{I,k}b_{I,k}\ .
\eeq
Since this interaction tends to align the fields $U_j$ and $U_{j+1}$ (if we think of them as four component unit vector
fields), it makes sense to introduce the slowly varying continuum fields $b_I(t,x,y)$, which are obtained from $b_{I,j}(t,x)$ by 
keeping only the modes near $k=0$. We have\footnote{For fermions one usually has a relation between the continuum fermion
field and the lattice fermion of the form $\psi(t,x,y) \approx \psi_j(t,x)/\sqrt{a_0}$, where $a_0$ is the lattice spacing. This
is because the bare scaling dimension of the fermion should increase by $\frac{1}{2}$ as we go from spatial dimension $D$
to spatial dimension $D+1$. On the other hand, the bare sigma model field is always dimensionless, therefore we do
not attach factors of $a_0$ to $U_j(t,x)$ to define the continuum field $U(t,x,y)$. This point is important because we
do not want to have factors of the lattice spacing showing up incorrectly in the theta term.}
\beq
	b_{I,j}(t,x) \approx b_I(t,x,y)= \frac{1}{\sqrt{N}} \int \frac{dk}{\left(\frac{2\pi}{N a_0}\right)} b_{I,k}(t,x) e^{iky}\ ,
\eeq
where $y= ja_0,$ and we have expressed the continuum field $b_I(t,x,y)$ as an integral over a continuous set of wavenumbers 
$k$. The continuum fields $b_I(t,x,y)$ then become the components of the continuum matrix field $U(t,x,y)$.
Back in real space, $S_{\perp}$ becomes the $y$ derivative term $(\pd_y U^{\dg})(\pd_y U)$ in the continuum limit.

Finally, the theta term comes from a careful evaluation of the alternating sum of Wess-Zumino terms. We have
\beq
	\sum_j (-1)^j S_{WZ}[U_j] \approx \frac{1}{2}\int dy\ \pd_y  S_{WZ}[U]\ ,
\eeq
where $U$ in $S_{WZ}[U]$ is the continuum field $U(t,x,y)$. It remains to evaluate $\pd_y  S_{WZ}[U]$. One method for 
evaluating this quantity is to simply use the definition of the derivative,
\beq
	\pd_y  S_{WZ}[U]= \lim_{\ep\to 0} \frac{S_{WZ}[U(t,x,y+\ep)] - S_{WZ}[U(t,x,y)]}{\ep}\ .
\eeq
We then expand $U(t,x,y+\ep) \approx U(t,x,y) + \ep \pd_y U(t,x,y)$ and use the formula for the variation 
of the Wess-Zumino term with $\delta U$ set equal to $\ep \pd_y U$. The variation of the WZ term is 
\beq
	\delta S_{WZ}[U]= \frac{1}{4\pi}\int d^2 x\ \ep^{\tmu\tnu} \text{tr}[ (U^{\dg}\pd_{\tmu}U) (U^{\dg}\pd_{\tnu}U)(U^{\dg}\delta U)]\ ,
\eeq
where $\tmu,\tnu= t,x$ only. Setting $\delta U = \ep \pd_y U$, we obtain for the $y$ derivative,
\beq
	\pd_y  S_{WZ}[U]= \frac{1}{4\pi}\int d^2 x\ \ep^{\tmu\tnu} \text{tr}[ (U^{\dg}\pd_{\tmu}U) (U^{\dg}\pd_{\tnu}U)(U^{\dg}\pd_y U)]\ .
\eeq
A small amount of algebra then gives the final result
\beq
	\sum_j (-1)^j S_{WZ}[U_j]= \pi S_{\theta}[U]\ ,
\eeq
where $S_{\theta}[U]$ is the theta term for the $O(4)$ NLSM from Eq.~\eqref{eq:thetaPC}. 
Note that the theta angle $\theta$ works out to be exactly $\pi$.

\subsection{Coupled wires construction of the 2D BSM model}

In this section we give a coupled wires construction of the 2D BSM model. Specifically, the construction presented here
yields our 2D BSM model with only the $y$-component of the field $B_{\mu}$ non-zero. As we have discussed, 
our coupled wires construction uses two $SU(2)_1$ WZW theories in each
unit cell $j$ in the stacking direction. We label the fields for the two copies of the WZW model in each unit cell as
$U_{\pm,j}$. Below, we will see how the ``A" and ``B" fields for the 2D BSM model emerge from these initial $\pm$ fields 
(they are not the same). In order to accommodate the inversion transformation in the stacking direction,
we take the wires to be numbered as $j= -\frac{N}{2},\dots, \frac{N}{2}-1$ (so there are still $N$ unit cells).
We take $N$ even and assume periodic 
boundary conditions in the stacking direction so that $j=\frac{N}{2}$ is identified with $j=-\frac{N}{2}$. 
The unperturbed action for the decoupled collection of wires is 
\beqa
	S_0 &=& \sum_j \Bigg\{ \sum_{s=\pm}  \frac{1}{8\pi}\int d^2 x\ \text{tr}[(\pd^{\mu}U^{\dg}_{s,j}) (\pd_{\mu} U_{s,j})]   \nnb \\
&+&   S_{WZ}[U_{+,j}] -  S_{WZ}[U_{-,j}] \Bigg\}\ ,
\eeqa
which consists of two $SU(2)_1$ WZW theories in each unit cell $j$, but with the $\pm$ copies having opposite signs on 
their respective WZ terms. We add two kinds of inter-wire coupling terms, which take the form
\begin{widetext}
\beq
	S_{\perp,1}= \frac{t_1}{2}\sum_j\int d^2 x \sum_I \left\{ b^*_{I,+,j}b_{I,-,j+1} + b^*_{I,-,j+1} b_{I,+,j} + b^*_{I,-,j}b_{I,+,j+1} + b^*_{I,+,j+1}b_{I,-,j} \right\}
\eeq
and
\beq
	S_{\perp,2}= -i\frac{t_2}{2}\sum_j (-1)^j\int d^2 x \sum_I \left\{ b^*_{I,+,j}b_{I,-,j+1} - b^*_{I,-,j+1} b_{I,+,j} -( b^*_{I,-,j}b_{I,+,j+1} - b^*_{I,+,j+1}b_{I,-,j}) \right\}\ .
\eeq
\end{widetext}

The hopping term $S_{\perp,1}$ is proportional to 
$\text{tr}[(U^{\dg}_{+,j} U_{-,j+1} + \text{h.c.}) +(U^{\dg}_{-,j} U_{+,j+1} + \text{h.c.})]$, while the term
$S_{\perp,2}$ is proportional to $\text{tr}\left[\left( (i U^{\dg}_{+,j} U_{-,j+1} + \text{h.c.}) - (iU^{\dg}_{-,j} U_{+,j+1} + \text{h.c.}) \right)\sigma^z\right]$. 
When $t_1>0$ the term $S_{\perp,1}$ will tend to align $U_{+,j}$ with $U_{-,j+1}$ and
$U_{-,j}$ with $U_{+,j+1}$. We therefore define the new fields $b_{I,A,j}$ and $b_{I,B,j}$ by
\beq
	b_{I,A,j}= \begin{cases}
	b_{I,+,j}, \ j = \text{even} \\
	b_{I,-,j}, \ j =\text{odd}
\end{cases}
\eeq
and
\beq
	b_{I,B,j}= \begin{cases}
	b_{I,-,j}, \ j = \text{even} \\
	b_{I,+,j}, \ j =\text{odd}
\end{cases}\ .
\eeq
It is these fields which have a nice continuum limit for the chosen hopping terms. In terms of these fields the hopping
terms take the simpler form
\begin{widetext}
\beq
	S_{\perp,1}= \frac{t_1}{2}\sum_j \int d^2 x \sum_I \left\{ b^{*}_{I,A,j}b_{I,A,j+1}  + b^{*}_{I,A,j+1}b_{I,A,j} + b^{*}_{I,B,j}b_{I,B,j+1} + b^{*}_{I,B,j+1}b_{I,B,j}    \right\}\ ,
\eeq
and
\beq
	S_{\perp,2}=- i\frac{t_2}{2}\sum_j \int d^2 x\sum_I \left\{ b^{*}_{I,A,j}b_{I,A,j+1} - b^{*}_{I,A,j+1}b_{I,A,j} - ( b^{*}_{I,B,j}b_{I,B,j+1} - b^{*}_{I,B,j+1}b_{I,B,j})      \right\}\ .
\eeq
\end{widetext}

Now we Fourier transform the ``A" and ``B" fields as in Eq.~\eqref{eq:FT},
and also make a specific choice of hopping parameters, $t_1= t\cos(B_y a_0)$ and $t_2= t\sin(B_y a_0)$. In terms of 
the Fourier-transformed fields the inter-wire coupling now takes the form (with $S_{\perp}= S_{\perp,1} + S_{\perp,2}$)
\beqa
	S_{\perp} &=& t \sum_k \left\{  \cos[(k -B_y) a_0]\sum_I b^{*}_{I,A,k}b_{I,A,k} \right. \nnb \\ 
&+& \left. \cos[(k+B_y) a_0]\sum_I b^{*}_{I,B,k}b_{I,B,k} \right\}\ .
\eeqa
It is clear that the additional imaginary hopping terms with amplitude $t_2$ have cause the minima of the cosine potentials
to shift from $k=0$ to $k=\pm B_y$.

Finally, we take the continuum limit in the stacking direction. For the ``A" fields we expand the cosine around $k= B_y$,
and for the ``B" fields around $k= -B_y$, which is where the potential energy 
(which is proportional to $-\cos[(k \pm B_y) a_0]$) has its minimum. 
The lattice fields now
take the approximate form
\begin{subequations}
\label{eq:continuum-bosons}
\beqa
	b_{I,A,j}(t,x) &\approx& e^{iB_y y} b_{I,A}(t,x,y) \\
	b_{I,B,j}(t,x) &\approx& e^{-iB_y y} b_{I,B}(t,x,y)\ ,
\eeqa
\end{subequations}
where the slowly varying continuum fields are now given by

\beqa
	b_{I,A}(t,x,y) &=& \frac{1}{\sqrt{N}} \int \frac{dk}{\left(\frac{2\pi}{N a_0}\right)} b_{I,A,k+B_y}(t,x) e^{iky} \\
	b_{I,B}(t,x,y) &=& \frac{1}{\sqrt{N}} \int \frac{dk}{\left(\frac{2\pi}{N a_0}\right)} b_{I,B,k-B_y}(t,x) e^{iky}\ ,
\eeqa
where the integration over wavenumbers $k$ is now centered at the modes with wavenumber $\pm B_y$ instead of at $k=0$. 
The term $S_{\perp}$ will give the terms $|(\pd_y - i B_y)b_{I,A}|^2$ and
$|(\pd_y + i B_y)b_{I,B}|^2$ in the continuum limit, so this construction gives the correct minimal coupling of the bosonic fields 
to the ``gauge field" $B_y$.

Now we look at how the alternating sums of WZ terms transform into the theta terms for the ``A" and ``B" copies
of the $O(4)$ NLSM. We first define the matrix lattice
fields $U_{A,j}$ and $U_{B,j}$, whose matrix elements are the lattice fields $b_{I,A,j}$ and $b_{I,B,j}$. In the continuum
limit these are expressed in terms of the continuum matrix fields $U_A(t,x,y)$ and $U_B(t,x,y)$ (whose matrix
elements are the continuum fields $b_{I,A}(t,x,y)$ and $b_{I,B}(t,x,y)$), as
\beqa
	U_{A,j}(t,x) &\approx& U_A(t,x,y) e^{i (B_y y)\sigma^z} \\
	U_{B,j}(t,x) &\approx& U_B(t,x,y) e^{-i( B_y y)\sigma^z}\ .
\eeqa
The matrix phase factors $e^{\pm i (B_y y)\sigma^z}$ attach the appropriate phase to the lattice bosons, as shown in 
Eq.~\eqref{eq:continuum-bosons}. 
Because of the form of the WZ term,
the matrix phase factors $e^{\pm i (B_y y)\sigma^z}$ completely cancel each other, and the evaluation of the theta terms 
from alternating sums of WZ terms proceeds exactly as in the case of one copy of the $O(4)$ NLSM. 
In addition, we have
\begin{align}
	\sum_j \left( S_{WZ}[U_{+,j}] -  S_{WZ}[U_{-,j}] \right) = & \nnb \\ 
	 \sum_j (-1)^j  S_{WZ}[U_{A,j}] &- \sum_j (-1)^j S_{WZ}[U_{B,j}]\ ,   
\end{align}
so the theta angles for the ``A" and ``B" copies of the $O(4)$ NLSM will have opposite sign.

\subsection{Symmetry transformations}

We now define transformations for the lattice bosonic fields $b_{I,\pm,j}$ under inversion $\ZI$ and
time-reversal $\ZT$ in such a way that in the continuum limit we get the transformations shown in 
Eq.~\eqref{eq:symI} and Eq.~\eqref{eq:symT}  for the fields $b_{I,A}$ and $b_{I,B}$ of the 2D BSM model. 
For time-reversal, we take
\beq
	 \T b_{I,\pm,j} \T^{-1} =  b_{I,\mp,j}\ .
\eeq
It is easy to see that the term $S_{\perp,1}$ has this symmetry. The term $S_{\perp,2}$ picks up a minus sign under the swap 
$+\to -$, but the factor of $i$ in that term is also negated since
$\T$ is anti-unitary. These two signs cancel each other, and so the term $S_{\perp,2}$ is also symmetric
under this time-reversal symmetry. We also see that $\T b_{I,A,j} \T^{-1}= b_{I,B,j},$ which then translates over to the 
correct continuum transformation $\T b_{I,A} \T^{-1}= b_{I,B}$, as can be seen from Eq.~\eqref{eq:continuum-bosons}.

Next we consider the action of inversion symmetry. We take $\I$ to act on the lattice fields as
\beq
	\I b_{I,\pm,j}(x) \I^{-1}= b_{I,\mp,-j}(-x)\ ,
\eeq
which is just an inversion about the origin $x=0,j=0$. Again, it is easy to see that 
$S_{\perp,1}$ has this inversion symmetry. Although it is not obvious, one 
can explicitly check that $S_{\perp,2}$ also has this symmetry. For example the terms $b^*_{I,+,1}b_{I,-,2}$
and $b^*_{I,-,-1}b_{I,+,-2}$, which are partners under inversion, appear in $S_{\perp,2}$ with the same sign.
We also see that $\I b_{I,A,j}(x) \I^{-1}= b_{I,B,-j}(-x)$ since $j \equiv -j$ mod $2$.
In the continuum limit this inversion symmetry then
translates into $\I b_{I,A}(\mb{x}) \I^{-1} = b_{I,B}(-\mb{x})$, as can be seen from Eq.~\eqref{eq:continuum-bosons},
and this is exactly the inversion transformation for the continuum fields in the 2D BSM model.

%Next we consider the action of inversion symmetry. It turns out that the correct definition of this symmetry involves a 
%bond-centered inversion for the stacking direction.  In the $x$-direction (which is already continuous),
%we just send $x\to -x$. Therefore we take $\I$ to act on the lattice fields as
%\beq
%	\I b_{I,\pm,j}(x) \I^{-1}= b_{I,\mp,-j -1}(-x)\ .
%\eeq
%This is inversion about the center of the bond between $j=-1$ and $j=0$. Again, it is easy to see that 
%$S_{1,\perp}$ has this inversion symmetry. Although $S_{\perp,2}$ picks up a minus sign under $+ \to -$, 
%this bond-centered inversion also changes the parity of $j$ (if $j$ is even, then $-j-1$ is odd, and vice-versa), 
%which gives another minus sign due to the factor of $(-1)^j$  in $S_{\perp,2}$. So
%$S_{\perp,2}$ also has this inversion symmetry. We also see that $\I b_{I,A,j}(x) \I^{-1}= b_{I,B,-j-1}(-x)$
%for the ``A" and ``B" lattice fields. Finally, we note that in the continuum limit this inversion symmetry will
%just translate into $\I b_{I,A}(\mb{x}) \I^{-1} = b_{I,B}(-\mb{x})$, since as the lattice spacing $a_0$ goes
%to zero the relation $y \to -y - a_0$ just becomes $y\to -y$. The latter inversion transformation is exactly the one we
%defined for the continuum fields in the 2D BSM model.

Finally, we discuss the emergence of the $U(1)_t$ translation symmetry for the continuum fields. 
We saw that after expanding the cosines near $k=\pm B_y$ and taking the continuum limit in the $y$ direction, the
term $S_{\perp}$ gave the kinetic terms $|(\pd_y - i B_y)b_{I,A}|^2$ and $|(\pd_y + i B_y)b_{I,B}|^2$ for the 
continuum fields $b_{I,A}$ and $b_{I,B}$. We can see from the
form of these terms that the continuum action is invariant under the transformation 
$b_{I,A} \to e^{i\xi} b_{I,A}$, $b_{I,B} \to e^{-i\xi} b_{I,B}$ 
and $B_y \to B_y + \pd_y \xi$. This transformation is exactly the 
$U(1)_t$ gauge transformation shown in Eq.~\eqref{eq:U1t} and discussed in the paragraphs following that equation
(in the special case where only the $y$-component of $B_{\mu}$ is non-zero).

\subsection{Discussion}

In this section we have shown how to construct our 2D BSM model from a quasi-1D coupled wires model. Let us now contrast 
the coupled wires model for the BSM phase with the coupled wires model for the DSM phase 
(derived in Ref.~\onlinecite{Ramamurthy2014}).

In the DSM case, we considered fermions on the square lattice at half-filling. The Bloch Hamiltonian
for the model in question featured two bands with energies $E_{\pm}(\mb{k})$ shown in Eq.~\eqref{eq:DSMenergy}. 
At half-filling, the low-energy excitations of that model were at the locations in the BZ where the two bands touched, i.e.,
at the locations where $E_{+}(\mb{k})= E_{-}(\mb{k})=0$.
For this reason we expanded the Bloch
Hamiltonian where $E_{\pm}(\mb{k})=0$ to obtain the low energy description of the system. If
the band was just a cosine, e.g., $\cos(k_y)$, then we would expand at $k_y = \pm \frac{\pi}{2}$ (so \emph{two} locations), 
which are the locations of the two Dirac points. From
this discussion it is clear why the form $m \pm t_y \cos(k_y)$ for the dispersion was appropriate for the construction of the
DSM model: the
addition of the intra-wire mass $m$ shifts the cosine \emph{vertically}, which changes the positions of the zeros of energy, 
and hence shifts the locations of the Dirac nodes in the BZ.

Now we compare to the BSM case. For bosons there is no notion of filling a band or of expanding a dispersion near band 
touchings. Instead, the appropriate method for finding the low energy description of the system was to expand the 
potential about its minimum. For a potential which is just a cosine, e.g., $-\cos(k_y)$, we expand around $k_y = 0$ (so 
a \emph{single} location). From this discussion it is apparent that in order to move the low-energy physics of the 
bosonic system away from $k_y=0$, we need to shift the minimum of the cosine potential, i.e., we need a \emph{horizontal}
shift of the cosine, as in $-\cos(k_y - B_y)$. In our coupled wires construction of the BSM model this horizontal shift
was accomplished using an imaginary inter-wire hopping term, not an intra-wire mass term as in the DSM case.

It seems that the essential difference between the coupled wires constructions of the DSM and BSM models comes
from the simple fact that fermions fill a band structure, while bosons do not. Therefore a different mechanism is needed in
the two cases to shift the low-energy physics to the points $(0,\pm B_y)$ in momentum space.

Finally, we note that our coupled wires model for the BSM can be driven into time-reversal or inversion breaking phases
by adding dimerization to the inter-wire tunneling terms. As we discussed in Sec.~\ref{sec:BSM-model} (see the discussion
in the paragraph above Eq.~\eqref{eq:below-theta-discussion}), the time-reversal and inversion breaking perturbations
to the BSM model correspond to correlated shifts of the theta angles $\theta_A$ and $\theta_B$ away from their original
values $\theta_A=-\theta_B=\pi$. In Ref.~\onlinecite{TanakaHu}, Tanaka and Hu have shown that incorporating dimerization
into the inter-wire interactions in the coupled wires construction (of Ref.~\onlinecite{SenthilFisher}) of the $O(4)$ NLSM at 
$\theta=\pi$ leads to an $O(4)$ NLSM with $\theta$ shifted away from $\pi$. It is therefore possible
to investigate the time-reversal and inversion breaking phases of the BSM model within its quasi-1D description in terms
of coupled wires, just by adding suitable dimerization to the inter-wire tunneling terms. However, we do not carry
out this analysis here as we have already investigated these phases within the continuum description in 
Sec.~\ref{sec:BSM-model} and we do not expect the results to be modified in an essential way.

\section{Conclusion}

We have constructed an effective theory and a coupled-wire model for a bosonic analog of a topological DSM, in which the Dirac cones of the DSM are replaced
with copies of the $O(4)$ NLSM with topological theta term and theta angle $\theta=\pm \pi$. %Just one copy of 
%an $O(4)$ NLSM with $\theta=\pi$ can describe the gapless surface of the 3D BTI. 
We computed the time-reversal
and inversion symmetry breaking electromagnetic responses of this BSM model, and showed that they are
twice the value of the responses obtained in the fermionic DSM case. We also examined the stability of our BSM model to many
kinds of perturbations, and found that the same composite $\ZTI$ symmetry which protects the local stability of the DSM
also plays an important role in the local stability of the BSM. Finally, we provided a quasi-1D construction of the BSM model
using an array of coupled 1D wires in which each individual wire is made up of two copies of the $SU(2)_1$ WZW conformal field
theory.

Along the way we have been able to clarify many 
aspects of the $O(4)$ NLSM with $\theta=\pi$ which have been discussed in the literature. In particular we provided
a detailed analysis of the stability of the BTI surface theory to symmetry-allowed perturbations, which were only briefly
discussed in Ref.~\onlinecite{VS2013}. We were also able to prove the results on the charges and statistics
of vortices in the $O(4)$ NLSM with theta term which were argued for in Refs.~\onlinecite{SenthilFisher,VS2013}. 
We also conjectured a relationship between the descriptions of the BTI surface discussed in this paper, in particular the 
dual vortex description of Ref.~\onlinecite{VS2013} and the description in terms of Abanov-Wiegmann fermions, 
and the recently proposed dual description in terms of $N=2$ QED$_3$ \cite{xu2015}. As we discussed in 
Sec.~\ref{sec:BTI-surface}, one interesting direction for future work would be to give a direct derivation of the 
$N=2$ QED$_3$ description of the BTI surface, starting from the description in terms of the $O(4)$ NLSM with $\theta=\pi$.

Another interesting direction for future work would be to explore bosonic analogues of Weyl semi-metals in three
spatial dimensions. In particular, it would be interesting to understand the requirements for the local stability of a bosonic
analogue of a Weyl semi-metal, since in the fermion case the local stability of the Weyl nodes does not depend on any
discrete symmetry~\cite{turner2013}. This is quite different from the DSM case in 2D, in which the composite symmetry
$\ZTI$ was necessary to ensure the local stability of the Dirac nodes. One possibility for a bosonic analogue of a Weyl
semi-metal would be to try replacing each Weyl node with a copy of the $O(5)$ NLSM with theta term and theta angle
$\theta=\pm \pi$.

Finally, there is still more to be learned about the $O(4)$ NLSM at $\theta=\pi$. The disordered (symmetry-preserving) phase 
of this model was first argued to be gapless in Ref.~\onlinecite{SenthilFisher}. Qualitative arguments about
the RG flows of this model also indicate the existence of a fixed point (representing the putative gapless
phase) at $\theta=\pi$ at a large but finite value of the coupling constant $g$~\cite{XuLudwig}. Very recently,
numerical simulations on a (fermionic) honeycomb lattice model whose low energy sector is described by the $O(4)$ NLSM with 
$\theta=\pi$ have shown that this model is indeed gapless~\cite{slagle2015exotic,you2015quantum}. It would be very 
interesting to understand how the vortex braiding processes we described in Appendix~\ref{app:path-integral}, which
at $\theta=\pi$ lead to destructive interference between the different field configurations summed over in the path integral of 
the $O(4)$ NLSM, lead to this gapless behavior. In addition, it would be interesting to calculate the scaling dimension
of the $O(4)$ field $\mb{N}$ at the disordered fixed point.

\begin{acknowledgements}

We thank M. Hermele for discussions, and especially Cenke Xu for several helpful discussions and explanations of recent work.
TLH and MFL acknowledge support from the ONR YIP Award N00014-15-1-2383. 
GYC is supported by the NSF under 
Grant No. DMR-1064319, the Brain Korea 21 PLUS Project of Korea Government, and Grant No. 2016R1A5A1008184 under NRF 
of Korea. We gratefully acknowledge the support of the Institute for Condensed Matter Theory at UIUC. 

\end{acknowledgements}

\appendix

\section{Canonical quantization of the $O(4)$ non-linear sigma model}
\label{app:canonical}

In this appendix we briefly discuss the canonical quantization of the $O(4)$ NLSM. We use these 
commutation relations in Sec.~\ref{sec:BTI-surface} to understand the effects of
symmetry-allowed perturbations on the surface theory of the BTI. Since the 
$O(4)$ NLSM is a constrained system, it is necessary to use the Dirac bracket formalism to obtain the canonical
commutators of this system \cite{Dirac,HRT}. Let $\psi_{i}$, $i= 1,\dots,M$, be the second class constraints of 
the system in question. Then the Dirac bracket is given by
\beqa
	& & \{f(\mb{x}),g(\mb{y})\}_{D} = \{f(\mb{x}),g(\mb{y})\} \\
&-& \sum_{i,j=1}^M \int d^2\mb{z}\ d^2\mb{z}'\ \{f(\mb{x}),\psi_i(\mb{z})   \}C^{-1}_{ij}(\mb{z},\mb{z}')\{\psi_j(\mb{z}'), g(\mb{y})\}\ , \nnb
\eeqa
where the $C_{ij}(\mb{z},\mb{z}')$, which are the matrix elements of a matrix with discrete indices $i,j$ and 
continuous spatial indices $\mb{z}$ and $\mb{z}'$, are given by
\beq
	C_{ij}(\mb{z},\mb{z}')= \{\psi_i(\mb{z}) ,\psi_j(\mb{z'})  \}\ ,
\eeq
and where $\{\ ,\ \}$ is the ordinary Poisson bracket. 

In the case of the $O(4)$ NLSM, one possible choice of coordinates and momenta is just the fields $N^a$ and 
their canonically conjugate momenta $\Pi^a= \frac{\pd \mathcal{L}}{\pd (\pd_t N^a)}$. In terms of these variables the 
Poisson bracket reads
\beqa
	& & \{f(\mb{x}),g(\mb{y})\}= \\
& & \sum_{a=1}^4 \int d^2 \mb{z}\   \left(\frac{\delta f(\mb{x})}{\delta N^a(\mb{z})}\frac{\delta g(\mb{y})}{\delta \Pi^a(\mb{z})} -  \frac{\delta f(\mb{x})}{\delta \Pi^a(\mb{z})}\frac{\delta g(\mb{y})}{\delta N^a(\mb{z})}  \right) \nnb\ , 
\eeqa
where $\frac{\delta}{\delta N^a(\mb{z})}$ is a functional derivative. This system has two second class constraints, which
take the form
\beqa
	\psi_1 &=& \sum_a N^a N^a - 1 \\
	\psi_2 &=& \sum_a N^a \Pi^a\ .
\eeqa
Using this data one finds, for example, that the Dirac bracket for $N^a$ and $\Pi^b$ is
\beq
	\{N^a(\mb{x}),\Pi^b(\mb{y})\}_{D} = \left( \delta^{ab} - N^a(\mb{x})N^b(\mb{y}) \right)\delta^{(2)}(\mb{x}-\mb{y})\ . 
\eeq
The rest of the Dirac brackets for this system are shown explicitly in Ref.~\onlinecite{Karabali}.
The commutator for the quantum field theory is then obtained by replacing $\{N^a(\mb{x}),\Pi^b(\mb{y})\}_{D}$
with $-i[N^a(\mb{x}),\Pi^b(\mb{y})]$ in the previous expression.

In this paper we discuss the $O(4)$ NLSM using the variables $b_1$ and $b_2$ defined in Eq.~\eqref{eq:b-variables}.
In the canonical formalism we now have the coordinates $b_I$ and $b^*_I$ and momenta 
$\pi_I = \frac{\pd \mathcal{L}}{\pd (\pd_t b_I)}$ and $\pi^*_I = \frac{\pd \mathcal{L}}{\pd (\pd_t b^*_I)}$ for $I=1,2$.
In these variables the second class constraints are
\beqa
	\psi_1 &=& \sum_I b^*_I b_I - 1 \\
	\psi_2 &=& \sum_I \left( b_I \pi_I + b^*_I \pi^*_I \right)\ .
\eeqa
The Dirac bracket for $b_I$ and $\pi_J$ takes the form
\beq
	\{b_I(\mb{x}),\pi_J(\mb{y})\}_{D} = \left(\delta_{IJ} - \frac{1}{2}b_I(\mb{x})b^*_J(\mb{y})\right)\delta^{(2)}(\mb{x}-\mb{y})\ .
\eeq
For the quantum theory this yields the commutation relation
\beq
	[b_I(\mb{x}),\pi_J(\mb{y})] = i\left(\delta_{IJ} - \frac{1}{2}b_I(\mb{x})b^{\dg}_J(\mb{y})\right)\delta^{(2)}(\mb{x}-\mb{y})\ ,
\eeq
where the function $b^*_I(\mb{x})$ has been replaced with the operator $b^{\dg}_I(\mb{x})$ on the Hilbert space. One can 
also show that the operators $b_I(\mb{x})$ and $b^{\dg}_J(\mb{x})$ all commute with each other. These are the
only commutation relations we require for this paper, but the others can also be derived using the Dirac bracket
formalism.

\section{Vortices in the $O(4)$ NLSM and their quantum numbers}
\label{app:vortices}

In this appendix we study vortex solutions of the equations of motion for the $O(4)$ NLSM, and we also perform
a collective coordinate quantization of the global excitations on the background of a single vortex. This allows us to show very
directly that vortices in the phase of $b_1$ carry charge $\frac{\theta}{2\pi}$ of $b_2$ and vice-versa, as
was argued in Ref.~\onlinecite{VS2013}. A more 
precise statement is that in the presence of a vortex in $b_1$, the charge spectrum of $b_2$ is shifted by
$\frac{\theta}{2\pi}$. Our analysis (in particular, the collective coordinate quantization) 
closely parallels the analysis in Ref.~\onlinecite{Karabali} of solitons in the $O(3)$
NLSM with Hopf term. In Ref.~\onlinecite{Karabali}, the authors showed that a soliton of topological charge $Q$ carries 
angular momentum $\frac{\theta}{2\pi}Q^2$, where $\theta$ is the coefficient of the Hopf term (the result for
$Q=1$ was originally worked out in Ref.~\onlinecite{WilczekZee}).

\subsection{Finite energy vortex solutions}

We start by discussing a class of finite energy vortex solutions to the NLSM equations of motion. To the best of our 
knowledge, these solutions have not appeared in the literature. They are, however, closely related to solitons in the 
$O(3)$ NLSM, due to the fact that they involve only three components of the $O(4)$ field. Exact soliton solutions
for the $O(3)$ NLSM were obtained long ago by Belavin and Polyakov \cite{BP1975}. Our vortex solutions, however, involve
different boundary conditions than those considered in the soliton case. Indeed, in the study of solitons in an 
$O(3)$ NLSM, with field $\mb{m}$, one imposes the boundary condition that $\mb{m}$ tends to a fixed configuration
$\mb{m}_0$ at spatial infinity. This boundary condition has the effect of compactifying 2D space to the sphere $S^2$. 
For the vortex configurations considered here, we will instead regard 2D space as a large disk of radius $R$, and only 
take $R$ to infinity at the end of the calculation.

If we vary the $O(4)$ NLSM action in Eq.~\eqref{eq:O4} with respect to $U$ and use
$\delta U^{\dg} = -U^{\dg}\delta U U^{\dg}$ (since $U$ is an $SU(2)$ matrix) we find the equation of motion 
\beq
	\Box U - U(\Box U^{\dg})U = 0\ ,
\eeq
where $\Box = \pd_t^2 - \nabla^2$. The theta term does not contribute to the equation of motion since its
variation is a total derivative. We work in polar coordinates $(r,\phi)$ for the plane, but with an upper cutoff $R$
for the radial direction, i.e., $r\in [0,R]$, and take $R\to\infty$ at the end of the calculation. 
Let $z= (b_1,b_2)^T$, where $b_1$ and $b_2$ are the elements of $U$ as shown in Eq.~\eqref{eq:b-variables}.
We make the time-independent vortex ansatz, 
\beq
	z= \begin{pmatrix}
	\cos(f(r))e^{i\alpha\phi} \\
	\sin(f(r))
\end{pmatrix} \label{eq:vortex-ansatz}
\eeq
where $\alpha \in \mathbb{Z}$ (so that the solution is single-valued)
and we take the boundary conditions $f(0)= \tfrac{\pi}{2}$ and $f(R)= 0$, so that the amplitude of $b_1$ vanishes
in the vortex core. One can actually take $\al$ to be any real number in what follows. Solutions with general values of $\al$ 
might be relevant for the study of braiding statistics of excitations in gauged NLSM's as considered in 
Ref.~\onlinecite{CenkeBraiding}.
Plugging this ansatz into the equations of motion yields a differential equation for $f(r)$
\beq
	f''(r) + \frac{1}{r}f'(r) +\alpha^2\frac{\sin(f(r))\cos(f(r))}{r^2}= 0\ , \label{eq:diffeq}
\eeq
whose exact solution for the given boundary conditions is
\beq
	f(r)= \text{am}[\log\left(\tfrac{R}{r}\right)^{|\alpha|},1] = -\tfrac{\pi}{2} + 2\tan^{-1}\left[\left(\tfrac{R}{r}\right)^{|\alpha|}\right]\ . \label{eq:soln}
\eeq
In this expression, $\text{am}[u,k]$ is the Jacobi Amplitude function. When $k=1$, this function reduces to a much more
manageable form. 

Next we show that this solution has finite energy. We will see that the energy of the solution is actually independent of the 
long-distance cutoff $R$. The topological term does not contribute to the energy, so we just have ($i=x,y$ and
 we sum over $i$)
\beqa
	E_{\al} &=& \int d^2\mb{x}\ \frac{1}{g}(\pd_i z^{\dg}) (\pd_i z) \nnb \\
	&=& \int  d^2\mb{x}\ \frac{1}{g}\left\{ (\pd_i f(r))^2 + \frac{\cos^2(f(r))}{r^2} \right\}\ .
\eeqa
Next we go to polar coordinates and use the fact that 
$(\pd_i f(r))^2 = \frac{4\alpha^2}{r^2}\frac{R^{2|\al|}r^{2|\al|}}{(r^{2|\al|}+ R^{2|\al|})^2}$ and $\frac{\cos^2(f(r))}{r^2} = \frac{4}{r^2}\frac{R^{2|\al|}r^{2|\al|}}{(r^{2|\al|}+ R^{2|\al|})^2}$ for the solution for $f(r)$ in Eq.~\eqref{eq:soln}
to find
\beqa
	E_{\al} &=& \frac{1}{g}\int_0^{2\pi}d\phi\ \int_0^R r dr\ \frac{4(\alpha^2 + 1)}{r^2}\frac{R^{2|\al|}r^{2|\al|}}{(r^{2|\al|}+ R^{2|\al|})^2} \nnb \\
	&=& \frac{2\pi}{g}\left( |\al| + \frac{1}{|\al|}\right)\ .
\eeqa
So we find that the vortex solution has finite energy, and that the energy is independent of the upper cutoff $R$.
The energy increases essentially linearly with the ``vortex strength" $\al$. For the case of $\alpha=1$, we just get 
$E_1= \frac{4\pi}{g}$. 

It is interesting to note that this theory admits finite energy vortex solutions without requiring coupling to a dynamical
gauge field, as is necessary in the case of an ordinary complex scalar field in 2D (see, for example, 
the discussion of the Abelian Higgs model in Ref.~\onlinecite{Shifman}). These vortex solutions are, however, somewhat
pathological, in the sense that the size of the vortex core grows without bound as the upper cutoff $R$ is pushed to infinity.
Vortex-anti-vortex pairs, however, do not have this problem. This is because the energy density of such a pair
falls of faster than $\frac{1}{r^2}$ at long distances, so these objects are well-defined when the system size is infinite.

%\textbf{MFL: Should we include this extra discussion?}
%
%This solution is easily extended to a co-dimension one ``vortex-hyperplane" solution for
%$O(D+2)$ NLSM's in $D+1$ spacetime dimensions. The equations of motion for the $D+2$ components $N^a$ of the NLSM
%field can be obtained as follows. We first find a family of $D+2$ anti-commuting hermitian matrices $\Gamma^a$ which 
%obey a Euclidean Clifford algebra $\{\Gamma^a,\Gamma^b\}= 2\delta^{ab}\mathbb{I}$ (of course the size of the matrices
%$\Gamma^a$ will depend on $D$). Using these we construct the matrix field $V= iN^a\Gamma^a$,
%which satisfies $V^{\dg}= -iN^a\Gamma^a$ and $V^{\dg}V= \mathbb{I}$. The 
%kinetic part of the Lagrangian can be written as
%\beq
%	\mathcal{L} \sim \text{tr}(\pd^{\mu}V^{\dg}\pd_{\mu}V)\ .
%\eeq
%We can then vary with respect to $V$ (using $\delta V^{\dg} = -V^{\dg}\delta V V^{\dg}$) to find the equation of motion
%\beq
%	\Box V - V(\Box V^{\dg})V = 0\ ,
%\eeq
%which can be reduced to
%\beq
%	\Box N^a - N^a (\mb{N}\cdot(\Box \mb{N})) = 0\ , 
%\eeq
%for $a= 1,\dots,D+2$. 
%
%Now we make the generalized ``vortex-hyperplane" ansatz
%\beqa
%	N^1 &=& \cos(f(r))\cos(\al\phi) \\
%	N^2 &=& \cos(f(r))\sin(\al\phi) \\
%	N^a &=& \frac{\sin(f(r))}{\sqrt{D}}\ , a= 3,\dots, D+2\ .
%\eeqa
%For this system of coordinates we have $\Box= \pd_t^2 - (\pd_r^2 + \frac{1}{r}\pd_r + \frac{1}{r^2}\pd_{\phi}^2 + \dots)$
%and one can easily check that the equation of motion for $f(r)$ that results from this vortex ansatz is just 
%Eq.~\eqref{eq:diffeq}. Therefore we have shown that this vortex solution can used in any dimension.

\subsection{$O(2)$ NLSM for phase excitations of $b_2$ on a vortex background}

We now study the global excitations of the phase of the boson $b_2$ on the background of a vortex in $b_1$. 
Note that the classical energy $E_{\al}$ of the vortex ansatz in Eq.~\eqref{eq:vortex-ansatz} is invariant under
the replacement $\sin(f(r)) \to \sin(f(r))e^{i\bar{\theta}_2}$ where $\bar{\theta}_2$ is any \emph{constant} phase. 
To study the global excitations about the vortex solution, we promote $\bar{\theta}_2$ to a time-dependent phase 
$\theta_2(t)$,  
\beq
	z= \begin{pmatrix}
	\cos(f(r))e^{i\phi} \\
	\sin(f(r))e^{i\theta_2(t)}
\end{pmatrix} \label{eq:O4ansatz}
\eeq
where $f(r)$ is the vortex solution from Eq.~\eqref{eq:soln} with $\al=1$. We then evaluate the action on this configuration
and quantize the motion of $\theta_2(t)$. This type of analysis is referred to as \emph{collective coordinate} quantization
(see Refs.~\onlinecite{adkins1983,Karabali}) and is useful for understanding how quantum fluctuations can lift the classical 
degeneracy of global fluctuations about the vortex solution. 

On this field configuration the theta term in the action 
reduces as
\beqa
	S_{\theta} &=& \frac{1}{24 \pi^2} \int d^3 x\ \ep^{\mu\nu\lambda} \text{tr}[(U^{\dg}\pd_{\mu} U)( U^{\dg}\pd_{\nu} U)( U^{\dg}\pd_{\lambda} U)] \nnb \\
	&\to& \frac{1}{2\pi}\int dt\ \pd_t\theta_2(t)\ ,
\eeqa
which is precisely the theta term for an $O(2)$ NLSM in $0+1$-d \cite{CenkeLine}.
The kinetic term in the action reduces to 
\beqa
	S_{kin} &=&  \int d^3 x\ \frac{1}{g}(\pd^{\mu} z^{\dg}) (\pd_{\mu} z) \nnb \\
	&\to&  \int dt\ \left\{\frac{2\pi R^2 J}{g}(\pd_t \theta_2)^2 - E_1\right\} \ ,
\eeqa
where $E_1= \frac{4\pi}{g}$ is the energy of the vortex solution and $J$ is 
the convergent integral
\beq
	J= \int_0^{\infty} dw\ e^{-2w}\text{sn}^2[w,1]= \frac{3}{2}-\ln(4)\ .
\eeq
In this expression $\text{sn}[w,1]= \sin(\text{am}[w,1])$ is one of the Jacobi Elliptic functions. 
An important point here is that it does not make physical sense to evaluate the action on a vortex solution with
infinite energy, therefore it is crucial for our analysis that the vortex solutions do have finite energy.

The full action for the phase excitation $\theta_2(t)$ is (neglecting the constant $E_1$)
\beq
	S_{\theta_2}= \int dt\ \left\{ \frac{2\pi R^2 J}{g}(\pd_t \theta_2)^2 -  \frac{\theta}{2\pi}\pd_t\theta_2 \right\} \ .
\eeq
This is exactly the action for an $O(2)$ NLSM with theta term in $0+1$ dimensions.
We can now canonically quantize the action for $\theta_2$. We define the canonical momentum 
\beq
	p_2= \frac{\pd \mathcal{L}_{core}}{\pd (\pd_t\theta_2)}= \frac{4\pi R^2 J}{g}(\pd_t \theta_2) -  \frac{\theta}{2\pi}\ ,
\eeq
from which we derive the Hamiltonian
\beq
	H_{core}= \frac{1}{2m}\left(p_2 + \frac{\theta}{2\pi}\right)^2\ , \label{eq:Hcore}
\eeq
where $m= \frac{4\pi R^2 J}{g}$ is the ``mass" of the degree of freedom inside the vortex. In canonical quantization we set
$p_2= -i\frac{\pd}{\pd \theta_2}$, and so we find that the eigenfunctions of the vortex Hamiltonian are
\beq
	\psi_n(\theta_2)= \frac{1}{\sqrt{2\pi}}e^{i n \theta_2}\ , n\in \mathbb{Z}
\eeq
with energies
\beq
	E_n= \frac{1}{2 m}\left(n+\frac{\theta}{2\pi}\right)^2\ .
\eeq
We see that there is generally a unique ground state except for when $\theta=\pi$, in which case the $n=0$ and $n=-1$
states are degenerate. The energies of these states do, however, all collapse to zero in the thermodynamic limit $R\to \infty$.

\subsection{Spectrum of charges}

Finally, we can look at the charge spectrum of $\theta_2(t)$ fluctuations on the background of a vortex in $b_1$.
We start by considering the conserved charge for boson species $2$,
\beq
	Q_2= \int d^2 \mb{x}\ \frac{i}{g}(\pd_t b_2^{*} b_2 - b_2^*\pd_t b_2)\ ,
\eeq
where the integration is taken over all of space. This is the conserved charge for the 
Noether current of the $O(4)$ NLSM associated with the invariance of the action under the symmetry $b_2 \to e^{i\chi}b_2$.
After canonical quantization, $Q_2$ will become the number operator for the $b_2$ bosons.
Evaluating this expression on our vortex solution gives
\beq
	Q_2= \frac{4\pi R^2 J}{g}(\pd_t \theta_2)\ ,
\eeq
and replacing $\pd_t\theta_2$ with the canonical momentum $p_2$ gives
\beq
	Q_2= p_2 + \frac{\theta}{2\pi}\ .
\eeq

This shows that the charge spectrum of $b_2$ is shifted by $ \frac{\theta}{2\pi}$ in the presence of a vortex in $b_1$,
which means that a half-charge of $b_2$ may be associated to vortices in $b_1$ at $\theta=\pi$. An analogous result
holds for vortices in $b_2$. It follows that a vortex in $b_1$
will carry half of \emph{any} $U(1)$ charge carried by $b_2$, 
for example the $U(1)_c$ and $U(1)_t$ charges considered in this paper.
Thus, we have been able to prove the result of Ref.~\onlinecite{VS2013},
which is that vortices on the surface of the BTI
carry charge $\pm \frac{1}{2}$, directly from the description of the surface in terms of the $O(4)$ NLSM with $\theta=\pi$.

\section{Theta term and the Minkowski space path integral for the $O(4)$ NLSM}
\label{app:path-integral}

In this Appendix we discuss the role of the theta term in the Minkowksi spacetime (i.e., \emph{real} time) path
integral of the $O(4)$ NLSM. 
Recall from Sec.~\ref{sec:BTI-surface} that in Euclidean spacetime (compactified to the
sphere $S^3$ via appropriate boundary conditions), 
the theta term was quantized due to the non-trivial homotopy group $\pi_3(S^3)= \mathbb{Z}$. In
that case the theta term contributed a phase $e^{i\theta n_I}$ to the Euclidean path integral, where $n_I\in \mathbb{Z}$
was the instanton number of the field configuration (see Eq.~\eqref{eq:inst-no}). 
It then followed that the time-reversal symmetric values of $\theta$ are $\theta= n\pi$, $n\in\mathbb{Z}$, at which the 
phase $e^{i\theta n_I}$ is real.
In Minkowski spacetime these arguments no longer hold, and it is illuminating to develop a separate understanding 
of the role of the theta term in the real time path integral. 

In this Appendix we show that in the real time path integral the theta term gives a weight $e^{i\theta}$ to 
spacetime configurations of the $O(4)$ field in which a vortex in the field $b_2$ makes a complete circuit around around a 
vortex in $b_1$. This result was anticipated by the Euclidean spacetime arguments of Senthil and Fisher 
(Ref.~\onlinecite{SenthilFisher}), but in this Appendix we derive this result using only the properties of the theta term in 
Minkowski spacetime. In addition, following an argument used by Wilczek and Zee in Ref.~\onlinecite{WilczekZee}
in their analysis of solitons in the $O(3)$ NLSM with Hopf term, our result implies that a bound state of a vortex
in $b_1$ and a vortex in $b_2$ carries intrinsic angular momentum $J=\frac{\theta}{2\pi}$. When $\theta=\pi$, we have
$J=\frac{1}{2}$, which means that the bound state is a fermion. This result was also argued for in Ref.~\onlinecite{VS2013}.

To start, we express the components $b_1$ and $b_2$ of the NLSM field $U$ in Hopf coordinates as in 
Sec.~\ref{sec:BTI-surface}. In these coordinates
the bosonic fields are expressed as $b_1= \sin(\eta)e^{i\vth_1}$ and $b_2= \cos(\eta)e^{i\vth_2}$
with $\eta\in [0,\frac{\pi}{2}]$ and $\vth_1,\vth_2\in [0,2\pi)$. The theta term can be written in the form
(compare to Eq.~\eqref{eq:theta-hopf})
\beq
	S_{\theta}[U]= \frac{1}{4\pi^2}\int d^3 x\ \ep^{\mu\nu\lam}\pd_{\mu}\left(\sin^2(\eta)\right)\pd_{\nu}\vth_1\pd_{\lam}\vth_2\ .
\eeq
Now we integrate by parts, for the moment ignoring boundary terms. Later we will comment on the boundary 
conditions necessary to justify ignoring these boundary terms. We get
\beq
	S_{\theta}[U]= \frac{1}{2\pi}\int d^3 x\ \sin^2(\eta)\left(\pd_{\mu}\vth_1 K^{\mu}_2 - \pd_{\mu}\vth_2 K^{\mu}_1   \right)\ , \label{eq:vortex-pair}
\eeq
where we have introduced the vortex currents $K^{\mu}_I= \frac{1}{2\pi}\ep^{\mu\nu\lam}\pd_{\nu}\pd_{\lam}\vth_I$
for vortices in the phase of the field $b_I$. If $\vth_I$ has vortices of vorticity $q_{I,j}$ ($q_{I,j} \in \mathbb{Z}$) at 
locations  $\mb{r}_{I,j}(t)$ (i.e., $\mb{r}_{I,j}(t)$ is the location of the vortex core), then
the components of the vortex current $K^{\mu}_I$ take the form
\beqa
	K^t_I &=& \sum_j q_{I,j}\delta^{(2)}(\mb{x}-\mb{r}_{I,j}(t)) \\
	\mb{K}_I &=& \sum_j q_{I,j} \mb{v}_{I,j}(t)\delta^{(2)}(\mb{x}-\mb{r}_{I,j}(t))\ , 
\eeqa
where $\mb{K}_I= (K^x_I,K^y_I)$ and $\mb{v}_{I,j}(t) = \frac{d \mb{r}_{I,j}(t)}{dt}$. 
Since $\sin(\eta)=0$ at the core of vortices in $b_1$, and $\sin(\eta)=1$ at the core of vortices in $b_2$, the theta term 
reduces further to
\beq
	S_{\theta}[U]= \frac{1}{2\pi}\int d^3 x\ \pd_{\mu}\vth_1 K^{\mu}_2 \ .
\eeq

We now show that the theta term gives a phase of $e^{i\theta}$ in the real time path integral whenever a vortex (of
strength $q=1$) in the phase of $b_2$ makes a complete circuit around a vortex (also of strength $q=1$) in the phase of
$b_1$.  We take the vortex in $\vth_1$ to be located at $\mb{r}_1(t)$, and the vortex in $\vth_2$ to be located at 
$\mb{r}_2(t)$, and we restrict the time integration in the action to be on the interval $[0,T]$, where $T$ is the time it takes for 
the vortices to complete their circuit. From the form of the components of the vortex current shown above, we find the result
\beq
	\int d^3 x\ \pd_{\mu}\vth_1 K^{\mu}_2 = \int_0^T dt \frac{d}{dt}\vth_1(t,\mb{r}_2(t))\ ,
\eeq
where the integrand is the \emph{total} time derivative of $\vth_1(t,\mb{r}_2(t))$,
\beq
	\frac{d}{dt}\vth_1(t,\mb{r}_2(t))= \pd_t\vth_1(t,\mb{r}_2(t)) + \mb{v}_{2}(t)\cdot\nabla\vth_1(t,\mb{x})\bigg|_{\mb{x}= \mb{r}_2(t)}\ .
\eeq
Here the function $\vth_1(t,\mb{r}_2(t))$ is the phase of $b_1$ evaluated at the core of the vortex in $b_2$.

Now we integrate the total time derivative of $\vth_1(t,\mb{r}_2(t))$ from $t=0$ to $t=T$ to obtain
\beq
	S_{\theta}[U]= \frac{1}{2\pi}\left(   \vth_1(T,\mb{r}_2(T)) - \vth_1(0,\mb{r}_2(0))       \right)\ .
\eeq
Finally, since the core of the vortex in $b_2$ makes
one full circuit around the core of the vortex in $b_1$ as $t$ varies from $0$ to $T$, we have
$\vth_1(T,\mb{r}_2(T)) - \vth_1(0,\mb{r}_2(0))  = 2\pi$. We then get $S_{\theta}[U]= 1$, which means that 
in the real time path integral we get a phase $e^{i\theta S_{\theta}[U]} = e^{i\theta}$ for every field configuration
in which a vortex in $b_2$ makes a complete circuit around a vortex in $b_1$. More generally, if a vortex of 
strength $q_2$ in $b_2$ makes a complete circuit around a vortex of strength $q_1$ in $b_1$, we get a phase
of $e^{i q_1 q_2\theta}$.

The result obtained above can also be used to investigate the intrinsic angular momentum and statistics of 
the bound state of vortices in $b_1$ and $b_2$. In Ref.~\onlinecite{WilczekZee}, the authors calculated the intrinsic angular 
momentum $J$ of a soliton in the $O(3)$ NLSM with Hopf term by calculating the action corresponding to an adiabatic
rotation of the soliton by $2\pi$. If the time it takes to complete the rotation is $T$, then the action should evaluate to
$S= 2\pi J + O(\frac{1}{T})$. The topological term in the action is responsible for the value of $J$, and the terms of order
$\frac{1}{T}$ are produced by the other terms in the action. From our result above, we immediately see that
$2\pi J= \theta$ for the vortex bound state, so $J= \frac{\theta}{2\pi}$. At $\theta=\pi$ we get $J= \frac{1}{2}$, 
which means that the vortex bound state is a fermion.

Finally, a few words are in order about the conditions necessary to justify ignoring the boundary terms produced when
we integrated the theta term by parts. First, the boundary terms in the time direction can be neglected if the
field configurations at the initial and final time are chosen to be the same. This is the usual choice in field theory, where
the path integral represents a matrix element of the form $\lan \psi | e^{-iHT}|\psi\ran$, in which the time
evolution operator $e^{-iHT}$ is sandwiched between the same initial and final state $|\psi\ran$ 
(usually the vacuum, or ground state).

Now we discuss the spatial boundary terms.
One way to ensure that the spatial boundary terms vanish is to require the phases
$\vth_1$ and $\vth_2$ to tend to constants at spatial infinity. This means that these two phases cannot wind at spatial
infinity, which means that if vortices are present in $\vth_1$ or $\vth_2$, there must also be an equal number of 
anti-vortices present to completely cancel the winding of the phase at spatial infinity. In other words, the sum over 
configurations of the $O(4)$ field $U$ in the path integral should be restricted to include only those configurations which 
contain an equal number of vortices and anti-vortices in the phase of each boson $b_I$. 
This requirement makes physical sense since, as we saw in 
Appendix~\ref{app:vortices}, isolated vortices have some undesirable properties (their core size grew without bound
as the system size was taken to infinity).  As we discussed in Appendix~\ref{app:vortices}, 
vortex-anti-vortex pairs do not have this problem.

\section{Abanov-Wiegmann fermions and the relation to the $O(4)$ NLSM with theta term}
\label{app:fermion-puzzle}

We mentioned in Sec.~\ref{sec:BTI-surface} that the Abanov-Wiegmann method seems to be more closely connected to an $O(4)$ NLSM
in the ordered (small $g$) phase, whereas we are interested in studying the disordered (large $g$) phase of the model. 
Nevertheless, our response calculation using Abanov-Wiegmann fermions completely agrees with the response calculation
of Ref.~\onlinecite{VS2013} using the dual vortex theory (which we reviewed in Sec.~\ref{sec:BTI-surface}). In this
Appendix we use the Abanov-Wiegmann formula to argue that the topological part of the electromagnetic response of
the $O(4)$ NLSM with $\theta=\pi$ must be exactly equal to the topological part of the response of the theory of four
massless fermions $\psi_a$, where $\psi_a$ are the four Abanov-Wiegmann fermions which can be coupled to the $O(4)$
field to produce an $O(4)$ NLSM at $\theta=\pi$.

As discussed above, the Abanov-Wiegmann method cannot produce an $O(4)$ NLSM in the disordered, or large $g$ phase,
because the expansion in powers of $M^{-1}$ would not be reliable at such low orders if $M$ was taken to be small. Let us
instead consider a completely different scenario, in which we start out with a system containing 
bosonic \emph{and} fermionic degrees of freedom. 
The ingredients in this theory are (i) an $O(4)$ NLSM in the disordered phase with
a theta angle $\theta=-\pi$, and (ii) the four massless fermions $\psi_a$ introduced in the discussion of the
Abanov-Wiegmann method in Sec.~\ref{sec:BTI-surface}. The action for these two decoupled theories takes the
form $S= S_b + S_f$ with
\beq
	S_b= \int d^3 x\ \left[ \frac{1}{g}(\pd^{\mu}N^a)(\pd_{\mu}N^a) \right] + \pi S_{\theta}[\mb{N}]\ ,
\eeq
and 
\beq
	S_f=  \int d^3 x\ i\bar{\Psi}\slashed{\pd}\Psi\ , \label{eq:massless-fermions}
\eeq
where $\Psi= (\psi_1,\psi_2,\psi_3,\psi_4)^T$. We now turn on a strong interaction
between these two theories of the form
\beq 
  	S_{int}= -M\sum_{a=1}^4\int d^3 x\ \bar{\Psi} N^a\Gamma^a\Psi\ ,
\eeq
with $M>0$ and large (so the coupling is strong).

If we integrate out the fermions in this theory (using the Abanov-Wiegmann formula), 
then the theta term for the $O(4)$ NLSM will 
be canceled (recall that the original theta angle was $-\pi$), and we are left with the action
\beq
	S= \int d^3 x\ \frac{1}{\tilde{g}}(\pd^{\mu}N^a)(\pd_{\mu}N^a)\ ,
\eeq
where $\tilde{g}$ is very small, since $\tilde{g}^{-1} = g^{-1} + \frac{M}{\text{const.}}$.  
The result is an $O(4)$ NLSM with no theta term which is in its ordered phase. 
We see that strong coupling to the four massless fermions $\psi_a$ has
completely destroyed the topological properties of the original $O(4)$ NLSM with $\theta=-\pi$.

Our interpretation of this is as follows. The theory of four massless fermions in Eq.~\eqref{eq:massless-fermions} (in
which the fermions carry the charges $q_a$ calculated in Sec.~\ref{sec:BTI-surface})
should be regarded, in some sense, as the \emph{inverse} of the $O(4)$ NLSM with $\theta=-\pi$, since strong coupling
between the two theories completely destroys the topological properties of the latter theory. In particular, the topological
part of the electromagnetic responses of these two theories should have opposite signs. 
Now the $O(4)$ NLSM with $\theta=\pi$
is also, in this same sense, the inverse of the $O(4)$ NLSM with $\theta=-\pi$. To see this, suppose we had two $O(4)$ NLSM's 
with theta term, with fields $\mb{N}$ and $\mb{M}$, with the first copy having $\theta=\pi$ and the second copy having
$\theta=-\pi$. Then a strong dot product coupling of the form $\mb{N}\cdot\mb{M}$ between these two theories will have the 
effect of setting $\mb{N}=\pm \mb{M}$, which 
will in turn cause the theta terms for the two theories to cancel. We therefore conclude that the topological
part of the electromagnetic response of the fermion theory in Eq.~\eqref{eq:massless-fermions} should be exactly 
\emph{equal} to the topological part of the response of the $O(4)$ NLSM with $\theta=\pi$. This explains why we were able to 
calculate the electromagnetic response of the $O(4)$ NLSM with $\theta=\pi$ by instead coupling the fermion theory 
in Eq.~\eqref{eq:massless-fermions} to the external field $A_{\mu}$.

%\bibliography{BSMrefs}

\end{document}